\documentclass[onecolumn,superscriptaddress,aps]{revtex4-2}
\usepackage{graphicx}
\usepackage{tikz}
\usepackage{amsmath}
\usepackage{amsthm}
\usepackage{amssymb}
\usepackage{latexsym}
\usepackage{array}
\usepackage{hyperref}
\usepackage{float}
\usepackage{amsfonts}
\usepackage{dsfont}
\usepackage{mathrsfs}
\usepackage{verbatim}
\usepackage{bbold}
\usepackage{lipsum}
\usepackage[normalem]{ulem}

\usepackage{bm}
\usepackage{times}

\usepackage{upgreek}

\usepackage{makecell}
\usepackage{adjustbox,lipsum}

\usepackage{algorithm}
\usepackage{algpseudocode}

\usepackage{color}

\usepackage{xcolor}
\usepackage[normalem]{ulem}

\graphicspath{{figures/}{./}}

\newcommand{\bra}[1]{\left<#1\right|}
\newcommand{\ket}[1]{\left|#1\right>}
\newcommand{\abs}[1]{\bigl|#1\bigr|}
\newcommand{\norm}[1]{\left\lVert#1\right\rVert}
\newcommand{\expt}[1]{\langle #1 \rangle}
\newcommand{\braket}[2]{\left<{#1}|{#2}\right>}
\newcommand{\ketbra}[2]{\ket{#1}\!\!\bra{#2}}

\newtheorem{theorem}{Theorem}
\newtheorem{proposition}{Proposition}
\newtheorem{lemma}{Lemma}
\newtheorem{corollary}{Corollary}
\theoremstyle{definition}
\newtheorem{definition}{Definition}
\theoremstyle{plain}

\newtheorem{remark}{Remark}

\DeclareMathOperator{\tr}{Tr}

\makeatletter
\renewcommand\part[1]{%
  \clearpage
  \onecolumngrid
  \section*{#1}
  
}
\makeatother

\makeatletter
\let\origaddcontentsline\addcontentsline
\renewcommand{\addcontentsline}[3]{}
\makeatother

\begin{document}

\part{}

\title{Logical entanglement deviation as a geometric diagnostic for quantum error correction}

\author{Kyoungho~Cho}
\affiliation{Institute for Convergence Research and Education in Advanced Technology, Yonsei University, Seoul 03722, Republic of Korea}
\affiliation{Department of Statistics and Data Science, Yonsei University, Seoul 03722, Republic of Korea}

\author{Ilkwon~Sohn}\email{d2estiny@kisti.re.kr}
\affiliation{Quantum Network Research Center, Korea Institute of Science and Technology Information, Daejeon 34141, Korea}

\author{Yongsoo~Hwang}\email{yhwang@etri.re.kr}
\affiliation{Electronics and Telecommunications Research Institute, Daejeon 34129, Korea}

\author{Jeongho~Bang}\email{jbang@yonsei.ac.kr}
\affiliation{Institute for Convergence Research and Education in Advanced Technology, Yonsei University, Seoul 03722, Republic of Korea}
\affiliation{Department of Quantum Information, Yonsei University, Incheon 21983, Republic of Korea}

\date{\today}

\begin{abstract}
Code distance in quantum error correction (QEC) specifies how many qubit losses can always be corrected. It is fixed by the smallest uncorrectable pattern, offering limited guidance on larger loss geometries. A safely erasable region reveals nothing about the stored state, so its entanglement with the remaining qubits stays fixed across pure logical inputs. We quantify departures from this stability through logical entanglement deviation (Logical-ED), the standard deviation of regional linear entropy over uniformly sampled pure logical inputs. Its profiles define a soft distance, the largest size or geometric scale up to which every selected region meets a fluctuation tolerance. For exact stabilizer codes encoding one logical qubit, Logical-ED detects access to exactly one logical Pauli observable but can miss full logical access. For any finite-dimensional subspace encoding, zero deviation certifies exact erasure correction if the surviving subsystem implements a nonscalar logical unitary. This condition makes the zero-tolerance geometric soft distance an exact erasure radius for eligible loss families. For one logical qubit under this condition, three regional purities give a rank-dependent bound on residual logical information in the erased subsystem. We verify logical-ED scaling for boundary-anchored strips in patches with up to 225 data qubits. In a $[[121,1,11]]$ patch, a $100$-qubit corner erasure is exactly correctable, whereas an $11$-qubit spanning erasure precludes exact recovery.
\end{abstract}

\maketitle

\section*{Introduction}

Fault-tolerant quantum computation stores quantum information across many physical qubits so that faults can be detected and corrected without reading the stored information itself~\cite{shor1995scheme,steane1996error,knill1997theory}. In leading architectures such as the surface code and its variants, this protection depends on geometry: error-detection checks involve nearby qubits on a two-dimensional layout~\cite{kitaev2003fault,dennis2002topological,fowler2012surface,terhal2015quantum,google2025below}.

The central measure of protection is the distance $d$, defined as the smallest number of physical qubits on which an undetectable operation can change the encoded information. Operationally, an $[[n,k,d]]$ code corrects arbitrary errors on up to $\lfloor(d-1)/2\rfloor$ unknown locations and erasures on up to $d-1$ known locations. The distance, however, is intentionally coarse. It is a worst-case guarantee determined by the smallest such logical operation, and does not characterize which logical observables are accessible from physical regions of different shapes. Even at fixed code distance, the correctability of a region can depend on its geometry relative to the code boundaries and the supports of logical operators. Moreover, real devices have imperfect encoding and error detection and can exhibit spatially correlated faults~\cite{harrington2025synchronous}, so protection may need to be described quantitatively rather than through a sharp distinction between perfect correction and failure. This motivates a diagnostic that can be evaluated region by region, with a precise account of what it says about the ability of local measurements to access logical information.

Suppose a quantum processor reports the loss of a region $A$, such as a cluster of qubits near the edge of a code block. The immediate question is whether the logical state can be recovered from the qubits that remain. If fewer than $d$ qubits are lost, recovery is guaranteed. For a larger cluster, the answer depends on which qubits were lost. Judging by the count alone could prompt a designer to add qubits even when the existing code already corrects that loss pattern. What matters is whether the lost region carried any information about the logical input. Exact recovery is possible precisely when the reduced state on $A$ is the same for every logical input, a condition called decoupling. Then, can entanglement tell us which regions of the code block are safe to lose?

Our key idea to approach this question is to examine the stability of entanglement as the logical input changes. For a pure encoded state, a physical region and the rest of the code share entanglement. If the region is safe to lose, its reduced state cannot depend on the encoded information, and its entanglement with the rest is therefore fixed across logical inputs~\cite{cleve1999share}. This motivates examining how the entanglement between a physical region and its complement varies with the encoded logical state. To quantify this dependence, we introduce logical entanglement deviation (logical-ED) as the standard deviation of regional linear entropy over Haar-random pure logical inputs. This diagnostic captures how strongly the entanglement between a region and the rest depends on the encoded state. A nonzero deviation reveals a failure of perfect decoupling. Both the question and its resolution can follow the physical geometry. Regions chosen to represent boundary neighborhoods, narrow fault corridors, or device modules turn entanglement fluctuations into a map of where this input dependence emerges. Moreover, purity can be estimated from repeated preparations using finite-copy measurements or randomized methods, offering a route to characterization without reconstructing every reduced state~\cite{ekert2002direct,elben2023toolbox}. The resulting profile connects the local indistinguishability condition for quantum error correction (QEC) with the spatial structures encountered in quantum hardware.

\begin{figure*}[t]
\centering
\includegraphics[width=1.00\linewidth]{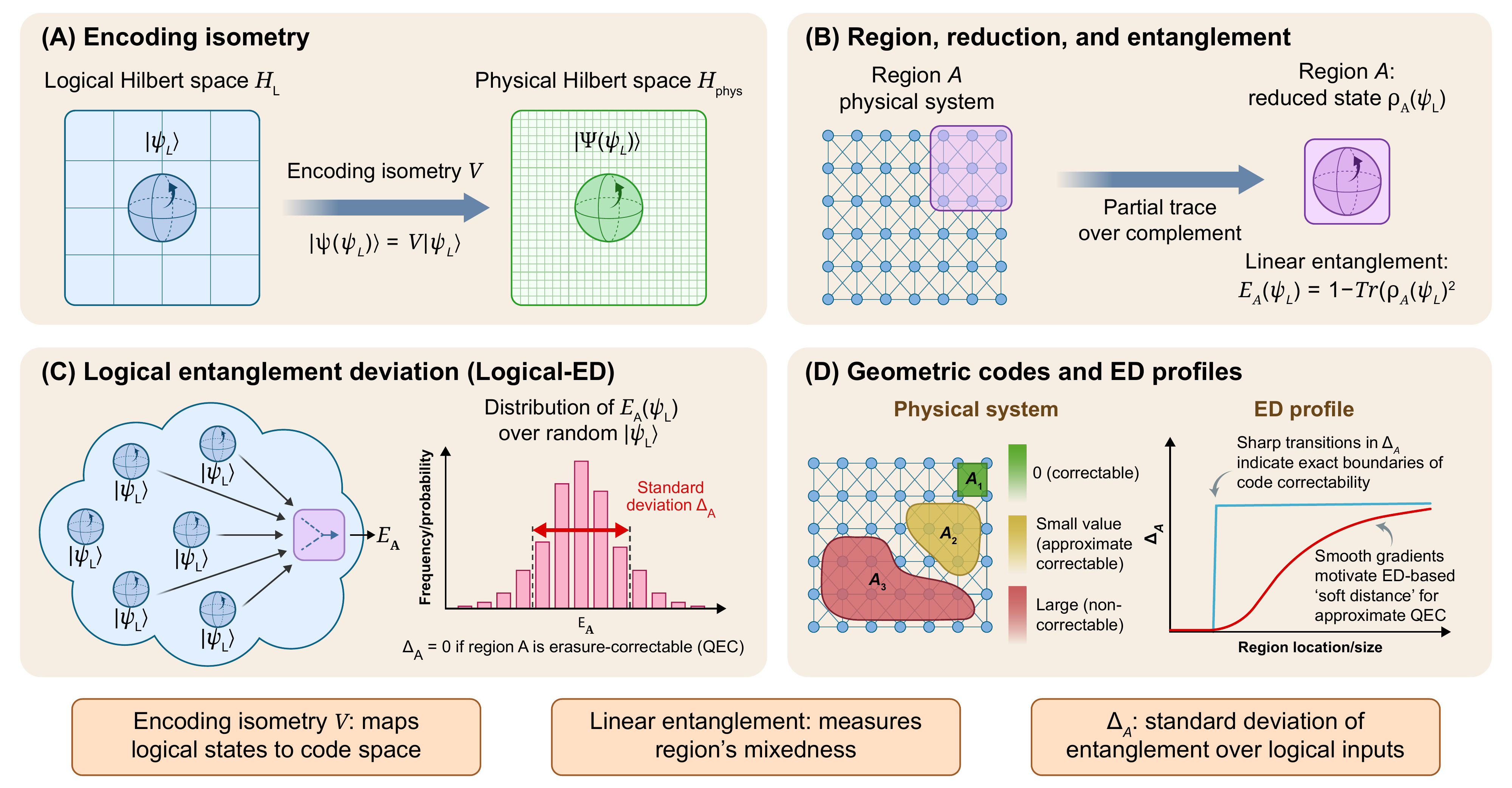}
\caption{\textbf{Logical entanglement deviation and correctability.} (a) An encoding isometry $\hat{V}:\mathcal{H}_L\to\mathcal{H}_{\rm phys}$ maps a logical input $\ket{\psi_L}$ to an encoded pure state $\ket{\Psi(\psi_L)}=\hat{V}\ket{\psi_L}$. (b) For a physical region $A$, the reduced state $\hat{\rho}_A(\psi_L)=\tr_{A^c}\bigl(\ketbra{\Psi(\psi_L)}{\Psi(\psi_L)}\bigr)$ defines the linear-entropy entanglement $E_A(\psi_L)=1-\tr{(\hat{\rho}_A(\psi_L)^2)}$. (c) Varying $\ket{\psi_L}$ over the logical state manifold generates an ``entanglement landscape'' (a distribution of $E_A$ values); its mean $\overline{E}_A$ and deviation $\Delta_A$ summarize typical entanglement and its input dependence. (d) Scanning $\Delta_A$ over a family of regions defines an ED profile. Its cumulative worst-case values yield the ED-soft distance $d_{\Delta}(\varepsilon)$ as a tolerance-dependent size radius (Definition~\ref{def:soft_distance}), or a geometric radius for a specified region family (Definition~\ref{def:geo_soft_distance}). If the surviving subsystem implements a nonscalar logical unitary for each candidate region, the zero-tolerance geometric radius certifies exact erasure correction.}
\label{fig:overview}
\end{figure*}

Our aim is to turn this spatial map into a guide to QEC design. The ED-soft distance summarizes the largest size up to which every region meets a chosen fluctuation tolerance; its geometric version follows the shapes relevant to a device. These profiles and radii help identify vulnerable regions and compare the geometric properties of different QEC code layouts under the same fault patterns. Turning this diagnostic into a correction certificate requires an additional safeguard. We show that for an arbitrary finite-dimensional subspace encoding, the diagnostic becomes exact whenever the surviving subsystem can implement a nonscalar logical unitary. Within a family satisfying this condition for every mask, zero deviation is equivalent to exact erasure correctability, and the zero-tolerance geometric soft distance gives the corresponding correction radius. A retained logical Pauli on a stabilizer code gives a concrete realization of this condition. The resulting design principle is to identify relevant fault families together with the logical operations that survive them, then use the entanglement profile to assess their protection. Surface-code calculations illustrate how logical access depends on the loss geometry, and an analytical syndrome criterion connects this dependence to the first possible ambiguity for Pauli errors. Fig.~\ref{fig:overview} summarizes the connection between entanglement profiles, fault geometry, and QEC design.

\section*{Results}\label{sec:results}

\subsection*{Definition of logical entanglement deviation}

We consider a code subspace $\mathcal{C}\subset\mathcal{H}_{\rm phys}$ with $\dim\mathcal{C}=d_L$, and an encoding isometry $\hat{V}:\mathcal{H}_L\to\mathcal{H}_{\rm phys}$ with $\hat{V}^\dagger \hat{V}=\mathds{1}_{L}$ and ${\rm Im}(\hat{V})=\mathcal{C}$. For a logical pure state $\ket{\psi_L}\in\mathcal{H}_L$, the encoded state is
\begin{eqnarray}
\ket{\Psi(\psi_L)} := \hat{V}\ket{\psi_L},
\quad
\hat{\rho}(\psi_L) := \ketbra{\Psi(\psi_L)}{\Psi(\psi_L)}.
\label{eq:results_encoded_state}
\end{eqnarray}
For a region $A$ of physical qudits (or qubits) and its complement $A^c$, let
\begin{eqnarray}
\hat{\rho}_A(\psi_L) := \tr_{A^c}\Bigl(\hat{\rho}(\psi_L)\Bigr).
\label{eq:results_rhoA}
\end{eqnarray}
We define the region entanglement functional by the linear entropy
\begin{eqnarray}
E_A(\psi_L) := 1 - \tr{\Bigl(\hat{\rho}_A(\psi_L)^2\Bigr)}.
\label{eq:results_linear_entropy}
\end{eqnarray}
Since $\hat{\rho}(\psi_L)$ is pure, $E_A(\psi_L)$ measures the bipartite entanglement between $A$ and $A^c$ (equivalently, it is a monotone function of the R\'{e}nyi-$2$ entropy).

Let $d\psi_L$ denote the normalized Haar-induced measure on pure states of $\mathcal{H}_L$. We define the logical entanglement mean and the logical entanglement deviation (logical-ED) by
\begin{eqnarray}
\overline{E}_A &:=& \int d\psi_L\, E_A(\psi_L),
\label{eq:results_Ebar_def}
\\
\Delta_A &:=& \sqrt{\int d\psi_L\, E_A(\psi_L)^2 - \overline{E}_A^{\,2}}.
\label{eq:results_Delta_def}
\end{eqnarray}
Thus, $\Delta_A$ quantifies how strongly the entanglement across the cut $A|A^c$ fluctuates when scanning over logical inputs. Importantly, $\Delta_A$ is a property of the pair (code, region): it depends on the subspace geometry and the region $A$, but not on a particular logical basis.

While $\overline{E}_A$ captures the \emph{typical} amount of bipartite entanglement generated by the encoding across the cut, $\Delta_A$ probes a genuinely different facet: the \emph{state dependence} of that entanglement. In particular, $\Delta_A$ is sensitive to logical directions whose reduced states have different purities, even when $\overline{E}_A$ remains large. For multi-logical systems ($k>1$), this deviation can be interpreted as a geometric analogue of spectral anisotropy in the code space: $\Delta_A$ is small when the purity across $A|A^c$ is nearly uniform over the logical manifold, and it is large when the code admits logical states with markedly different purities relative to $A$. Because the raw deviation inherits the regional-purity scale, $\Delta_A/\overline{P}_A$, where $\overline{P}_A:=1-\overline{E}_A$ is the mean purity, provides a relative fluctuation for comparing regions with different mean purities. We retain $\Delta_A$ as the primary, absolute measure; since $\overline{P}_A>0$, normalization does not change whether the deviation vanishes.

Two invariances are immediate. First, $\Delta_A$ is invariant under \emph{logical} unitaries: replacing $\hat{V}\to \hat{V}\hat{U}_L$ simply reparametrizes $\psi_L$ under Haar measure. Second, $\Delta_A$ is invariant under \emph{physical local} unitaries $\hat{U}_A\otimes\hat{U}_{A^c}$, since $E_A(\psi_L)$ depends only on the spectrum of $\hat{\rho}_A(\psi_L)$. In addition, purity of the encoded global state implies the pointwise complement identity
\begin{eqnarray}
E_A(\psi_L)=E_{A^c}(\psi_L),
\quad
\Delta_A=\Delta_{A^c}.
\label{eq:results_complement_symmetry}
\end{eqnarray}
The first equality follows because $\hat{\rho}_A(\psi_L)$ and $\hat{\rho}_{A^c}(\psi_L)$ have identical nonzero spectra. This symmetry is intrinsic to a spectral bipartite-entanglement diagnostic and also shows that $\Delta_A$ cannot, by itself, assign a direction to logical-information flow.

\subsection*{Correctable regions have vanishing logical-ED}

We now connect $\Delta_A$ to the defining QEC notion of correctability. For an erasure of region $A$, correctability means that there exists a recovery channel $\mathcal{R}$ acting on $A^c$ such that for all logical states $\hat{\sigma}_L$,
\begin{eqnarray}
\mathcal{R}\circ \tr_{A}\Bigl(\hat{V} \hat{\sigma}_L \hat{V}^\dagger\Bigr) \,=\, \hat{V} \hat{\sigma}_L \hat{V}^\dagger.
\label{eq:results_erasure_correctable_def}
\end{eqnarray}
Exact erasure correction is equivalent to a local indistinguishability condition on $A$: all reduced states $\hat{\rho}_A(\psi_L)$ are identical across the code space~\cite{knill1997theory,cleve1999share}. This immediately forces $E_A(\psi_L)$ to be constant, and hence the deviation $\Delta_A$ to vanish.

\begin{theorem}[Vanishing logical-ED on correctable regions]\label{thm:vanishing}
Let $\hat{V}:\mathcal{H}_L\to\mathcal{H}_{\rm phys}$ be an encoding isometry and let $A$ be a physical region. If the erasure of $A$ is exactly correctable on the code space ${\rm Im}(\hat{V})$, then there exists a fixed density operator $\hat{\omega}_A$ such that
\begin{eqnarray}
\hat{\rho}_A(\psi_L) = \hat{\omega}_A \quad \text{for all pure }\ket{\psi_L}\in\mathcal{H}_L.
\label{eq:thm_rhoA_constant}
\end{eqnarray}
Consequently, $E_A(\psi_L)$ is constant over $\psi_L$ and the logical-ED vanishes: $\Delta_A = 0$.
\end{theorem}

\begin{proof}[Proof sketch]---Let $\hat{P}:=\hat{V}\hat{V}^\dagger$ be the code projector. For the erasure channel $\tr_A$, the Kraus operators $\bra{a}_A\otimes\mathds{1}_{A^c}$ have pairwise products given by matrix units on $A$. Exact erasure correction then implies the Knill--Laflamme relations $\hat{P}(\ket{a}\bra{b}_A\otimes\mathds{1}_{A^c})\hat{P}=\alpha_{ab}\hat{P}$ for all $a,b$~\cite{knill1997theory}. By linearity, every observable $\hat{O}_A$ supported on $A$ acts as a scalar on the code space, $\hat{P}(\hat{O}_A\otimes \mathds{1}_{A^c})\hat{P}=c(\hat{O}_A)\hat{P}$. Hence all code states have identical expectation values for all operators supported on $A$, which forces their reduced density matrices on $A$ to coincide, yielding Eq.~(\ref{eq:thm_rhoA_constant}). Since $E_A(\psi_L)$ depends only on $\hat{\rho}_A(\psi_L)$, it is constant and $\Delta_A=0$. A complete proof is given in Sec.~II of the Supplementary Information.
\end{proof}

Theorem~\ref{thm:vanishing} can be read as a sharp one-way ``no-leakage'' statement: when $A$ is correctable, the discarded subsystem is forced into a fixed state $\hat{\omega}_A$ that is independent of the logical input. In other words, the erasure channel's complementary output carries zero information about the code, and $\Delta_A$ vanishes as an entanglement-level consequence of the Knill--Laflamme decoupling mechanism. Complementary-channel formulations likewise underlie approximate QEC~\cite{kretschmann2008information,beny2010general}. Here a controlled approximate version holds in the same direction. Define the channel from the logical system to $A$ and a constant channel by
\begin{eqnarray}
\mathcal{N}_A(\hat{\sigma}_L)&:=&\tr_{A^c}\!\left(\hat{V}\hat{\sigma}_L\hat{V}^{\dagger}\right), \nonumber \\
\mathcal{C}_{\omega_A}(\hat{\sigma}_L)&:=&\tr{(\hat{\sigma}_L)}\hat{\omega}_A, \nonumber \\
\eta_A&:=&\inf_{\hat{\omega}_A}\norm{\mathcal{N}_A-\mathcal{C}_{\omega_A}}_{\diamond}.
\label{eq:results_decoupling_distance}
\end{eqnarray}
Supplementary Note~S6 proves the dimension-independent bound
\begin{eqnarray}
\Delta_A\le \eta_A.
\label{eq:results_decoupling_to_ED}
\end{eqnarray}
Thus approximate decoupling controls logical-ED; equivalently, an observed $\Delta_A$ lower-bounds the decoupling error as $\eta_A\ge\Delta_A$. The reverse type of estimate---an upper bound on $\eta_A$ from small logical-ED---is false in general, so small logical-ED alone is not an operational recovery guarantee.

Theorem~\ref{thm:vanishing} yields an immediate distance-based corollary. For an $[[n,k,d]]$ code, erasure of any $\abs{A}\le d-1$ qubits is correctable (since the erasure location is known), hence $\Delta_A=0$ for all such regions. This shows that logical-ED is consistent with the standard distance picture, while also opening a new direction: beyond the ``hard'' threshold $\abs{A}=d-1$, the values of $\Delta_A$ for larger or topologically nontrivial regions encode additional geometry-dependent structure not captured by $d$ alone.

\begin{corollary}[Distance-implied vanishing]\label{cor:distance}
For any exact $[[n,k,d]]$ code and any region $A$ with $\abs{A}\le d-1$, we have $\Delta_A=0$.
\end{corollary}

\begin{proof}[Proof sketch]---An $[[n,k,d]]$ code corrects erasures of size up to $d-1$ because the location is known and the distance-$d$ condition implies the Knill--Laflamme relations for any operator supported on such a region. Applying Theorem~\ref{thm:vanishing} gives $\Delta_A=0$.
\end{proof}

\begin{remark}\label{rem:nonconverse}
The converse does \emph{not} hold in general: $\Delta_A=0$ only implies that the purity $\tr{(\hat{\rho}_A(\psi_L)^2)}$ is constant over $\psi_L$, but does not guarantee that $\hat{\rho}_A(\psi_L)$ is input independent. For example, the isometry $\hat{V}\ket{\psi_L}=\ket{\psi_L}_A\ket{0}_{A^c}$ gives $E_A(\psi_L)=0$ and $\Delta_A=0$ although $A$ contains the entire logical state. Together with Eq.~(\ref{eq:results_complement_symmetry}), this shows that logical-ED is a spectral-anisotropy witness, not a complete leakage or correctability measure. The exact limitations and the one-way approximate bound~(\ref{eq:results_decoupling_to_ED}) are proved in Supplementary Notes~S3 and S6.
\end{remark}

\subsection*{Soft distance and ED profiles for geometric codes}

The integer distance $d$ induces a sharp partition: regions below the correctability threshold have no logical information, while regions above may. Logical-ED suggests a tolerance-dependent refinement of purity uniformity. 

\begin{definition}[ED-soft distance]\label{def:soft_distance}
For a code on $n$ physical qubits and a tolerance $\varepsilon\ge 0$, the \emph{ED-soft distance} is
\begin{eqnarray}
\begin{gathered}
d_{\Delta}(\varepsilon) := \max\Bigl\{ s\in\{0,\ldots,n\}: \forall A\ \bigl(\abs{A}\le s\bigr)\Rightarrow \Delta_A\le \varepsilon\Bigr\}.
\end{gathered}
\label{eq:results_soft_distance}
\end{eqnarray}
It is the largest number of qubits up to which every region has logical-ED at most $\varepsilon$.
\end{definition}

This is a ``purity-uniformity radius'' in the space of regions. Equivalently, the cumulative worst-case profile $\Delta_{\max}(s):=\max_{\abs{A}\le s}\Delta_A$ determines the radius as the largest admissible integer $s$ with $\Delta_{\max}(s)\le\varepsilon$. This cumulative profile is nondecreasing even when the ED of individual regions is not monotone in their size. Here, we include the empty region only as a zero-size baseline: $\Delta_{\emptyset}=0$ makes $s=0$ always admissible, so the maximum remains defined even when no positive size satisfies the cumulative condition. The finite domain ensures saturation at $n$ when every region lies within the tolerance. The next proposition makes the connection to the standard distance explicit.

\begin{proposition}[Distance-implied lower bound on the ED-soft distance]\label{prop:soft_distance_lb}
For any exact $[[n,k,d]]$ code and any tolerance $\varepsilon\ge 0$, the ED-soft distance of Definition~\ref{def:soft_distance} satisfies
\begin{eqnarray}
d_{\Delta}(\varepsilon)\ \ge\ d-1.
\label{eq:prop_soft_distance_lb}
\end{eqnarray}
In particular, $d_{\Delta}(0)\ge d-1$. Moreover, $d_{\Delta}(\varepsilon)$ is nondecreasing in $\varepsilon$.
\end{proposition}

\begin{proof}[Proof sketch]---By Corollary~\ref{cor:distance}, every region $A$ with $\abs{A}\le d-1$ obeys $\Delta_A=0$, hence $\Delta_A\le\varepsilon$ for any $\varepsilon\ge 0$. Therefore $s=d-1$ is feasible in the maximization in Definition~\ref{def:soft_distance}, which gives $d_{\Delta}(\varepsilon)\ge d-1$. Monotonicity in $\varepsilon$ is immediate since enlarging the tolerance can only enlarge the feasible set. A complete discussion (including possible non-saturation of $d_{\Delta}(0)$ due to purity-only diagnostics) is given in Supplementary Note~S3.
\end{proof}

For an encoding isometry perturbed by coherent control errors, or an isometric encoding that provides only approximate error correction, $d_{\Delta}(\varepsilon)$ in Definition~\ref{def:soft_distance} remains well defined and measures the largest size up to which all regions satisfy the chosen purity-uniformity tolerance. Here, a coherent error changes the implemented encoding, as in $\hat{V}^{\prime}=\hat{U}_{\rm err}\hat{V}$ with unitary $\hat{U}_{\rm err}$, while preserving its isometric character; general dissipative noise is not included in this description. The diagnostic $\Delta_A$ and tolerance $\varepsilon$ are real valued, whereas $d_{\Delta}(\varepsilon)$ is an integer-valued, nondecreasing staircase; ``soft'' refers to the tunable tolerance. A larger value records a wider window of purity uniformity at that tolerance. Moreover, the implication runs from approximate decoupling to small logical-ED [Eq.~(\ref{eq:results_decoupling_to_ED})], not conversely. Accordingly, $d_{\Delta}(\varepsilon)$ should not be read as an operational recovery distance without an additional decoupling or distinguishability bound. For exact one-logical-qubit stabilizer encodings, Supplementary Notes~S3 and S7 give a surface-code corridor example in which the geometric soft distance certifies the largest correctable scale within that loss family, using a retained logical axis and a tolerance below a verified ED gap (Theorem~\ref{thm:protected_axis}).

In geometric architectures this uniformity viewpoint remains useful because relevant logical supports are tied to specific shapes rather than arbitrary subsets. For instance, in the planar surface code, narrow \emph{corridors} can connect appropriate boundary components and support a logical Pauli. By adapting the region constraint in Definition~\ref{def:soft_distance} to strip-like or annular families, one obtains geometry-adapted soft distances $\ell_{\Delta}(\varepsilon)$ that track the onset of one-axis, boundary-to-boundary or hole-encircling spectral anisotropy within the chosen family.

For geometric codes on a lattice, it is often more meaningful to parametrize regions by geometric features (diameter, width, connectivity) rather than by cardinality alone. 

\begin{definition}[Geometric ED-soft distance]\label{def:geo_soft_distance}
Let $\mathcal{L}\subset\mathbb{R}_{\ge0}$ be a finite set of numerical geometric scales containing $0$, ordered by their usual numerical values. A scale $\ell$ may represent, for example, a strip width or patch diameter measured in lattice units. Let $\mathfrak{A}(\ell)$ be a family of physical regions satisfying $\mathfrak{A}(0)=\{\emptyset\}$ and $\mathfrak{A}(\ell_1)\subseteq\mathfrak{A}(\ell_2)$ whenever $\ell_1\le\ell_2$. For a tolerance $\varepsilon\ge0$, the \emph{geometric ED-soft distance} is
\begin{eqnarray}
\ell_{\Delta}(\varepsilon)
:= \max\Bigl\{ \ell\in\mathcal{L}:\ \forall A\in \mathfrak{A}(\ell)\Rightarrow \Delta_A\le \varepsilon\Bigr\}.
\label{eq:results_geo_soft_distance}
\end{eqnarray}
It is the largest geometric scale for which every region in the selected cumulative family has logical-ED at most $\varepsilon$.
\end{definition}

Unlike the all-region size radius in Definition~\ref{def:soft_distance}, Definition~\ref{def:geo_soft_distance} depends on the selected region family. In the planar surface code, for example, a region supporting no nontrivial logical Pauli operator is erasure-correctable, hence $\Delta_A=0$ by Theorem~\ref{thm:vanishing}. In contrast, regions that support exactly one nontrivial logical Pauli class---such as a single boundary-to-boundary strip in the planar code---can exhibit nonzero $\Delta_A$, reflecting logical-state-dependent reduced-state purity.

This observation motivates working not only with a single number but with region-family statistics. Given a family $\mathfrak{A}$ (e.g., connected balls of diameter $\le \ell$, boundary neighborhoods of depth $\ell$, or strips of width $w$ connecting boundary components), one may consider the worst-case fluctuation within that family, $\Delta^{(\mathfrak{A})}_{\max}:=\max_{A\in\mathfrak{A}}\Delta_A$. Tracking $\Delta^{(\mathfrak{A})}_{\max}$ as a function of a geometric parameter provides a compact ``profile'' that is tuned to the failure modes of the architecture: strip families are natural for detecting boundary-to-boundary logical strings, while annular families are natural for hole-encircling loop operators in punctured surface codes. Although our main theorems are stated for arbitrary regions, this family viewpoint is where logical-ED is expected to be most useful in practice, because it lets one translate qualitative geometric intuition into a quantitative, scan-able diagnostic.

This leads to a central conceptual object: the \emph{ED profile} of a code, defined by scanning $\Delta_A$ over a chosen set of regions (patches, strips, annuli, boundary neighborhoods). Whereas $d$ is determined by a single extremal logical operator, the ED profile resolves where logical-state-dependent spectral anisotropy occurs as region geometry changes. Once $\abs{A}\ge d$, distance alone does not distinguish whether such anisotropy is widespread or confined to specific shapes and boundary connections; the ED profile supplies this narrower, complementary information. The ``Exact soft-distance examples'' subsection of Supplementary Note~S3 gives a complete $512$-region calculation for the rotated $[[9,1,3]]$ patch, yielding $d_\Delta(\varepsilon)=2$ for $\varepsilon<1/(2\sqrt{45})$ and $d_\Delta(\varepsilon)=9$ at or above this tolerance. A complementary $d=11$ corridor example extracts a family-restricted geometric radius from the existing ED profile. These examples make the discrete staircases in Definitions~\ref{def:soft_distance} and~\ref{def:geo_soft_distance}, and their purity-uniformity interpretation, explicit. The operational erasure subsection below shows when the corridor radius also certifies correction of the specified loss patterns.

\subsection*{Numerical simulation: geometry-resolved logical-ED in rotated surface codes}

We next evaluate logical-ED in a code-level setting that is large enough to be inaccessible to direct state-vector simulation. We use odd-distance rotated planar surface-code patches with parameters $[[d^2,1,d]]$ and analyze $d=5,7,9,11,13,15$~\cite{dennis2002topological,fowler2012surface}. The calculation is exact and uses only the binary stabilizer representation~\cite{gottesman1997stabilizer}, building on the supported-subgroup description of stabilizer entanglement~\cite{fattal2004entanglement,zhao2026graph}.

For an $[[n,1]]$ stabilizer code with stabilizer group $\mathcal{S}$, choose logical Pauli representatives $\overline{L}_{\alpha}$ for $\alpha\in\{x,y,z\}$ and set $\overline{L}_0=\mathds{1}$. For a region $A$, define the coset-support counts $N_{\alpha}(A):=\abs{\{g\in\mathcal{S}:\operatorname{supp}(g\overline{L}_{\alpha})\subseteq A\}}$.
Writing a logical input by its Bloch vector $\bm{r}=(r_x,r_y,r_z)$, Pauli orthogonality gives
\begin{eqnarray}
\tr{\bigl(\hat{\rho}_A(\bm{r})^2\bigr)} = 2^{-\abs{A}}\left[N_0(A)+N_x(A)r_x^2+N_y(A)r_y^2+N_z(A)r_z^2\right].
\label{eq:surface_purity_counts}
\end{eqnarray}
Using $\mathbb{E}(r_\alpha^2)=1/3$, $\mathbb{E}(r_\alpha^4)=1/5$, and $\mathbb{E}(r_\alpha^2r_\beta^2)=1/15$ for $\alpha\neq\beta$, we obtain
\begin{eqnarray}
\Delta_A = 2^{-\abs{A}}\sqrt{\frac{2}{45}\left[3\sum_{\alpha=x,y,z}N_\alpha(A)^2 - \left(\sum_{\alpha=x,y,z}N_\alpha(A)\right)^2\right]}.
\label{eq:surface_delta_counts}
\end{eqnarray}
The counts require no enumeration of $2^{n-1}$ stabilizers. If $t_A$ is the dimension of the stabilizer subgroup supported entirely in $A$, then $N_0(A)=2^{t_A}$; each $N_\alpha(A)$ is either $0$ or $N_0(A)$, determined by one affine row-space test over $\mathbb{F}_2$. Denoting by $q_A$ the number of supported nontrivial logical cosets, closure under Pauli multiplication gives $q_A\in\{0,1,3\}$, and Eq.~(\ref{eq:surface_delta_counts}) reduces to
\begin{eqnarray}
\Delta_A=
\begin{cases}
\displaystyle \frac{2^{t_A-\abs{A}+1}}{\sqrt{45}}, & q_A=1,\\[2mm]
0, & q_A=0\ \text{or}\ 3.
\end{cases}
\label{eq:surface_delta_q}
\end{eqnarray}

\begin{proposition}[Exact stabilizer logical-access trichotomy]\label{prop:stabilizer_trichotomy}
For an $[[n,1]]$ stabilizer code and a region $A$, exactly one of the following holds:
\begin{enumerate}
\item[(i)] $q_A=0$: $A$ supports no nontrivial logical Pauli, the erasure of $A$ is correctable, and $\Delta_A=0$;
\item[(ii)] $q_A=1$: $A$ supports exactly one logical Pauli axis, the erasure of $A$ is not correctable, and $\Delta_A>0$;
\item[(iii)] $q_A=3$: $A$ supports the full logical Pauli algebra, the erasure of $A^c$ is correctable (so the logical state is recoverable from $A$), yet $\Delta_A=0$.
\end{enumerate}
Thus, within this setting, logical-ED detects precisely the intermediate one-axis logical-access regime.
\end{proposition}

\begin{proof}[Proof sketch]---Modulo stabilizers and phases, the logical Pauli group of one encoded qubit is the vector space $\mathbb{F}_2^2$. Logical classes admitting representatives in $A$ form a subgroup, whose possible orders are $1$, $2$, and $4$, giving $q_A=0,1,3$. The stabilizer cleanability criterion identifies $q_A=0$ with exact correctability of the erasure of $A$~\cite{bravyi2009nogo}. If $q_A=3$, representatives of two anticommuting logical generators lie in $A$; any nontrivial logical Pauli supported in the disjoint region $A^c$ would commute with both, which is impossible for a one-qubit Pauli. Hence $q_{A^c}=0$, so erasure of $A^c$ is correctable. The values of $\Delta_A$ follow from Eq.~(\ref{eq:surface_delta_q}). A complete proof is given in Supplementary Note~S5.
\end{proof}

We call a region \emph{ED-active} when $\Delta_A>0$. For the exact $[[n,1]]$ stabilizer codes considered here, this is equivalent to $q_A=1$ by Proposition~\ref{prop:stabilizer_trichotomy}, rather than to $q_A>0$.

\begin{figure*}[t]
\centering
\includegraphics[width=0.95\linewidth]{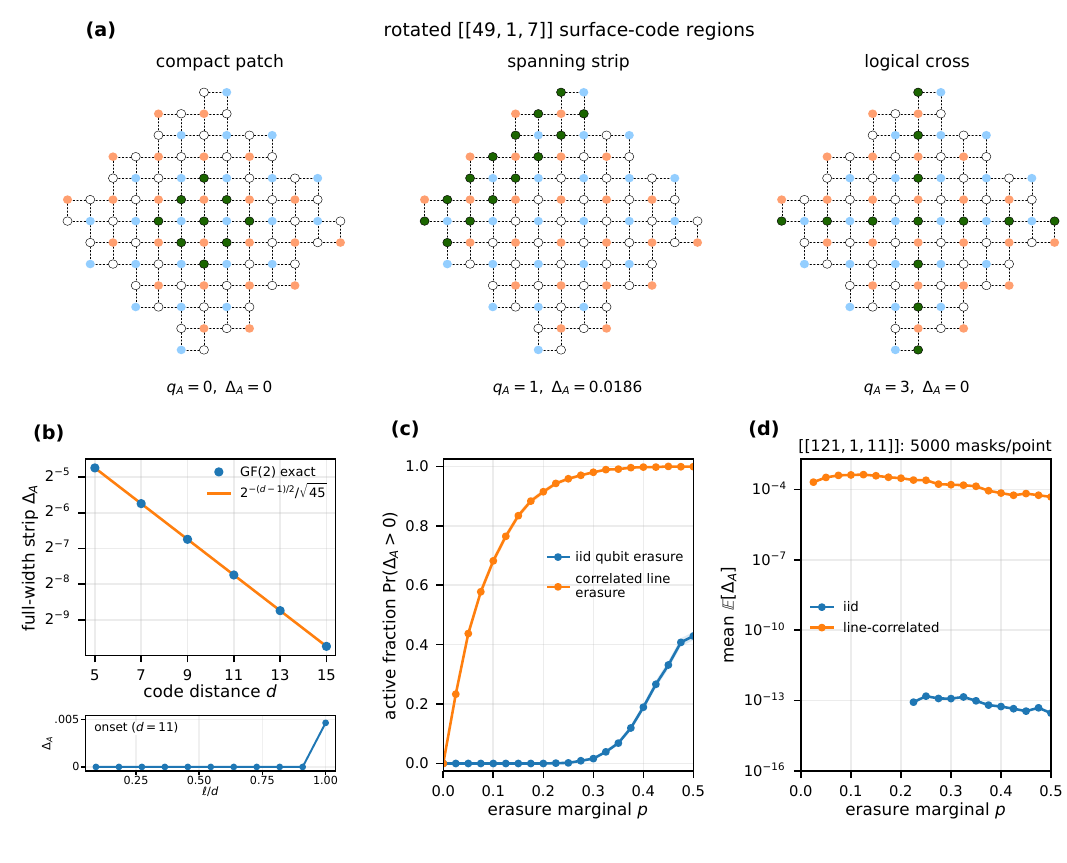}
\caption{\textbf{Large-scale logical-ED simulation for rotated planar surface codes.} \textbf{(a)} Three regions in a $[[49,1,7]]$ patch illustrate the exact coset classification. A compact patch supports no nontrivial logical coset ($q_A=0$), a boundary-spanning horizontal strip supports one logical coset ($q_A=1$), and a logical cross supports all three Pauli cosets ($q_A=3$). Eq.~(\ref{eq:surface_delta_q}) gives nonzero logical-ED only in the one-axis case; the cross is the expected purity-only blind spot. \textbf{(b)} Exact ED for boundary-anchored, full-width strips at distances $d=5,7,9,11,13,15$. The numerical points coincide with the closed form $2^{-(d-1)/2}/\sqrt{45}$ and are independent of strip depth for $1\le w<d$. The inset shows a width-$3$ boundary-anchored corridor in the $d=11$ patch: $\Delta_A$ turns on only when its length reaches the opposite boundary. \textbf{(c)} Fraction of ED-active masks, defined by $\Delta_A>0$ (equivalently $q_A=1$ here), $\Pr(\Delta_A>0)$, for independent and line-correlated erasures on the $[[121,1,11]]$ patch. This is not the total uncorrectable-erasure fraction $\Pr(q_A>0)$. Shaded bands are 95\% Wilson intervals. \textbf{(d)} Empirical mean $\mathbb{E}[\Delta_A]$ for the same masks on a logarithmic scale. Each point in panels (c,d) uses $5000$ independently generated masks. The line-correlated model has the same single-qubit marginal $p$ as the independent model but preferentially creates boundary-spanning logical corridors.}
\label{fig:sim}
\end{figure*}

Compact nonspanning patches realize case (i), a corridor supporting one logical string realizes case (ii), and a region supporting both independent logical axes realizes case (iii). The logical cross in Fig.~\ref{fig:sim}(a) is therefore an exact instance of the purity-only limitation emphasized in Remark~\ref{rem:nonconverse}.

For a full-width horizontal strip of depth $1\le w<d$ anchored to the top or bottom boundary, the rank calculation gives $t_A=dw-(d+1)/2$ and $q_A=1$, so that
\begin{eqnarray}
\Delta_{\rm strip}(d)=\frac{2^{-(d-1)/2}}{\sqrt{45}},
\label{eq:surface_strip_scaling}
\end{eqnarray}
independent of the strip depth. The raw deviation inherits the exponentially small regional-purity scale; nevertheless, the relative purity fluctuation is distance-independent, $\Delta_{\rm strip}/(1-\overline{E}_{\rm strip})=1/\sqrt{20}$. Consequently, comparisons based on a fixed absolute tolerance $\varepsilon$ should control for region size or be supplemented by this relative-purity normalization. Fig.~\ref{fig:sim}(b) verifies the boundary-law scaling through $d=15$ ($225$ data qubits); its inset shows the sharp onset for a boundary-anchored corridor, whose ED remains zero until the corridor first spans the patch.

To probe a hardware-relevant setting for known-location losses~\cite{zhang2026logical}, we then interpret $A$ as a static erased-qubit mask on the $[[121,1,11]]$ patch. We compare independent qubit erasure with probability $p$ to a common-line model: a horizontal or vertical orientation is chosen with equal probability and each complete line is erased with probability $p$. The two ensembles have the same single-qubit marginal $p$. They differ both in spatial correlations and in the distribution of the total number of erasures. In the line model the number $K$ of erased lines is binomial, and ED is active exactly for $0<K<d$, giving the closed form $\Pr_{\rm line}(\Delta_A>0)=1-(1-p)^d-p^d$. For $d=11$, this exact expression gives ED-active probabilities of $0.104662$ and $0.686189$ at $p=0.01$ and $p=0.10$, respectively. At $p=0.01$, independent erasures instead obey the distance-implied upper bound $\Pr_{\rm ind}(\Delta_A>0)\le\Pr[\operatorname{Bin}(121,0.01)\ge11]<4.66\times10^{-8}$, because every mask with at most $10$ erased qubits is correctable; here $\operatorname{Bin}(121,0.01)$ is the number of independently erased qubits. The two models give sharply different active probabilities even at this low marginal erasure rate. At $p=0.10$, the $5000$-mask samples give ED-active fractions of $0$ for independent erasures (95\% Wilson upper bound $7.68\times10^{-4}$) and $0.6822$ for line-correlated erasures. The $p=0.01$ comparison uses exact probabilities and an analytic upper bound. By Proposition~\ref{prop:stabilizer_trichotomy}, ED-active masks are exactly the $q_A=1$ masks. Therefore $\Pr(\Delta_A>0)$ is neither the probability $\Pr(q_A>0)$ of an uncorrectable erasure nor a decoder-failure probability, because it excludes the $q_A=3$ sector. The simulations compare one-axis logical access across two static erasure ensembles. Additional code distances, Monte Carlo uncertainties, validation, numerical tables, and algorithms are provided in the Supplementary Information.

\subsection*{From retained logical unitaries to erasure certificates.}

The stabilizer trichotomy suggests a way to remove the blind spot of logical-ED. The retained operation can be more general than a logical Pauli. Let $B=A^c$ be the surviving subsystem. A logical unitary $\hat U_L$ is retained if a unitary $\hat U_B$ acting only on $B$ realizes it exactly on the code.

\begin{theorem}[Erasure certificate from a retained logical unitary]\label{thm:protected_axis}
Let $\hat V:\mathcal H_L\to\mathcal H_A\otimes\mathcal H_B$ be a finite-dimensional encoding isometry. Suppose there exist a unitary $\hat U_B$ and a nonscalar unitary $\hat U_L$ such that
\begin{eqnarray}
(\mathds{1}_A\otimes\hat U_B)\hat V=\hat V\hat U_L.
\label{eq:results_protected_axis_guard}
\end{eqnarray}
Then
\begin{eqnarray}
\Delta_A=0 \quad\Longleftrightarrow\quad \text{erasure of } A\text{ is exactly correctable}.
\label{eq:results_protected_axis_certificate}
\end{eqnarray}
If this condition holds for every region in a nested cumulative family $\mathfrak A(\ell)$ as in Definition~\ref{def:geo_soft_distance}, then $\ell_\Delta(0)=\ell_{\rm er}$, where
\begin{eqnarray}
\ell_{\rm er}:=\max\{\ell\in\mathcal L:\text{every }A\in\mathfrak A(\ell)\text{ is erasure-correctable}\}.
\label{eq:results_family_erasure_radius}
\end{eqnarray}
The implementing unitaries may depend on the region.
\end{theorem}

\begin{proof}[Proof sketch]---Since the partial trace over $B$ is invariant under $\hat U_B$, $\mathcal N_A$ is invariant under conjugation of its input by $\hat U_L$. It therefore kills coherences between distinct eigenspaces of $\hat U_L$. If $\Delta_A=0$, continuity and the full support of Haar measure imply that every pure logical input has the same output purity. For unit vectors $\ket{u}$ and $\ket{v}$ in distinct eigenspaces, the equal superposition produces the average of their two outputs. The purity of that average is smaller than the average purity by one quarter of the squared Hilbert--Schmidt distance between the two outputs. Constant purity forces the two outputs to coincide. Comparing every unit vector in an eigenspace with a vector in another eigenspace then shows that all pure inputs within eigenspaces have the same output. Because cross-eigenspace coherences vanish, the same holds for every logical input. Exact erasure correction follows from local indistinguishability, and the reverse implication is Theorem~\ref{thm:vanishing}. Applying the equivalence to each candidate region gives the radius identity. Supplementary Note~S7 gives the complete proof.
\end{proof}

For a single logical qubit, each eigenspace of a nonscalar $\hat U_L$ is one-dimensional. Three input purities then give a quantitative bound on decoupling.

\begin{corollary}[Decoupling bound from three purities]\label{cor:three_purity_decoupling}
Assume Eq.~(\ref{eq:results_protected_axis_guard}) for one logical qubit. Let $\ket{0}$ and $\ket{1}$ be the logical eigenstates of $\hat U_L$, and let $\ket{+}=(\ket{0}+\ket{1})/\sqrt{2}$. Write $P_j=\tr(\hat\rho_j^2)$ for $j=0,1,+$, where $\hat\rho_j=\mathcal N_A(\ketbra{j}{j})$. Define
\begin{eqnarray}
G_A:=\frac{P_0+P_1}{2}-P_+
=\frac14\tr\bigl[(\hat\rho_0-\hat\rho_1)^2\bigr].
\label{eq:results_three_purity_gap}
\end{eqnarray}
With $r_A=\operatorname{rank}(\hat\rho_0-\hat\rho_1)$, the decoupling error in Eq.~(\ref{eq:results_decoupling_distance}) satisfies
\begin{eqnarray}
\eta_A=\frac12\norm{\hat\rho_0-\hat\rho_1}_1
\le\sqrt{r_A G_A}.
\label{eq:results_three_purity_bound}
\end{eqnarray}
In particular, $G_A=0$ is equivalent to exact erasure correctability.
\end{corollary}

The midpoint of the two eigenstate outputs attains the minimum over constant channels in Eq.~(\ref{eq:results_decoupling_distance}). Supplementary Note~S7 gives the proof, including inputs entangled with a reference, and an analytic nonstabilizer example. For measured purities, an upper confidence bound $G_A^{\rm U}$ on $G_A$ and a known rank bound $R_A\ge r_A$ give $\eta_A\le\sqrt{R_A G_A^{\rm U}}$ at the same confidence level. The dimension $\dim\mathcal H_A$ always gives a valid choice of $R_A$, though this can make the estimate loose.

A simple sufficient condition in an $[[n,1]]$ stabilizer code is that one physical representative $\hat{\overline P}$ of a nontrivial logical Pauli be supported outside every candidate region. We call this a protected-axis family. Any logical Pauli supported in $A$ then commutes with the retained representative. Thus $q_A\in\{0,1\}$, and the full-access sector that is invisible to logical-ED is excluded.

For the $[[121,1,11]]$ patch, consider $A_{\ell,h}=\{(x,y):0\le x<\ell,\ 0\le y<h\}$. The horizontal logical $\hat{\overline{Z}}$ on row $y=10$ survives every width-$3$ corridor $A_{\ell,3}$. For $\ell\le10$, the surviving column $x=10$ also supports $\hat{\overline{X}}$, so these regions are correctable; at $\ell=11$, the erased corridor contains a horizontal logical $\hat{\overline{Z}}$ and has $q_A=1$. Thus $\mathfrak{A}(\ell)=\{A_{r,3}:0\le r\le\ell\}$ has $\ell_\Delta(0)=\ell_{\rm er}=10$. In site counts, the $30$-qubit region $A_{10,3}$ is exactly correctable, whereas the $33$-qubit region $A_{11,3}$ is not. The contrast is sharper for $A_{10,10}$ and $A_{11,1}$: $100$ erased qubits are correctable in the former pattern, while $11$ are not in the latter. These pattern-specific statements supplement the worst-case guarantee of $10$ known erasures; they do not increase the arbitrary unknown-location error radius.

The logical-access sectors also have an exact recovery interpretation. Write $B=A^c$ and $\mathcal{N}_B(\hat{\rho}_L)=\tr_A(\hat{V}\hat{\rho}_L\hat{V}^\dagger)$, and let $\hat{\Phi}_{RL}=\ketbra{\Phi}{\Phi}$ be a normalized maximally entangled state of the logical qubit and a reference. For the maximally mixed logical input, use the Bell-overlap convention
\begin{eqnarray}
F_e^\star(A):=\max_{\mathcal{R}_{B\to L}} \bra{\Phi} \bigl({\rm id}_R\otimes(\mathcal{R}_{B\to L}\circ\mathcal{N}_B)\bigr)(\hat{\Phi}_{RL})
\ket{\Phi},
\label{eq:results_optimal_erasure_fidelity_def}
\end{eqnarray}
where the recovery is an unrestricted channel and the erased mask is known.

\begin{proposition}[Optimal entanglement fidelity for a known erasure]\label{prop:optimal_erasure_fidelity}
For an exact $[[n,1]]$ stabilizer code,
\begin{eqnarray}
F_e^\star(A)=
\begin{cases}
1,&q_A=0,\\
1/2,&q_A=1,\\
1/4,&q_A=3.
\end{cases}
\label{eq:results_optimal_erasure_fidelity}
\end{eqnarray}
\end{proposition}

\begin{proof}[Proof sketch]---The dimensions $\kappa_C:=\log_2(1+q_C)$ of the supported logical Pauli spaces, for $C=A,B$, satisfy $\kappa_A+\kappa_B=2$. In the one-axis case both sides share the same logical axis, and the stabilizer expansion makes $\mathcal{N}_B$ a measurement of that axis followed by preparation of orthogonal states. Recovery can retain its classical value and implement complete logical dephasing, attaining Bell overlap $1/2$; no recovery acting only on $B$ can raise this separable reference-output state above Bell overlap $1/2$. Exact correction gives $1$ for $q_A=0$, while $q_A=3$ makes $\mathcal{N}_B$ constant, forcing overlap $1/4$. Supplementary Note~S7 gives the channel formula and optimality proofs.
\end{proof}

For the flagged line-mask ensemble, the empty, nonempty proper, and full masks have $q_A=0,1,3$, respectively. Optimizing recovery separately for each known mask therefore gives
\begin{eqnarray}
\mathbb{E}_{\rm line}[F_e^\star]
=\frac12+\frac12(1-p)^d-\frac14p^d.
\label{eq:results_line_optimal_fidelity}
\end{eqnarray}
At $d=11$ and $p=0.01$ this is $0.9476691271$. For independent erasures at the same marginal rate, the distance guarantee instead yields
$\mathbb{E}_{\rm ind}[F_e^\star]\ge1-\frac34\Pr[\operatorname{Bin}(121,0.01)\ge11]>0.999999965$.
The latter is a conservative bound, not an exact independent-mask average. These quantities provide an operational comparison for static known loss without a specified decoder. In general, the ED-active probability omits full-access masks and remains distinct from the uncorrectable-erasure probability.

The resulting design principle is to specify the loss families and identify which logical operation survives each candidate loss. Within a protected-axis family of an exact $[[n,1]]$ stabilizer code, the normalized deviation $\delta_A:=\Delta_A/\overline P_A$ has a fixed gap $1/\sqrt{20}$ between $q_A=0$ and $q_A=1$. In this setting, $F_e^\star(A)=1-\sqrt{5}\,\delta_A$, as shown in Supplementary Note~S7. The general retained-unitary theorem gives an exact zero-tolerance certificate. A positive threshold for exact classification needs a verified ED gap, and a claim of approximate recovery needs a quantitative recovery bound.

\subsection*{Decoder-free error correctability analysis.}

The region-based logical-ED analysis above is not a decoder-success probability. To complement it with an idealized, error-space notion of correction capacity, we exhaustively classify the Pauli errors using only perfect stabilizer syndromes and stabilizer degeneracy. Let $\mathcal{E}_w$ be the phase-free weight-$w$ Pauli set and let $\mathcal{E}_{< w}$ be its accumulated library, so that
\begin{eqnarray}
\abs{\mathcal{E}_w}=\binom{n}{w}3^w,
\quad
\mathcal{E}_{< w}=\bigcup_{a=0}^{w}\mathcal{E}_a.
\label{eq:results_error_library}
\end{eqnarray}
An error $E\in\mathcal{E}_w$ is called \emph{syndrome-unambiguous} at its own weight cutoff when every $F\in\mathcal{E}_{< w}$ with the same syndrome differs from $E$ only by a stabilizer. We denote this weight-resolved fraction by $R_f(w)$. A same-syndrome error of strictly greater weight does not render a lower-weight error ambiguous at the latter's own cutoff. The distance guarantee gives $R_f(w)=1$ for $w\le t:=\lfloor(d-1)/2\rfloor$, but does not determine the residual fraction beyond $t$.

The mathematical bridge to logical-ED is the common supported-logical obstruction. For a stabilizer group $\mathcal{S}$ and its Pauli normalizer $\mathcal{N}(\mathcal{S})$,
\begin{eqnarray}
{\rm syn}(E)={\rm syn}(F), 
\quad
E^\dagger F\notin\mathcal{S} \Longleftrightarrow E^\dagger F\in\mathcal{N}(\mathcal{S})\setminus\mathcal{S}.
\label{eq:results_syndrome_collision}
\end{eqnarray}
Thus every harmful same-syndrome collision differs by a nontrivial logical Pauli supported on $A_{EF}:=\operatorname{supp}(E^\dagger F)$, with $\abs{A_{EF}}\le2w$. Conversely, every logical Pauli supported on at most $2w$ qubits can be split into two errors of weights at most $w$ that form such a collision. At fixed support this is exactly the erasure obstruction: collision-free Pauli fibers on $A$ are equivalent to $A$ supporting no nontrivial logical Pauli. For an $[[n,1]]$ code, $q_A=1$ therefore gives both a harmful collision and $\Delta_A>0$, while the $q_A=3$ sector contains harmful collisions but remains ED-inactive. Moreover, the first cumulative collision occurs at $w_\star=\lceil d/2\rceil$. Whenever a minimum logical support is ED-active, as in the one-axis minimum-string or loop geometry relevant here, $d_\Delta(0)=d-1$ and hence
\begin{eqnarray}
w_\star
=\left\lceil\frac{d_\Delta(0)+1}{2}\right\rceil
=\left\lfloor\frac{d-1}{2}\right\rfloor+1.
\label{eq:results_soft_distance_collision_scale}
\end{eqnarray}
Logical-ED and the zero-tolerance soft distance of Definition~\ref{def:soft_distance} thus localize the geometry at which one-axis ambiguity first becomes possible, whereas $R_f(w)$ measures how prevalent syndrome-unambiguous errors remain across all enumerated supports.

To display the numerical survey, we use the cumulative version of this syndrome-fiber criterion. At a maximum included weight $a$, all errors in $\mathcal{E}_{a}$ are compared with the same candidate library $\mathcal{E}_{< a}$. Define
\begin{eqnarray}
R_{\rm cum}(a) := \frac{1}{\abs{\mathcal{E}_{< a}}}\sum_{E \in \mathcal{E}_{< a}}\mathbf{1}\left[\forall F \in \mathcal{E}_{< a}, {\rm syn}(F)={\rm syn}(E)\Rightarrow E^\dagger F\in\mathcal{S}\right].
\label{eq:results_cumulative_survey}
\end{eqnarray}
Unlike $R_f(w)$, which counts only the exact-weight layer at its own cutoff, $R_{\rm cum}(a)$ counts every included layer after classification at the common cutoff $a$. Lower-weight errors can therefore become ambiguous within an enlarged candidate library; this does not mean that a decoder favoring lower-weight errors necessarily fails on them. The plotted quantity $1-R_{\rm cum}(a)$ is the fraction of the cumulative library belonging to syndrome fibers with more than one logical coset.

\begin{figure*}[t]
\centering
\includegraphics[width=0.90\linewidth]{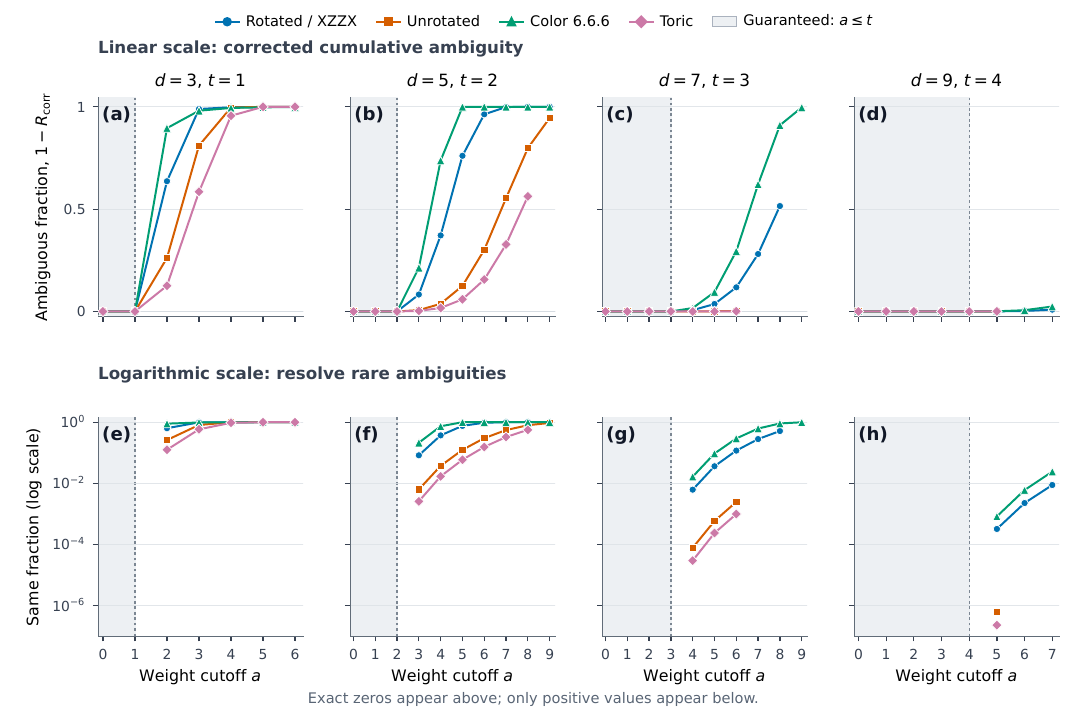}
\caption{\textbf{Syndrome ambiguity beyond the distance guarantee.} The cumulative candidate library contains all phase-free Pauli errors of weights $0,\ldots,a$, with $\abs{\mathcal{E}_{\le a}}=\sum_{j=0}^{a}\binom{n}{j}3^j$. An error is ambiguous when its syndrome fiber in this same library contains more than one logical coset, after stabilizer degeneracy is identified. The fraction plotted is $1-R_{\rm cum}(a)$ [Eq.~(\ref{eq:results_cumulative_survey})], distinct from the exact-layer fraction $1-R_f(w)$. \textbf{(a)--(d)} Linear axes show the overall ambiguity; \textbf{(e)--(h)} the same positive values on logarithmic axes resolve rare collisions. Exact zeros are retained in the upper row and are not replaced by an artificial floor in the lower row. The shaded region $a\le t=\lfloor(d-1)/2\rfloor$ has zero ambiguity by the distance guarantee. Its right edge and the dashed vertical line are placed exactly at $a=t$, the last guaranteed integer cutoff; the first cutoff beyond the guarantee is $a=t+1$. Rotated and XZZX surface-code values coincide in this uniform Pauli survey and share one curve. Values beyond the guaranteed interval come from the completed cumulative classifications; lines guide the eye, and series stop where displayed data end. Missing cutoffs are not treated as zero. The toric patches encode two logical qubits; the other displayed families encode one, and block sizes also differ at fixed $d$. The figure therefore compares idealized combinatorial profiles rather than resource-normalized decoder performance. Aggregate computational resources were $424$ CPU cores, $4.5$~TB RAM, $83$~TB SSD, $73$~TB HDD, and $100$~Gb/s networking.}
\label{fig:capacity_survey}
\end{figure*}

Fig.~\ref{fig:capacity_survey} separates the overall growth of ambiguity [linear panels (a)--(d)] from its initially small values [logarithmic panels (e)--(h)] for rotated and unrotated surface, XZZX surface~\cite{bonillaataides2021xzzx}, toric, and color codes~\cite{bombin2006topological}. The shaded interval $a\le t$ is guaranteed to have zero ambiguity; the remaining displayed values resolve how prevalent the supported-logical obstruction becomes beyond this interval. For example, at $d=9$ the first post-guarantee cutoff $a=5$ gives ambiguous fractions between $2.52\times10^{-7}$ and $8.83\times10^{-4}$ across the plotted families, all of which would appear close to zero on a linear axis. At $d=7$ and $a=6$, the unrotated patch retains a cumulative unambiguous fraction of approximately $0.997523$, illustrating that the first possible harmful collision does not imply widespread ambiguity. To correct a $6$-qubit error, we need a $d=13$ code in terms of code distance metric. Compared with $d=13$ code, which requires $337$ physical qubits and achieves a syndrome-unambiguous fraction of $100\%$, $d=7$ code uses only $97$ qubits while retaining the syndrome-unambiguous fraction of $99.75\%$ under the same error-weight cutoff. This corresponds to a $71\%$ reduction in physical qubit requirements at the cost of only a 0.25\% decrease in syndrome distinguishability, demonstrating a favorable trade-off between resource efficiency and error distinguishability.

The survey requires densely populated error libraries. For the unrotated $d=7$ patch ($n=85$), enumeration through weight $6$ covers $\sum_{j=1}^{6}\binom{85}{j}3^j\simeq3.27\times10^{11}$ nonidentity Pauli errors. Extending the unrotated $d=9$ patch ($n=145$) from the completed cutoff $a=5$ to $a=6$ would require $8.60\times10^{12}$ candidates and exceeded the available memory and storage; the missing cutoff is not plotted. This survey invokes no recovery heuristic or error prior and omits syndrome-extraction faults. Its fractions are combinatorial diagnostics, not threshold estimates or logical-error probabilities, and equal-$d$ comparisons are not normalized for block size or encoded dimension. The supported-logical bridge and computational scope are discussed in Supplementary Note~S5.

\subsection*{Moment representation and a practical analytic handle.}

The definition of $\Delta_A$ involves Haar averages over logical states, which can be evaluated analytically using standard Haar moment identities~\cite{mele2024haar}. In particular, $\overline{E}_A$ requires the second Haar moment on $\mathcal{H}_L$, while $\Delta_A$ requires the fourth moment. We express both quantities compactly using symmetric-subspace projectors, in close analogy to fluctuation measures for gate performance~\cite{cho2025EPD}.

The key technical simplification is that the purity is a two-copy observable~\cite{ekert2002direct}. Introducing the swap operator $\hat{S}_A$ that exchanges subsystem $A$ between two copies (and acts as the identity on $A^c$), one has the identity
\begin{eqnarray}
\tr{\bigl(\hat{\rho}_A(\psi_L)^2\bigr)} \,=\, \tr{\Bigl(\hat{S}_A\,\hat{\rho}(\psi_L)^{\otimes 2}\Bigr)}.
\label{eq:swap_identity_results}
\end{eqnarray}
Combining Eq.~(\ref{eq:swap_identity_results}) with the second Haar moment immediately yields an analytic expression for the mean entanglement,
\begin{eqnarray}
\overline{E}_A \,=\, 1- \frac{\tr{\Bigl(\hat{S}_A\,\hat{V}^{\otimes 2}\hat{\Pi}^{(2)}_{\rm sym}\hat{V}^{\dagger\otimes 2}\Bigr)}}{\tr{\bigl(\hat{\Pi}^{(2)}_{\rm sym}\bigr)}},
\label{eq:Ebar_swap_proj}
\end{eqnarray}
and an analogous contraction formula for the second moment $\int d\psi_L\,E_A(\psi_L)^2$ from the fourth moment. This is precisely why the linear-entropy choice is convenient: both the mean and deviation reduce to finite-copy permutation operators, enabling exact symbolic evaluation for small codes and numerical or sampling-based evaluation when an explicit encoder circuit is available. The finite design order does not by itself imply dimension-independent sample complexity.

\begin{lemma}[Haar moment projector identity]\label{lem:haar_projector}
Let $d\psi$ be the Haar measure on a $d$-dimensional Hilbert space $\mathcal{H}$. Then, for any integer $k\ge 1$,
\begin{eqnarray}
\int d\psi\, \bigl(\ketbra{\psi}{\psi}\bigr)^{\otimes k}
= \frac{\hat{\Pi}^{(k)}_{\rm sym}}{\tr{\bigl(\hat{\Pi}^{(k)}_{\rm sym}\bigr)}},
\label{eq:haar_projector_identity}
\end{eqnarray}
where $\hat{\Pi}^{(k)}_{\rm sym}$ is the projector onto the fully symmetric subspace of $\mathcal{H}^{\otimes k}$.
\end{lemma}

\begin{proof}[Proof sketch]---The operator $\hat{M}:=\int d\psi\,(\ketbra{\psi}{\psi})^{\otimes k}$ commutes with $\hat{U}^{\otimes k}$ for all unitaries $\hat{U}$ by Haar invariance. By Schur--Weyl duality (equivalently, Schur's lemma on irreducible components of the $\mathrm{U}(d)$ action), $\hat{M}$ must be proportional to the identity on each irreducible subrepresentation; since $\hat{M}$ is supported entirely on the symmetric subspace, it equals $c\,\hat{\Pi}^{(k)}_{\rm sym}$. The constant $c$ is fixed by taking the trace: $\tr{(\hat{M})}=\int d\psi\,\tr{((\ketbra{\psi}{\psi})^{\otimes k})}=1$, hence $c=1/\tr{(\hat{\Pi}^{(k)}_{\rm sym})}$.

The normalization factor admits a closed form. Since the fully symmetric subspace of $\mathcal{H}^{\otimes k}$ has dimension $\binom{d+k-1}{k}$, we have
\begin{eqnarray}
\tr{\bigl(\hat{\Pi}^{(k)}_{\rm sym}\bigr)} = \binom{d+k-1}{k}.
\label{eq:sym_dim}
\end{eqnarray}
In particular, for a logical space of dimension $d_L$, the second and fourth Haar moments are normalized by $\tr{(\hat{\Pi}^{(2)}_{\rm sym})}=d_L(d_L+1)/2$ and $\tr{(\hat{\Pi}^{(4)}_{\rm sym})}=\binom{d_L+3}{4}$.

Concretely, letting $\hat{\Pi}^{(k)}_{\rm sym}$ be the projector onto the fully symmetric subspace of $\mathcal{H}_L^{\otimes k}$,
\begin{eqnarray}
\int d\psi_L\, \bigl(\ketbra{\psi_L}{\psi_L}\bigr)^{\otimes k}
= \frac{\hat{\Pi}^{(k)}_{\rm sym}}{\tr{\bigl(\hat{\Pi}^{(k)}_{\rm sym}\bigr)}}.
\label{eq:results_sym_proj}
\end{eqnarray}
Using swap operators that implement $\tr{(\hat{\rho}_A^2)}$ on two copies, Eq.~(\ref{eq:results_sym_proj}) yields closed expressions for $\overline{E}_A$ and for the second moment $\int d\psi_L\, E_A(\psi_L)^2$ in terms of $\hat{V}$ and the swap on $A$ (Methods). These formulas make logical-ED amenable to symbolic evaluation for small codes and provide a controlled starting point for estimator-specific numerical workflows. Derivations are summarized in Methods and detailed in Supplementary Note~S3.
\end{proof}

\section*{Discussion and outlook}\label{sec:discussion}

What further protection can be certified when faults follow the geometry of a device? We addressed this question by resolving logical access beyond the minimum support summarized by code distance. Treating an encoding as a family of pure code states, logical-ED measures the variation of reduced-state purity over logical inputs, while regional profiles resolve its dependence on shape, connectivity, and boundary access. We established logical-basis invariance and complement symmetry, proved that exact erasure correctability forces vanishing deviation, and obtained a quantitative bound from approximate decoupling. These results recover the distance guarantee and define tolerance-dependent size and geometric radii for purity uniformity (Definitions~\ref{def:soft_distance} and~\ref{def:geo_soft_distance}). For one-logical-qubit stabilizer codes, the exact logical-access trichotomy enables evaluation through binary stabilizer ranks. The retained-unitary theorem adds a conditional converse for arbitrary finite-dimensional subspace codes. Zero regional ED is equivalent to exact erasure correctability whenever the surviving subsystem exactly implements a nonscalar logical unitary. A retained logical Pauli provides a concrete way to meet this condition in stabilizer codes. The analytical syndrome criterion connects the support of logical operators to the first possible harmful collision among Pauli errors. Finally, the Haar-moment representation gives analytic mean and deviation formulas and identifies the finite moments needed for empirical characterization.

For an exact stabilizer code encoding one qubit, zero logical-ED has two possible meanings. The erased region may reveal nothing about the logical state, or it may hold the entire logical qubit. If the surviving qubits still support a logical Pauli, only the first possibility remains and exact recovery is possible. In the $[[121,1,11]]$ patch, both logical Pauli generators survive the loss of a $100$-qubit corner block. Yet erasing a spanning row of just $11$ qubits exposes a logical Pauli and prevents exact recovery. A corridor three qubits wide likewise becomes uncorrectable when it first spans the patch. For a flagged erasure, the optimal entanglement fidelity is $1$, $1/2$, or $1/4$ according to whether the erased region reveals nothing, supports exactly one logical Pauli, or holds the full logical qubit. Even at the same per-qubit erasure probability $p=0.01$, the mean optimal fidelity on this patch is approximately $0.947669$ for complete-line losses and at least $0.999999965$ for independent losses.

These results suggest a practical criterion for designing quantum error-correcting codes. First specify the loss patterns a device can flag and identify, for each pattern, a nonscalar logical unitary that the surviving qubits can implement exactly. For such a family, the zero-tolerance geometric soft distance gives the largest scale up to which every candidate erasure is exactly correctable. Regional purities can be estimated from repeated preparations~\cite{ekert2002direct,hu2025shallow,elben2023toolbox}. For one logical qubit, the purities for two eigenstates of the retained unitary and their equal superposition also give a rank-dependent upper bound on logical information leakage [Eq.~(\ref{eq:results_three_purity_bound})]. The resulting picture can guide a design decision while the layout is still open to change. A designer can reinforce a boundary where a loss would expose logical information, or establish that the proposed code already corrects the loss patterns of concern. In the latter case, avoiding an unnecessary round of redesign could allow designers to test the layout sooner with noisy gates and repeated syndrome measurements, a concrete step toward fault-tolerant quantum computing.

\section*{Methods}

\subsection*{Code model, notation, and correctability}

The erasure theorems use an arbitrary finite-dimensional logical space and a finite-dimensional physical bipartition. When we specify an $[[n,k,d]]$ qubit code, the physical and logical spaces are $(\mathbb C^2)^{\otimes n}$ and $(\mathbb C^2)^{\otimes k}$, respectively. An encoding isometry $\hat{V}:\mathcal{H}_L\to\mathcal{H}_{\rm phys}$ defines the code projector $\hat{P}=\hat{V}\hat{V}^\dagger$. A region $A$ is (exactly) erasure-correctable if there exists a recovery map $\mathcal{R}$ acting on $A^c$ satisfying Eq.~(\ref{eq:results_erasure_correctable_def}) for all logical inputs. We use the standard equivalence between erasure correctability and the Knill--Laflamme condition applied to an operator basis supported on $A$~\cite{knill1997theory}. This equivalence is the key ingredient in the proof of Theorem~\ref{thm:vanishing}.

\subsection*{Logical ensemble and entanglement functional}

For a pure logical input $\ket{\psi_L}$, the encoded state is $\ket{\Psi(\psi_L)}=\hat{V}\ket{\psi_L}$. For a region $A$, the reduced state $\hat{\rho}_A(\psi_L)$ is defined in Eq.~(\ref{eq:results_rhoA}), and the linear entropy $E_A(\psi_L)$ is defined in Eq.~(\ref{eq:results_linear_entropy}). We define the logical mean and deviation by Eqs.~(\ref{eq:results_Ebar_def})--(\ref{eq:results_Delta_def}) using Haar measure on $\mathcal{H}_L$. In practice, Haar averages can be replaced by unitary designs on the logical space: a $2$-design suffices for $\overline{E}_A$ and a $4$-design for $\Delta_A$ (or for the second moment of $E_A$), mirroring standard moment-estimation logic in randomized protocols~\cite{brandao2016local,mele2024haar}. Locally randomized measurements provide a route to estimating the required regional purities~\cite{hu2025shallow,elben2023toolbox}; the measurement cost depends on the region and estimator.

\subsection*{Swap-operator representation and Haar moment evaluation}

A convenient expression for the purity $\tr{(\hat{\rho}_A(\psi_L)^2)}$ uses the swap operator $\hat{S}_A$ acting on two copies of the physical Hilbert space by swapping the subsystem $A$ while leaving $A^c$ untouched. For the encoded pure state $\hat{\rho}(\psi_L)=\ketbra{\Psi(\psi_L)}{\Psi(\psi_L)}$ we have
\begin{eqnarray}
\tr{\Bigl(\hat{\rho}_A(\psi_L)^2\Bigr)} = \tr{\Bigl(\hat{S}_A\, \hat{\rho}(\psi_L)^{\otimes 2}\Bigr)}.
\label{eq:methods_swap_purity}
\end{eqnarray}
Taking the Haar average over $\ket{\psi_L}$ and using Eq.~(\ref{eq:results_sym_proj}) yields
\begin{eqnarray}
\int d\psi_L\, \tr{\Bigl(\hat{\rho}_A(\psi_L)^2\Bigr)} = \frac{ \tr{\Bigl(\hat{S}_A\, \bigl(\hat{V}^{\otimes 2}\hat{\Pi}^{(2)}_{\rm sym}\hat{V}^{\dagger\otimes 2}\bigr)\Bigr)} }{\tr{\bigl(\hat{\Pi}^{(2)}_{\rm sym}\bigr)}}.
\label{eq:methods_mean_purity}
\end{eqnarray}
For completeness, pull the swap back to two logical copies and define
\begin{eqnarray}
\hat{W}_A:=\hat{V}^{\dagger\otimes2}\hat{S}_A\hat{V}^{\otimes2},
\quad
D_m:=\tr{\bigl(\hat{\Pi}^{(m)}_{\rm sym}\bigr)}.
\label{eq:methods_WA_def}
\end{eqnarray}
If $P_A(\psi_L):=\tr{(\hat{\rho}_A(\psi_L)^2)}$, then its first two Haar moments are
\begin{eqnarray}
\mu_A&:=&\int d\psi_L\,P_A(\psi_L)
=\frac{\tr{\bigl(\hat{W}_A\hat{\Pi}^{(2)}_{\rm sym}\bigr)}}{D_2},
\label{eq:methods_muA}
\\
\nu_A&:=&\int d\psi_L\,P_A(\psi_L)^2
=\frac{\tr{\bigl(\hat{W}_A^{(12)}\hat{W}_A^{(34)}\hat{\Pi}^{(4)}_{\rm sym}\bigr)}}{D_4},
\label{eq:methods_nuA}
\end{eqnarray}
where $\hat{W}_A^{(12)}$ and $\hat{W}_A^{(34)}$ act on the indicated pairs of four logical copies. Since $E_A=1-P_A$,
\begin{eqnarray}
\overline{E}_A=1-\mu_A,
\quad
\Delta_A=\sqrt{\nu_A-\mu_A^2}.
\label{eq:methods_exact_ED_moments}
\end{eqnarray}
These are the explicit finite-copy contraction formulas used in the analysis; their derivation and permutation expansion are provided in Supplementary Note~S3.

\subsection*{Surface-code simulation details for Fig.~\ref{fig:sim}}

For odd distance $d$, data qubits are placed on a $d\times d$ square array. Interior weight-$4$ plaquettes are assigned alternating $X$- and $Z$-type checks, and weight-$2$ boundary checks terminate the corresponding smooth and rough boundaries, giving an $[[d^2,1,d]]$ rotated planar patch. We use a horizontal row as $\overline{Z}$ and a vertical column as $\overline{X}$. The $n-1$ independent stabilizer generators are stored as rows of a binary symplectic matrix $G\in\mathbb{F}_2^{(n-1)\times 2n}$.

For a region $A$, let $G_{A^c}$ denote the columns of $G$ associated with the $X$ and $Z$ components on the complement. The number of stabilizers supported in $A$ is $N_0(A)=2^{t_A}$ with
\begin{eqnarray}
t_A=(n-1)-\operatorname{rank}_{\mathbb{F}_2}(G_{A^c}).
\label{eq:methods_surface_tA}
\end{eqnarray}
For a logical symplectic vector $\ell_\alpha$, the affine system $uG_{A^c}=\ell_{\alpha,A^c}$ is consistent if and only if $\ell_{\alpha,A^c}$ lies in the row space of $G_{A^c}$. Hence $N_\alpha(A)=N_0(A)$ when the system is consistent and $N_\alpha(A)=0$ otherwise. We evaluate these ranks and row-space tests by bit-packed Gaussian elimination over $\mathbb{F}_2$; no physical state vector or stabilizer-group enumeration is constructed. A direct $9$-qubit state-vector check at $d=3$ agrees with the per-input linear entropy from Eq.~(\ref{eq:surface_purity_counts}) to a maximum absolute error below $9\times10^{-16}$. Eq.~(\ref{eq:surface_delta_counts}) follows analytically from the Haar moments of the Bloch components.

For the erasure ensembles, $p$ is scanned from $0$ to $0.5$ in increments of $0.025$ and $5000$ independent masks are generated at every point. In the independent ensemble each data qubit is included in $A$ with probability $p$. In the line-correlated ensemble a horizontal or vertical orientation is selected uniformly, after which each of the $d$ complete lines is included independently with probability $p$; consequently every individual qubit still has marginal erasure probability $p$. Panel (c) reports the fraction of masks with $\Delta_A>0$ and 95\% Wilson score intervals, while panel (d) reports the empirical mean of $\Delta_A$. The additional comparison at $p=0.01$ in the main text uses the exact common-line probability and the binomial upper bound implied by the distance condition for independent erasures. Supplementary Note~S5 describes the numerical procedures and their validation. Access to the underlying data and code is described in the availability statements below.

\section*{Declarations}

\medskip\noindent
{\bf Data Availability.}---The datasets generated during the current study are not currently publicly available, as uses beyond non-commercial academic research may require institutional review under the policies of the authors' affiliated institutions. The datasets are available from the corresponding author on reasonable request for non-commercial academic research purposes.

\medskip\noindent
{\bf Code Availability.}---The code used in this study is not currently publicly available, as uses beyond non-commercial academic research may require institutional review under the policies of the authors' affiliated institutions. The code is available from the corresponding author on reasonable request for non-commercial academic research purposes. The computational methodology is described in detail in the Methods of the main text and in the Supplementary Information.

\medskip\noindent
{\bf Author contributions.}---IS, YH, and JB conceived the initial idea, which was subsequently developed further with KC. KC and JB were primarily responsible for the mathematical calculations. All authors verified the proofs and theoretical arguments. IS, YH, and JB performed the numerical analyses and validation. JB and KC wrote the manuscript, and all authors reviewed and revised it. Correspondence and requests for materials should be addressed to IS, YH, and JB.

\medskip\noindent
{\bf Competing interests.}---The authors declare no competing financial or non-financial interests.

\medskip\noindent
{\bf Acknowledgements.}---Not applicable.

\medskip\noindent
{\bf Funding information.}---This work was supported by the Institute of Information and Communications Technology Planning and Evaluation grant funded by the Korean government (RS-2019-II190003, ``Research and Development of Core Technologies for Programming, Running, Implementing and Validating of Fault-Tolerant Quantum Computing System''), the Korean ARPA-H Project through the Korea Health Industry Development Institute (KHIDI), funded by the Ministry of Health \& Welfare, Republic of Korea (RS-2025-25456722), and the Ministry of Trade, Industry and Resources (MOTIR), Korea, under the project ``Industrial Technology Infrastructure Program'' (RS-2024-00466693). This work was also supported by Korea Institute of Science and Technology Information (KISTI) ((KISTI)K26L1M3C5-01). We acknowledge the Yonsei University Quantum Computing Project Group for providing support and access to the Quantum System One, which is operated at Yonsei University.



\clearpage
\newpage
\onecolumngrid

\makeatletter
\let\addcontentsline\origaddcontentsline
\makeatother

\newcommand{\suppnote}[1]{%
  \refstepcounter{section}%
  \setcounter{subsection}{0}%
  \setcounter{subsubsection}{0}%
  \section*{Supplementary Note~\Roman{section}. #1}%
}
\newcommand{\suppappendix}[1]{%
  \refstepcounter{section}%
  \setcounter{subsection}{0}%
  \setcounter{subsubsection}{0}%
  \section*{Supplementary Appendix~\Alph{section}. #1}%
}
\renewcommand{\thetheorem}{S\arabic{theorem}}
\renewcommand{\theproposition}{S\arabic{proposition}}
\renewcommand{\thelemma}{S\arabic{lemma}}
\renewcommand{\thecorollary}{S\arabic{corollary}}
\renewcommand{\thedefinition}{S\arabic{definition}}
\renewcommand{\theremark}{S\arabic{remark}}

\part{Supplementary Information}

This Supplementary Information develops the linear-entropy formulation underlying logical entanglement deviation (logical-ED). We give a complete proof that exact erasure correctability forces a logical-state-independent reduced density matrix and hence vanishing logical-ED, derive explicit Haar second- and fourth-moment formulas, and analyze the tolerance-dependent ED-soft distance and geometric region families. We prove the exact $q_A\in\{0,1,3\}$ logical-access trichotomy for one-logical-qubit stabilizer codes and identify logical-ED as a detector of the one-axis sector, including its complement-symmetry and full-access blind spots. We prove the converse for finite-dimensional subspace encodings when a non-scalar logical unitary is exactly implementable on the retained subsystem. We give a three-purity test for a single logical qubit and an analytic example beyond stabilizer codes. Exact stabilizer-coset calculations for rotated planar surface-code patches up to $225$ data qubits illustrate geometric ED profiles under static independent and correlated erasures. We also show that Pauli-error syndrome ambiguity and the ED-soft-distance construction share a supported-logical obstruction, and distinguish the corresponding error-space and region-space diagnostics. We also give direct validation, implementation and scaling details, extensions to qudit and subsystem codes, and a rigorous one-way bound from operational approximate decoupling to small logical-ED.

\tableofcontents

\section{Background on entangling-power deviation and its logical-ED analogue}

\subsection{Linear entropy as a natural entanglement functional}

Throughout the main text we quantify the bipartite entanglement between a physical region $A$ and its complement $A^c$ in a \emph{pure} encoded state by the \emph{linear entropy}
\begin{eqnarray}
E_A(\psi_L)  := 1-\tr{\Bigl(\hat{\rho}_A(\psi_L)^2\Bigr)} ,
\label{eq:S1_linear_entropy_def}
\end{eqnarray}
where $\hat{\rho}_A(\psi_L)=\tr_{A^c}\!\left(\ketbra{\Psi(\psi_L)}{\Psi(\psi_L)}\right)$ and $\ket{\Psi(\psi_L)}=\hat{V}\ket{\psi_L}$ is the encoded code state.
For a pure bipartite state $\ket{\Psi}\in\mathcal{H}_A\otimes\mathcal{H}_{A^c}$, $E_A$ is a monotone function of the Rényi-$2$ entanglement entropy
\begin{eqnarray}
S_2(\hat{\rho}_A)  := -\log\tr{\left(\hat{\rho}_A^2\right)} ,
\label{eq:S1_Renyi2_def}
\end{eqnarray}
and hence provides a faithful proxy for entanglement across the cut $A|A^c$.
In particular,
\begin{eqnarray}
0 \ \le\ E_A(\psi_L)\ \le\ 1-\frac{1}{d_A},
\label{eq:S1_linear_entropy_bounds}
\end{eqnarray}
where $d_A=\dim(\mathcal{H}_A)$.
The lower bound is attained if and only if $\ket{\Psi(\psi_L)}$ factorizes across $A|A^c$ (no entanglement), and the upper bound is attained by maximally entangled states when $d_A\le d_{A^c}$.

There are two practical reasons for choosing $E_A$ over, e.g., the von Neumann entropy $S(\hat{\rho}_A)=-\tr{(\hat{\rho}_A\log\hat{\rho}_A)}$.

\smallskip\noindent
{\bf (i) Polynomial structure and moment methods.}
The purity $\tr{(\hat{\rho}_A^2)}$ is a degree-2 polynomial in $\hat{\rho}_A$, and therefore $E_A(\psi_L)$ is a low-degree polynomial functional of the encoded pure state. As a consequence, Haar averages of $E_A(\psi_L)$ and its powers reduce to \emph{finite} Haar moments, which admit closed expressions in terms of symmetric-subspace projectors and permutation (swap) operators. In particular, the mean $\int d\psi_L\,E_A(\psi_L)$ depends only on a \emph{second} Haar moment, while the second moment $\int d\psi_L\,E_A(\psi_L)^2$ requires a \emph{fourth} Haar moment. This structure is precisely what makes fluctuation diagnostics analytically tractable.

\smallskip\noindent
{\bf (ii) Direct experimental/statistical accessibility.}
Purity is naturally expressed using a swap operator on two copies~\cite{ekert2002direct}. For any physical state $\hat{\rho}$ on $\mathcal{H}_A\otimes\mathcal{H}_{A^c}$,
\begin{eqnarray}
\tr{\Bigl(\hat{\rho}_A^2\Bigr)}  = \tr{\Bigl(\hat{S}_A\,\hat{\rho}^{\otimes 2}\Bigr)},
\label{eq:S1_swap_purity}
\end{eqnarray}
where $\hat{S}_A$ swaps subsystem $A$ between the two copies and acts trivially on $A^c$. This representation underlies both theoretical simplifications and a variety of purity-estimation schemes (two-copy interference, randomized measurements, etc.)~\cite{elben2019statistical}, and it is well matched to a fluctuation-based quantity that requires evaluating second moments of $E_A(\psi_L)$.

\subsection{The conceptual lift to logical-ED}

In our previous work on \emph{entangling power deviation} (EPD) for bipartite unitaries~\cite{cho2025EPD}, the central idea was to characterize not only the \emph{average} entanglement generated by a unitary gate $\hat{U}$~\cite{zanardi2000entangling} but also its \emph{input dependence}. Concretely, for a bipartite system $\mathcal{H}_{1}\otimes\mathcal{H}_{2}$ and product inputs $\ket{\psi_1}\otimes\ket{\psi_2}$ drawn from the product Haar measure, we defined the output entanglement (again using linear entropy)
\begin{eqnarray}
E_U(\psi_1,\psi_2) := 1-\tr{\Bigl(\hat{\rho}_1(\psi_1,\psi_2)^2\Bigr)},
\quad
\hat{\rho}_1=\tr_{2}\!\left(\hat{U}\ketbra{\psi_1\psi_2}{\psi_1\psi_2}\hat{U}^\dagger\right),
\label{eq:S1_gate_ent_lin}
\end{eqnarray}
and introduced
\begin{eqnarray}
{\rm EP}(\hat{U})  := \int d\psi_1 d\psi_2\, E_U(\psi_1,\psi_2),
\quad
{\rm EPD}(\hat{U})  := \sqrt{\int d\psi_1 d\psi_2\, E_U(\psi_1,\psi_2)^2-{\rm EP}(\hat{U})^2}.
\label{eq:S1_gate_EPD_def}
\end{eqnarray}
The role of EPD was to quantify \emph{how non-uniformly} the gate entangles typical inputs, i.e., whether entanglement generation is evenly spread across inputs or concentrated on special directions~\cite{cho2025EPD}.

The present work transplants this fluctuation philosophy from \emph{gates} to \emph{codes}. The conceptual replacement is:

\smallskip\noindent
\begin{itemize}
\item gate $\hat{U}$ acting on a bipartite input $\longrightarrow$ encoder (isometry) $\hat{V}$ preparing a family of encoded \emph{pure} code states $\ket{\Psi(\psi_L)}=\hat{V}\ket{\psi_L}$;
\item product input ensemble on $\mathcal{H}_1\otimes\mathcal{H}_2$ $\longrightarrow$ Haar ensemble on the logical Hilbert space $\mathcal{H}_L$;
\item entanglement across the \emph{input bipartition} $\mathcal{H}_1|\mathcal{H}_2$ $\longrightarrow$ entanglement across a \emph{physical region cut} $A|A^c$ inside $\mathcal{H}_{\rm phys}$.
\end{itemize}

More explicitly, many encoders are implemented as a unitary $\hat{U}_{\rm enc}$ acting on $k$ logical qubits (or qudits) and $n-k$ ancillas prepared in a fixed reference state $\ket{0}^{\otimes (n-k)}$. In this realization,
\begin{eqnarray}
\hat{V}\ket{\psi_L}  = \hat{U}_{\rm enc}\left(\ket{\psi_L}\otimes\ket{0}^{\otimes (n-k)}\right).
\label{eq:S1_V_from_Uenc}
\end{eqnarray}
Thus, for any chosen physical region $A$ we can view $E_A(\psi_L)$ as the linear-entropy entanglement created \emph{by the encoding circuit} across the \emph{physical} bipartition $A|A^c$, with the restricted input ensemble $\ket{\psi_L}\otimes\ket{0}^{\otimes (n-k)}$. The logical-ED introduced in the main text,
\begin{eqnarray}
\Delta_A  := \sqrt{\int d\psi_L\, E_A(\psi_L)^2 - \left(\int d\psi_L\, E_A(\psi_L)\right)^2},
\label{eq:S1_logical_EPD_def}
\end{eqnarray}
is therefore an \emph{EPD-type fluctuation measure} for the encoding transformation, specialized to (i) a physically meaningful output cut and (ii) the natural logical Haar ensemble.

\subsection{Why ``deviation'' is the right object for QEC: uniformity over logical inputs}

The key distinction between gate-EPD~\cite{cho2025EPD} and logical-ED is the \emph{operational goal}. For gate entangling power, one asks how much entanglement is generated from generic \emph{product} inputs; in that setting, our previous results show that perfectly uniform nontrivial entanglers are impossible for the product ensemble (heuristically, entanglement generation necessarily depends on input directions)~\cite{cho2025EPD}. In QEC, the goal is different: we require that \emph{local} regions should not reveal the logical state. In the ideal (exact) erasure-correctable regime, this goal is expressed by local indistinguishability on $A$:
\begin{eqnarray}
\hat{\rho}_A(\psi_L) \ \ \text{is independent of}\ \ \ket{\psi_L}.
\label{eq:S1_local_indist}
\end{eqnarray}
If Eq.~(\ref{eq:S1_local_indist}) holds, then \emph{every} function of $\hat{\rho}_A(\psi_L)$ is also independent of $\psi_L$, and in particular the purity $\tr{(\hat{\rho}_A(\psi_L)^2)}$ is constant. Hence the linear entropy $E_A(\psi_L)$ is constant and its standard deviation vanishes:
\begin{eqnarray}
\hat{\rho}_A(\psi_L)=\hat{\omega}_A\ \ \forall \psi_L
\quad\Longrightarrow\quad
\Delta_A=0.
\label{eq:S1_correctable_implies_D0}
\end{eqnarray}
This is exactly the direction exploited in Theorem~1 of the main text.

In this restricted sense, logical-ED measures one necessary spectral consequence of QEC: it quantifies how strongly the \emph{purity}, and hence the linear-entropy entanglement, depends on the logical state. Exact correctability forces the full reduced state to be logical-state independent and therefore forces $\Delta_A=0$. The converse is false: $\Delta_A=0$ fixes only one spectral functional and does not imply local indistinguishability. Likewise, small $\Delta_A$ means only that the purity varies weakly; by itself it does not bound an operational recovery error. A precise one-way implication from approximate decoupling to small logical-ED is given in Supplementary Note~S6. Thus logical-ED is used below as a one-sided, geometry-resolved diagnostic of logical-state-dependent spectral anisotropy, rather than as a complete test of correctability.

\subsection{Complement symmetry and the intrinsic scope of logical-ED}

\begin{proposition}[Complement symmetry]\label{prop:S1_complement_symmetry}
For every pure-state encoding isometry $\hat{V}$ and every physical region $A$,
\begin{eqnarray}
\overline{E}_A=\overline{E}_{A^c},
\quad
\Delta_A=\Delta_{A^c}.
\label{eq:S1_complement_symmetry}
\end{eqnarray}
More generally, for every logical pure state and every integer $m\geq 1$,
\begin{eqnarray}
\tr{\!\left(\hat{\rho}_A(\psi_L)^m\right)}
=
\tr{\!\left(\hat{\rho}_{A^c}(\psi_L)^m\right)}.
\label{eq:S1_complement_moments}
\end{eqnarray}
\end{proposition}

\begin{proof}---For each encoded pure state $\hat{V}\ket{\psi_L}$, the Schmidt decomposition across $A|A^c$ shows that $\hat{\rho}_A(\psi_L)$ and $\hat{\rho}_{A^c}(\psi_L)$ have the same nonzero eigenvalues. Equation~(\ref{eq:S1_complement_moments}) follows pointwise in $\psi_L$. Hence $E_A(\psi_L)=E_{A^c}(\psi_L)$ pointwise, and equality of the Haar means and variances gives Eq.~(\ref{eq:S1_complement_symmetry}).
\end{proof}

Proposition~\ref{prop:S1_complement_symmetry} is important for interpretation. If erasure of $A$ is correctable, then $A^c$ contains sufficient information to recover the logical state, yet $\Delta_A=\Delta_{A^c}=0$. Logical-ED is therefore a property of the cut and cannot, by itself, distinguish an unauthorized region from its authorized complement.

An extreme counterexample makes the limitation explicit. Consider
\begin{eqnarray}
\hat{V}\ket{\psi_L}=\ket{\psi_L}_A\otimes\ket{0}_{A^c}.
\label{eq:S1_identity_encoding_counterexample}
\end{eqnarray}
Then $A$ contains the entire logical state, but $\hat{\rho}_A(\psi_L)=\ketbra{\psi_L}{\psi_L}$ is pure for every input, so $E_A(\psi_L)=0$ and $\Delta_A=0$. In fact all spectral moments in Eq.~(\ref{eq:S1_complement_moments}) are constant. Consequently, neither logical-ED nor any collection of reduced-state spectral moments is a complete logical-information or correctability diagnostic.

On the other hand, Theorem~1 has the rigorous contrapositive
\begin{eqnarray}
\Delta_A>0
\quad\Longrightarrow\quad
\text{the erasure of $A$ is not exactly correctable}.
\label{eq:S1_nonzero_witness}
\end{eqnarray}
Thus nonzero logical-ED is a valid one-sided witness of noncorrectability, whereas zero logical-ED is inconclusive without additional structure.

\subsection{Ensemble choice and basis invariance}

Finally, we comment on the role of the logical Haar ensemble. Averaging over $d\psi_L$ makes $\overline{E}_A$ and $\Delta_A$ invariant under changes of logical basis, i.e., under $\hat{V}\to \hat{V}\hat{U}_L$ for any logical unitary $\hat{U}_L$. This property ensures that logical-ED captures basis-independent spectral-anisotropy features of the code and region $A$, rather than artifacts of a chosen logical representation. For exact evaluation, Haar averages can be replaced by logical-state designs: a $2$-design suffices for $\overline{E}_A$ and a $4$-design for the second moment entering $\Delta_A$~\cite{mele2024haar}, consistent with the moment formulas in Supplementary Note~S3. This design-order statement does not itself imply a dimension-independent sampling cost.


\section{Complete proof of Theorem~1 (vanishing logical-ED on correctable regions)}

\subsection{Setup and statement}

Let the physical Hilbert space be decomposed as
\begin{eqnarray}
\mathcal{H}_{\rm phys}  = \mathcal{H}_A \otimes \mathcal{H}_{A^c},
\quad
d_A := \dim(\mathcal{H}_A).
\label{eq:S2_phys_split}
\end{eqnarray}
Let $\mathcal{H}_L$ be the logical Hilbert space and let $\hat{V}:\mathcal{H}_L\to\mathcal{H}_{\rm phys}$ be an encoding isometry, $\hat{V}^\dagger \hat{V}=\hat{\mathds{1}}_L$. Denote the code subspace by $\mathcal{C}:={\rm Im}(\hat{V})$ and the corresponding projector by
\begin{eqnarray}
\hat{P}  := \hat{V}\hat{V}^\dagger .
\label{eq:S2_code_projector}
\end{eqnarray}
For a logical pure state $\ket{\psi_L}$, the encoded state is $\ket{\Psi(\psi_L)}=\hat{V}\ket{\psi_L}$ and the reduced state on region $A$ is
\begin{eqnarray}
\hat{\rho}_A(\psi_L)  := \tr_{A^c}\!\Bigl(\ketbra{\Psi(\psi_L)}{\Psi(\psi_L)}\Bigr).
\label{eq:S2_rhoA_def}
\end{eqnarray}
The region entanglement functional used in the main text is the linear entropy
\begin{eqnarray}
E_A(\psi_L)  := 1-\tr{\Bigl(\hat{\rho}_A(\psi_L)^2\Bigr)},
\label{eq:S2_EA_def}
\end{eqnarray}
and the logical entanglement deviation (logical-ED) is
\begin{eqnarray}
\Delta_A
 :=
\sqrt{
\int d\psi_L\, E_A(\psi_L)^2
-
\left(\int d\psi_L\, E_A(\psi_L)\right)^2
},
\label{eq:S2_Delta_def}
\end{eqnarray}
where $d\psi_L$ is Haar measure on $\mathcal{H}_L$.

We recall the operational notion of (exact) erasure correctability~\cite{cleve1999share}. The erasure of region $A$ is described by the partial trace channel $\mathcal{N}_A := \tr_A$, which maps operators on $\mathcal{H}_{\rm phys}$ to operators on $\mathcal{H}_{A^c}$. We say that the erasure of $A$ is \emph{exactly correctable on the code} if there exists a CPTP recovery map $\mathcal{R}$ acting on $A^c$ such that
\begin{eqnarray}
(\mathcal{R}\circ \tr_A)\bigl(\hat{V}\hat{\sigma}_L \hat{V}^\dagger\bigr) = \hat{V}\hat{\sigma}_L \hat{V}^\dagger
\quad
\text{for all density operators }\hat{\sigma}_L\ \text{on}\ \mathcal{H}_L .
\label{eq:S2_erasure_correctable}
\end{eqnarray}
We now prove the theorem stated in the main text.

\begin{theorem}[Theorem~1 (main text)]\label{thm:S2_main}
Assume that the erasure of $A$ is exactly correctable on the code ${\rm Im}(\hat{V})$, in the sense of Eq.~(\ref{eq:S2_erasure_correctable}). Then there exists a fixed density operator $\hat{\omega}_A$ on $\mathcal{H}_A$ such that
\begin{eqnarray}
\hat{\rho}_A(\psi_L) = \hat{\omega}_A
\quad
\text{for all pure }\ket{\psi_L}\in\mathcal{H}_L .
\label{eq:S2_rhoA_constant}
\end{eqnarray}
Consequently, $E_A(\psi_L)$ is constant in $\psi_L$ and $\Delta_A=0$.
\end{theorem}

\subsection{Kraus operators for erasure and the Knill--Laflamme condition.}

Choose an orthonormal basis $\{\ket{a}\}_{a=1}^{d_A}$ for $\mathcal{H}_A$. The partial trace channel $\tr_A$ admits the Kraus representation
\begin{eqnarray}
\tr_A(\hat{\rho}) = \sum_{a=1}^{d_A} \hat{K}_a \hat{\rho} \hat{K}_a^\dagger,
\quad
\hat{K}_a  := \bra{a}_A \otimes \hat{\mathds{1}}_{A^c}.
\label{eq:S2_Kraus_trace}
\end{eqnarray}
Each $\hat{K}_a$ maps $\mathcal{H}_A\otimes \mathcal{H}_{A^c}\to \mathcal{H}_{A^c}$, and one checks that
\begin{eqnarray}
\sum_{a=1}^{d_A} \hat{K}_a^\dagger \hat{K}_a
 =
\sum_{a=1}^{d_A}
\Bigl(
\ket{a}\bra{a}_A \otimes \hat{\mathds{1}}_{A^c}
\Bigr)
 = \hat{\mathds{1}}_{A}\otimes\hat{\mathds{1}}_{A^c}.
\label{eq:S2_completeness}
\end{eqnarray}
Moreover, the ``error overlaps'' $\hat{K}_a^\dagger \hat{K}_b$ are precisely matrix units on $A$ tensored with identity on $A^c$:
\begin{eqnarray}
\hat{K}_a^\dagger \hat{K}_b
 =
\Bigl(\ket{a}\bra{b}_A \otimes \hat{\mathds{1}}_{A^c}\Bigr).
\label{eq:S2_KdagK}
\end{eqnarray}

We now invoke the standard Knill--Laflamme (KL) condition for exact quantum error correction on a subspace~\cite{knill1997theory,nielsen_chuang}. For completeness, we state it in the form needed here.

\begin{lemma}[Knill--Laflamme condition for a channel]\label{lem:S2_KL}
Let $\mathcal{N}(\hat{\rho})=\sum_\mu \hat{E}_\mu \hat{\rho} \hat{E}_\mu^\dagger$ be a CPTP map (possibly mapping between different output spaces), and let $\hat{P}$ be the projector onto a code subspace $\mathcal{C}$. If there exists a recovery channel $\mathcal{R}$ such that $(\mathcal{R}\circ\mathcal{N})(\hat{\rho})=\hat{\rho}$ for all density operators $\hat{\rho}$ supported on $\mathcal{C}$, then there exists a matrix $\alpha=(\alpha_{\mu\nu})$ such that
\begin{eqnarray}
\hat{P} \hat{E}_\mu^\dagger \hat{E}_\nu \hat{P} = \alpha_{\mu\nu}\, \hat{P}
\quad
\text{for all }\mu,\nu.
\label{eq:S2_KL_condition}
\end{eqnarray}
\end{lemma}

\begin{proof}---We provide a complete proof of the necessity direction. By linearity, the hypothesis that $(\mathcal{R}\circ\mathcal{N})(\hat{\rho})=\hat{\rho}$ for all density operators $\hat{\rho}$ supported on the code implies the stronger statement
\begin{eqnarray}
(\mathcal{R}\circ\mathcal{N})(\hat{X}) = \hat{X}
\quad
\text{for all operators }\hat{X}\ \text{with }\hat{X}=\hat{P}\hat{X}\hat{P}.
\label{eq:S2_linear_extension}
\end{eqnarray}
Fix Kraus decompositions $\mathcal{N}(\hat{\rho})=\sum_\mu \hat{E}_\mu \,\hat{\rho}\,\hat{E}_\mu^\dagger$ and $\mathcal{R}(\hat{\sigma})=\sum_\alpha \hat{R}_\alpha \,\hat{\sigma}\,\hat{R}_\alpha^\dagger$.
Then $\Phi:=\mathcal{R}\circ\mathcal{N}$ has Kraus operators $\hat{F}_{\alpha\mu}:=\hat{R}_\alpha \hat{E}_\mu$, so that
\begin{eqnarray}
\Phi(\hat{X})  = \sum_{\alpha,\mu}\hat{F}_{\alpha\mu}\,\hat{X}\,\hat{F}_{\alpha\mu}^\dagger .
\label{eq:S2_Phi_Kraus}
\end{eqnarray}
We now use a Choi-rank argument restricted to the code. Let $\{\ket{i}\}_{i=1}^{d_{\mathcal{C}}}$ be an orthonormal basis of the code subspace $\mathcal{C}$, and define the (unnormalized) maximally entangled vector $\ket{\Omega}:=\sum_{i=1}^{d_{\mathcal{C}}}\ket{i}_R\otimes \ket{i}$ on $R\otimes \mathcal{H}_{\rm phys}$, where $R\simeq \mathcal{C}$ is a reference system.
Since $\ket{\Omega}\bra{\Omega}$ is supported on $R\otimes \mathcal{C}$, Eq.~(\ref{eq:S2_linear_extension}) implies
\begin{eqnarray}
(\operatorname{id}_R\otimes \Phi)\bigl(\ketbra{\Omega}{\Omega}\bigr)
 =
\ketbra{\Omega}{\Omega}.
\label{eq:S2_Choi_identity}
\end{eqnarray}
On the other hand, using Eq.~(\ref{eq:S2_Phi_Kraus}),
\begin{eqnarray}
(\operatorname{id}_R\otimes \Phi)\bigl(\ketbra{\Omega}{\Omega}\bigr)
 =
\sum_{\alpha,\mu}
(\hat{\mathds{1}}_R\otimes \hat{F}_{\alpha\mu})\ketbra{\Omega}{\Omega}(\hat{\mathds{1}}_R\otimes \hat{F}_{\alpha\mu}^\dagger).
\label{eq:S2_Choi_sum}
\end{eqnarray}
The operator on the left-hand side of Eq.~(\ref{eq:S2_Choi_identity}) is rank one. Therefore, each vector $(\hat{\mathds{1}}_R\otimes \hat{F}_{\alpha\mu})\ket{\Omega}$ must lie in the one-dimensional range of $\ketbra{\Omega}{\Omega}$, i.e., there exist scalars $c_{\alpha\mu}\in\mathbb{C}$ such that
\begin{eqnarray}
(\hat{\mathds{1}}_R\otimes \hat{F}_{\alpha\mu})\ket{\Omega} = c_{\alpha\mu}\,\ket{\Omega} \quad \text{for all }\alpha,\mu.
\label{eq:S2_Kraus_on_Omega}
\end{eqnarray}
Expanding $\ket{\Omega}$ in the code basis and comparing coefficients in Eq.~(\ref{eq:S2_Kraus_on_Omega}) yields
\begin{eqnarray}
\hat{F}_{\alpha\mu}\ket{i} = c_{\alpha\mu}\,\ket{i} \quad \text{for all code basis vectors }\ket{i}.
\label{eq:S2_F_scalar_on_code}
\end{eqnarray}
Equivalently, $\hat{F}_{\alpha\mu}$ acts as a scalar on the code subspace:
\begin{eqnarray}
\hat{F}_{\alpha\mu}\,\hat{P}  = c_{\alpha\mu}\,\hat{P},
\quad
\hat{P}\,\hat{F}_{\alpha\mu}^\dagger  = c_{\alpha\mu}^\ast\,\hat{P}.
\label{eq:S2_F_scalar_projector}
\end{eqnarray}

Finally, using that $\sum_\alpha \hat{R}_\alpha^\dagger \hat{R}_\alpha=\hat{\mathds{1}}$ (trace preservation of $\mathcal{R}$), we compute on the code:
\begin{eqnarray}
\hat{P}\hat{E}_\mu^\dagger \hat{E}_\nu \hat{P}
&=&
\hat{P}\hat{E}_\mu^\dagger\Bigl(\sum_\alpha \hat{R}_\alpha^\dagger \hat{R}_\alpha\Bigr)\hat{E}_\nu \hat{P}
 =
\sum_\alpha
(\hat{P}\hat{E}_\mu^\dagger \hat{R}_\alpha^\dagger)(\hat{R}_\alpha \hat{E}_\nu \hat{P})
\nonumber\\
&=&
\sum_\alpha
(\hat{P}\hat{F}_{\alpha\mu}^\dagger)(\hat{F}_{\alpha\nu}\hat{P})
 =
\sum_\alpha
(c_{\alpha\mu}^\ast \hat{P})(c_{\alpha\nu}\hat{P})
 =
\Bigl(\sum_\alpha c_{\alpha\mu}^\ast c_{\alpha\nu}\Bigr)\hat{P}.
\label{eq:S2_KL_from_c}
\end{eqnarray}
Thus Eq.~(\ref{eq:S2_KL_condition}) holds with $\alpha_{\mu\nu}:=\sum_\alpha c_{\alpha\mu}^\ast c_{\alpha\nu}$.
\end{proof}

\noindent

{\bf Comment.} Lemma~\ref{lem:S2_KL} is the ``necessity'' direction of the KL theorem; it is standard and is proved in many references (see, e.g., Refs.~\cite{knill1997theory,nielsen_chuang}). We use it with $\hat{E}_\mu=\hat{K}_a$ from Eq.~(\ref{eq:S2_Kraus_trace}).

Applying Lemma~\ref{lem:S2_KL} to the erasure channel $\mathcal{N}_A=\tr_A$ and its Kraus operators $\{\hat{K}_a\}$, exact erasure correctability (Eq.~(\ref{eq:S2_erasure_correctable})) implies the existence of complex numbers $\alpha_{ab}$ such that
\begin{eqnarray}
\hat{P} \hat{K}_a^\dagger \hat{K}_b \hat{P}  = \alpha_{ab}\,\hat{P}
\quad
\text{for all }a,b\in\{1,\dots,d_A\}.
\label{eq:S2_KL_for_trace}
\end{eqnarray}
Using Eq.~(\ref{eq:S2_KdagK}), this becomes the \emph{scalar action} property for all matrix units on $A$:
\begin{eqnarray}
\hat{P}\Bigl(\ket{a}\bra{b}_A\otimes \hat{\mathds{1}}_{A^c}\Bigr)\hat{P}
 = \alpha_{ab}\,\hat{P}
\quad
\text{for all }a,b.
\label{eq:S2_scalar_matrix_units}
\end{eqnarray}

Equation~(\ref{eq:S2_scalar_matrix_units}) is the algebraic expression of local indistinguishability on $A$: on the code subspace, every operator supported entirely on $A$ acts as a scalar multiple of the identity.

\subsection{From scalar action to a state-independent reduced density matrix}

We now show that Eq.~(\ref{eq:S2_scalar_matrix_units}) forces the reduced density matrix on $A$ to be independent of the logical input.

Let $\hat{\rho}$ be an arbitrary density operator supported on the code, i.e.,
\begin{eqnarray}
\hat{\rho}  = \hat{P}\hat{\rho}\hat{P},
\quad \tr{(\hat{\rho})}=1.
\label{eq:S2_code_state_general}
\end{eqnarray}
(For our theorem, one can take $\hat{\rho}=\ketbra{\Psi(\psi_L)}{\Psi(\psi_L)}$.) Define the reduced state $\hat{\rho}_A:=\tr_{A^c}(\hat{\rho})$. For each pair $(a,b)$ we compute the $(b,a)$ matrix element of $\hat{\rho}_A$ in the basis $\{\ket{a}\}$:
\begin{eqnarray}
\bra{b}\hat{\rho}_A\ket{a}
&=&
\tr{\Bigl(\ket{a}\bra{b}_A\,\hat{\rho}_A\Bigr)}
\label{eq:S2_rhoA_matrix_1}
\\
&=&
\tr{\Bigl(\bigl(\ket{a}\bra{b}_A\otimes \hat{\mathds{1}}_{A^c}\bigr)\,\hat{\rho}\Bigr)}.
\label{eq:S2_rhoA_matrix_2}
\end{eqnarray}
Using $\hat{\rho}=\hat{P}\hat{\rho}\hat{P}$ and cyclicity of trace, we insert $\hat{P}$ on both sides of the operator:
\begin{eqnarray}
\tr{\Bigl(\bigl(\ket{a}\bra{b}_A\otimes \hat{\mathds{1}}_{A^c}\bigr)\,\hat{\rho}\Bigr)}
&=&
\tr{\Bigl(\hat{P}\bigl(\ket{a}\bra{b}_A\otimes \hat{\mathds{1}}_{A^c}\bigr)\hat{P}\,\hat{\rho}\Bigr)}.
\label{eq:S2_insert_P}
\end{eqnarray}
Now apply Eq.~(\ref{eq:S2_scalar_matrix_units}):
\begin{eqnarray}
\tr{\Bigl(\hat{P}\bigl(\ket{a}\bra{b}_A\otimes \hat{\mathds{1}}_{A^c}\bigr)\hat{P}\,\hat{\rho}\Bigr)}
&=&
\tr{\Bigl(\alpha_{ab} \hat{P}\,\hat{\rho}\Bigr)}
 =
\alpha_{ab}\,\tr{(\hat{P}\hat{\rho})}
 =
\alpha_{ab}\,\tr{(\hat{\rho})}
 =
\alpha_{ab}.
\label{eq:S2_matrix_element_fixed}
\end{eqnarray}
Combining Eqs.~(\ref{eq:S2_rhoA_matrix_1})--(\ref{eq:S2_matrix_element_fixed}), we have shown that
\begin{eqnarray}
\bra{b}\hat{\rho}_A\ket{a}
 =
\alpha_{ab}
\quad
\text{for all code states }\hat{\rho}=\hat{P}\hat{\rho}\hat{P}.
\label{eq:S2_rhoA_entries_alpha}
\end{eqnarray}
In particular, the reduced density matrix $\hat{\rho}_A$ is uniquely fixed by the constants $\alpha_{ab}$ and therefore is independent of the logical input state.

Define $\hat{\omega}_A$ by its matrix elements
\begin{eqnarray}
\bra{b}\hat{\omega}_A\ket{a}  := \alpha_{ab}
\quad
\text{for all }a,b.
\label{eq:S2_define_omega}
\end{eqnarray}
Then Eq.~(\ref{eq:S2_rhoA_entries_alpha}) implies
\begin{eqnarray}
\hat{\rho}_A  = \hat{\omega}_A
\quad
\text{for all code states }\hat{\rho}.
\label{eq:S2_rhoA_equals_omega}
\end{eqnarray}

\medskip\noindent
{\bf $\hat{\omega}_A$ is a density operator.}
We briefly verify that $\hat{\omega}_A$ is a valid density matrix. First, taking $\hat{O}_A=\hat{\mathds{1}}_A$ in Eq.~(\ref{eq:S2_scalar_matrix_units}) (or simply noting $\tr{(\hat{\rho}_A)}=1$) shows
\begin{eqnarray}
\tr{(\hat{\omega}_A)}  = 1.
\label{eq:S2_trace_one}
\end{eqnarray}
Second, Hermiticity follows because $\hat{\omega}_A$ arises as the reduced state of some code state (e.g., fix any $\ket{\psi_L^0}$ and set $\hat{\omega}_A:=\hat{\rho}_A(\psi_L^0)$), and hence $\hat{\omega}_A^\dagger=\hat{\omega}_A$. Finally, positivity can be seen directly from expectation values: for any $\ket{v}\in\mathcal{H}_A$,
\begin{eqnarray}
\bra{v}\hat{\omega}_A\ket{v}
 =
\tr{\Bigl((\ketbra{v}{v}_A\otimes \hat{\mathds{1}}_{A^c})\,\hat{\rho}\Bigr)}
\ \ge\ 0,
\label{eq:S2_omega_positive}
\end{eqnarray}
because $\ketbra{v}{v}_A\otimes \hat{\mathds{1}}_{A^c}\succeq 0$ and $\hat{\rho}\succeq 0$. Thus $\hat{\omega}_A\succeq 0$.

\subsection{Conclusion of the theorem: constancy of $E_A$ and vanishing of $\Delta_A$}

We now apply the above to the encoded pure states $\hat{\rho}(\psi_L)=\ketbra{\Psi(\psi_L)}{\Psi(\psi_L)}$.
Equation~(\ref{eq:S2_rhoA_equals_omega}) gives the key structural conclusion:
\begin{eqnarray}
\hat{\rho}_A(\psi_L)  = \hat{\omega}_A
\quad
\text{for all }\ket{\psi_L}\in\mathcal{H}_L,
\label{eq:S2_rhoA_const_final}
\end{eqnarray}
which is Eq.~(\ref{eq:S2_rhoA_constant}) in the theorem statement.

Since $E_A(\psi_L)$ depends on $\psi_L$ only through $\hat{\rho}_A(\psi_L)$, it follows immediately that $E_A(\psi_L)$ is constant over logical inputs:
\begin{eqnarray}
E_A(\psi_L)
 =
1-\tr{\Bigl(\hat{\rho}_A(\psi_L)^2\Bigr)}
 =
1-\tr{\Bigl(\hat{\omega}_A^2\Bigr)}
\quad
\text{for all }\psi_L.
\label{eq:S2_EA_constant}
\end{eqnarray}
Therefore, its variance under the Haar measure vanishes and hence the logical-ED is zero:
\begin{eqnarray}
\Delta_A  = 0.
\label{eq:S2_Delta_zero}
\end{eqnarray}
This completes the proof of Theorem~\ref{thm:S2_main}. \qed

\subsection{A useful equivalent form: scalar action of all operators on $A$}

For later use (and to emphasize the operator-algebraic content), we record an equivalent extension of Eq.~(\ref{eq:S2_scalar_matrix_units}) from matrix units to arbitrary operators on $A$. Any $\hat{O}_A\in\mathcal{B}(\mathcal{H}_A)$ can be expanded as $\hat{O}_A=\sum_{a,b} O_{ab}\ket{a}\bra{b}$. Multiplying Eq.~(\ref{eq:S2_scalar_matrix_units}) by $O_{ab}$ and summing yields
\begin{eqnarray}
\hat{P}\bigl(\hat{O}_A\otimes \hat{\mathds{1}}_{A^c}\bigr)\hat{P}
 =
\left(\sum_{a,b} O_{ab}\alpha_{ab}\right)\hat{P}
 =
\tr{\bigl(\hat{\omega}_A \hat{O}_A\bigr)}\,\hat{P}.
\label{eq:S2_scalar_all_ops}
\end{eqnarray}
Thus, on the code subspace, every operator supported on $A$ is proportional to the identity, with proportionality constant given by the $\hat{\omega}_A$-expectation value. Equation~(\ref{eq:S2_scalar_all_ops}) is often used as a concise formulation of local indistinguishability on correctable regions.


\section{Haar-moment formulas and technical lemmas for the ED-soft distance}

\subsection{Definitions and equivalent forms}

For a fixed physical region $A\subset\{1,\dots,n\}$, the logical entanglement deviation (logical-ED) is
\begin{eqnarray}
\Delta_A := \sqrt{\int d\psi_L\, E_A(\psi_L)^2 - \left(\int d\psi_L\, E_A(\psi_L)\right)^2},
\quad
E_A(\psi_L):=1-\tr{\bigl(\hat{\rho}_A(\psi_L)^2\bigr)},
\label{eq:S3_Delta_def_recall}
\end{eqnarray}
with $\hat{\rho}_A(\psi_L)=\tr_{A^c}\!\left(\hat{V}\ketbra{\psi_L}{\psi_L}\hat{V}^\dagger\right)$ and Haar measure $d\psi_L$ on the logical Hilbert space $\mathcal{H}_L$.

\subsection{Explicit second- and fourth-moment formulas}

Let $d_L:=\dim(\mathcal{H}_L)$ and let $\hat{S}_A$ swap region $A$ between two physical copies while acting trivially on $A^c$. Define the two-copy logical operator
\begin{eqnarray}
\hat{M}_A
:=
(\hat{V}^\dagger)^{\otimes 2}\hat{S}_A \hat{V}^{\otimes 2}
\quad\text{on }\mathcal{H}_L^{\otimes 2}.
\label{eq:S3_MA_def}
\end{eqnarray}
Writing $P_A(\psi_L):=\tr{(\hat{\rho}_A(\psi_L)^2)}$, the swap identity gives
\begin{eqnarray}
P_A(\psi_L)
=
\tr{\!\left[\hat{M}_A
\bigl(\ketbra{\psi_L}{\psi_L}\bigr)^{\otimes 2}\right]}.
\label{eq:S3_purity_MA}
\end{eqnarray}
For $r\geq1$, the Haar projector identity~\cite{mele2024haar} is
\begin{eqnarray}
\int d\psi_L\, \bigl(\ketbra{\psi_L}{\psi_L}\bigr)^{\otimes r} = \frac{\hat{\Pi}^{(r)}_{\rm sym}}{\binom{d_L+r-1}{r}},
\quad
\hat{\Pi}^{(r)}_{\rm sym} = \frac{1}{r!}\sum_{\pi\in S_r}\hat{W}_\pi,
\label{eq:S3_Haar_projector}
\end{eqnarray}
where $\hat{W}_\pi$ permutes the $r$ logical copies. Therefore
\begin{eqnarray}
\mu_A
:=
\int d\psi_L\,P_A(\psi_L)
=
\frac{\tr{\!\left(\hat{M}_A\hat{\Pi}^{(2)}_{\rm sym}\right)}}{\binom{d_L+1}{2}},
\label{eq:S3_mean_purity_explicit}
\end{eqnarray}
and, with $\hat{M}_A^{(12)}$ and $\hat{M}_A^{(34)}$ acting on logical copies $(1,2)$ and $(3,4)$,
\begin{eqnarray}
\nu_A
:=
\int d\psi_L\,P_A(\psi_L)^2
=
\frac{\tr{\!\left(\hat{M}_A^{(12)}\hat{M}_A^{(34)}
\hat{\Pi}^{(4)}_{\rm sym}\right)}}{\binom{d_L+3}{4}}.
\label{eq:S3_second_purity_explicit}
\end{eqnarray}
Since $E_A=1-P_A$, subtraction of a constant does not change the variance. Hence
\begin{eqnarray}
\overline{E}_A=1-\mu_A,
\quad
\Delta_A^2=\nu_A-\mu_A^2.
\label{eq:S3_ED_explicit}
\end{eqnarray}
Equations~(\ref{eq:S3_mean_purity_explicit})--(\ref{eq:S3_ED_explicit}) are the explicit contraction formulas invoked in the main text. They also show directly that a logical state $2$-design suffices for $\overline{E}_A$ and a logical state $4$-design suffices for $\Delta_A$.

\begin{proof}[Derivation]---Equation~(\ref{eq:S3_purity_MA}) is the two-copy swap identity pulled back through the encoder. Averaging it and applying Eq.~(\ref{eq:S3_Haar_projector}) with $r=2$ gives Eq.~(\ref{eq:S3_mean_purity_explicit}). Squaring Eq.~(\ref{eq:S3_purity_MA}) places one copy of $\hat{M}_A$ on logical copies $(1,2)$ and the other on $(3,4)$; the $r=4$ case of Eq.~(\ref{eq:S3_Haar_projector}) then gives Eq.~(\ref{eq:S3_second_purity_explicit}). Finally, $\operatorname{Var}(1-P_A)=\operatorname{Var}(P_A)$, proving Eq.~(\ref{eq:S3_ED_explicit}).
\end{proof}

We recall the ED-soft distance introduced in the main text: for a tolerance $\varepsilon\ge 0$,
\begin{eqnarray}
d_{\Delta}(\varepsilon)
 :=
\max\Bigl\{
s\in\{0,1,\dots,n\}:\ 
\forall A\ \bigl(\abs{A}\le s\bigr)\Rightarrow \Delta_A\le \varepsilon
\Bigr\}.
\label{eq:S3_soft_distance_def}
\end{eqnarray}
Equivalently, for $s\in\{0,1,\dots,n\}$ define the cumulative worst-case ED by
\begin{eqnarray}
\Delta_{\max}(s)
 :=
\max\Bigl\{\Delta_A:\ A\subseteq\{1,\dots,n\},\ \abs{A}\le s\Bigr\},
\label{eq:S3_Delta_max_def}
\end{eqnarray}
(with the convention $\Delta_{\max}(0)=0$ since the only region is $A=\emptyset$).
Then
\begin{eqnarray}
d_{\Delta}(\varepsilon)
 =
\max\Bigl\{
s\in\{0,1,\dots,n\}:\ \Delta_{\max}(s)\le \varepsilon
\Bigr\}.
\label{eq:S3_soft_distance_equiv}
\end{eqnarray}
Equation~(\ref{eq:S3_soft_distance_equiv}) is often useful in practice: $d_\Delta(\varepsilon)$ is the largest size scale up to which the \emph{worst-case} region entanglement fluctuation is bounded by $\varepsilon$. The finite integer domain makes the maximum well defined even if every region satisfies the tolerance. The diagnostic $\Delta_A$ and its tolerance $\varepsilon$ are real valued, whereas $d_\Delta(\varepsilon)$ counts physical sites. Here ``soft'' refers to relaxing the ED-uniformity criterion with a tunable tolerance; it does not mean that the resulting radius is continuous. Because $\Delta_{\max}(s)$ includes \emph{all} smaller regions, a vanishing ED on a large region cannot erase an earlier violation.

For geometric codes it is natural to restrict to a region family controlled by a geometric scale. Let $\mathfrak{A}(\ell)$ be a chosen set of regions on a finite ordered scale set $\mathcal{L}$ (e.g., diameter-bounded regions or width-bounded strips at the positive scales), with an initial scale $\ell_0$ satisfying $\mathfrak{A}(\ell_0)=\{\emptyset\}$. For a radius interpretation we assume $\mathfrak{A}(\ell_1)\subseteq\mathfrak{A}(\ell_2)$ whenever $\ell_1\le\ell_2$; a nonnested collection should instead be reported directly as a profile. We define the corresponding \emph{geometric} soft distance by
\begin{eqnarray}
\ell_{\Delta}(\varepsilon)
 :=
\max\Bigl\{
\ell\in\mathcal{L}:\ \forall A\in \mathfrak{A}(\ell)\Rightarrow \Delta_A\le \varepsilon
\Bigr\},
\label{eq:S3_geo_soft_distance_def}
\end{eqnarray}
and similarly introduce $\Delta_{\max}^{(\mathfrak{A})}(\ell):=\max_{A\in\mathfrak{A}(\ell)}\Delta_A$, with the maximum defining $\ell_\Delta$ taken over the same finite scale set. Other finite ordered scales, such as the distinct diameters of a finite metric graph, can be used in the same way. The global quantity $d_\Delta$ summarizes all geometries by site count; the selected family and its scale specify the geometric meaning of $\ell_\Delta$. Neither unrestricted radius alone certifies operational recovery, since ED can vanish on logically informative regions. Supplementary Note~S7 identifies a protected-axis condition under which the family-restricted zero-tolerance radius is an exact erasure-correction certificate.

\subsection{Distance implies a vanishing window for logical-ED}

We first formalize a standard implication relating code distance to erasure correctability.

\begin{lemma}[Distance implies correctability of all small erasures]\label{lem:S3_distance_erasure}
Let $\mathcal{C}$ be an exact $[[n,k,d]]$ subspace code with projector $\hat{P}$.
Then for any region $A$ with $\abs{A}\le d-1$, the erasure of $A$ is exactly correctable on $\mathcal{C}$.
\end{lemma}

\begin{proof}---One definition of distance $d$ for a subspace code is that the code detects all errors supported on fewer than $d$ physical sites:
\begin{eqnarray}
\hat{P}\hat{E}\hat{P}  = c_E\, \hat{P}
\quad
\text{for all operators }\hat{E}\text{ supported on }\abs{{\rm supp}(\hat{E})}\le d-1.
\label{eq:S3_distance_detect_def}
\end{eqnarray}
Fix a region $A$ with $\abs{A}\le d-1$ and let $\{\hat{E}_\mu\}$ be any operator basis supported on $A$ (e.g., matrix units on $\mathcal{H}_A$ tensored with identity on $A^c$). Then Eq.~(\ref{eq:S3_distance_detect_def}) implies
\begin{eqnarray}
\hat{P} \hat{E}_\mu^\dagger \hat{E}_\nu \hat{P}  = \alpha_{\mu\nu}\, \hat{P}
\quad
\forall \mu,\nu,
\label{eq:S3_KL_for_A}
\end{eqnarray}
which is precisely the Knill--Laflamme condition for correcting the erasure supported on $A$. Hence $A$ is erasure-correctable.
\end{proof}

Combining Lemma~\ref{lem:S3_distance_erasure} with Theorem~1 (main text) (proved in Supplementary Note~S2), we obtain the key implication that underlies Proposition~\ref{prop:S3_prop2}.

\begin{lemma}[Distance implies vanishing logical-ED for all small regions]\label{lem:S3_distance_vanish}
For any exact $[[n,k,d]]$ code and any region $A$ with $\abs{A}\le d-1$, one has $\Delta_A=0$.
\end{lemma}

\begin{proof}---By Lemma~\ref{lem:S3_distance_erasure}, any $\abs{A}\le d-1$ erasure is exactly correctable. By Theorem~1 (main text), exact correctability implies $\hat{\rho}_A(\psi_L)$ is independent of $\psi_L$, hence $E_A(\psi_L)$ is constant and $\Delta_A=0$. 
\end{proof}

\subsection{Standard-distance lower bound on the ED-soft distance}

We now prove the proposition announced in the main text in a form that is both general and logically complete.

\begin{proposition}[Standard-distance lower bound on $d_\Delta(\varepsilon)$]\label{prop:S3_prop2}
Let $\mathcal{C}$ be an exact $[[n,k,d]]$ code and let $d_\Delta(\varepsilon)$ be defined as in Eq.~(\ref{eq:S3_soft_distance_def}). Then for all $\varepsilon\ge 0$,
\begin{eqnarray}
d_\Delta(\varepsilon)\ \ge\ d-1.
\label{eq:S3_prop2_lb}
\end{eqnarray}
In particular, $d_\Delta(0)\ge d-1$.

Moreover, if there exists at least one region $A_\star$ with $\abs{A_\star}\le d$ such that $\Delta_{A_\star}>0$, then necessarily
\begin{eqnarray}
d_\Delta(0) = d-1.
\label{eq:S3_prop2_eq_cond}
\end{eqnarray}
\end{proposition}

\begin{proof}---The lower bound~(\ref{eq:S3_prop2_lb}) follows directly from Lemma~\ref{lem:S3_distance_vanish}: for all $\abs{A}\le d-1$ we have $\Delta_A=0\le \varepsilon$, hence $s=d-1$ is admissible in the maximization defining $d_\Delta(\varepsilon)$.

For the second statement, suppose $\Delta_{A_\star}>0$ for some $\abs{A_\star}\le d$. Then $s=d$ is \emph{not} admissible for $\varepsilon=0$ because the defining condition in Eq.~(\ref{eq:S3_soft_distance_def}) requires $\Delta_A\le 0$ for \emph{all} $\abs{A}\le d$. Since $d_\Delta(0)\ge d-1$ from the first part, we conclude $d_\Delta(0)=d-1$.
\end{proof}

\medskip\noindent
{\bf Remark (on the possibility $d_\Delta(0)>d-1$).}
The inequality $d_\Delta(0)\ge d-1$ is universal for exact codes. However, the equality $d_\Delta(0)=d-1$ is \emph{not automatic} because $\Delta_A$ probes only the fluctuation of the linear entropy (equivalently, the purity) of $\hat{\rho}_A(\psi_L)$. It is possible (in principle) for $\hat{\rho}_A(\psi_L)$ to depend on $\psi_L$ while maintaining constant purity, in which case $\Delta_A=0$ despite the region being logically informative. This is the same structural subtlety noted in the main text: $\Delta_A=0$ does not imply $\hat{\rho}_A(\psi_L)$ is constant.

A simple illustrative mechanism is unitary covariance on $A$:
if there exists a family of unitaries $\hat{U}_A(\psi_L)$ such that
$\hat{\rho}_A(\psi_L)=\hat{U}_A(\psi_L)\,\hat{\omega}_A\,\hat{U}_A(\psi_L)^\dagger$
for a fixed spectrum $\hat{\omega}_A$, then $\tr{(\hat{\rho}_A(\psi_L)^2)}$ is constant and $\Delta_A=0$ while $\hat{\rho}_A(\psi_L)$ varies.
As a code-level example, in the pure $[[5,1,3]]$ perfect code every region of at most two qubits is correctable and every region of at least three qubits has a correctable complement. Complement symmetry therefore gives $\Delta_A=0$ for every $A$, so $d_\Delta(0)=5$ even though $d-1=2$.
This does not affect the operational direction used in our main theorem (correctable $\Rightarrow \Delta_A=0$). It does mean, however, that $d_\Delta(\varepsilon)$ records ED-uniformity at the chosen tolerance but does not by itself certify correctability or approximate recovery; in particular, $d_\Delta(0)$ need not coincide with $d-1$.

\medskip\noindent
{\bf Remark (scale dependence and normalization).}
The absolute value of $\Delta_A$ inherits the purity scale of the subsystem and can become small solely because the mean purity is small. Define
\begin{eqnarray}
\overline{P}_A:=1-\overline{E}_A,
\quad
\delta_A:=\frac{\Delta_A}{\overline{P}_A}.
\label{eq:S3_relative_ED}
\end{eqnarray}
Since $\overline{P}_A>0$ in finite dimension, $\delta_A$ is well defined and removes this overall scale. One may define a normalized ED-soft distance by replacing $\Delta_A$ with $\delta_A$ in Eq.~(\ref{eq:S3_soft_distance_def}). Raw and normalized profiles answer different questions and should not be interchanged. In general neither is an operational recovery metric by itself; Supplementary Note~S7 gives an exact, family-restricted exception under a logical-support condition.

\subsection{Basic structural properties of $d_\Delta(\varepsilon)$ and $\ell_\Delta(\varepsilon)$}

We collect several properties that are immediate but useful when interpreting the soft distance.

\begin{lemma}[Monotonicity in tolerance]\label{lem:S3_monotone_eps}
The soft distance $d_\Delta(\varepsilon)$ is nondecreasing in $\varepsilon$:
if $0\le \varepsilon_1\le \varepsilon_2$, then
\begin{eqnarray}
d_\Delta(\varepsilon_1)\ \le\ d_\Delta(\varepsilon_2).
\label{eq:S3_monotone_eps}
\end{eqnarray}
The same holds for the geometric version $\ell_\Delta(\varepsilon)$.
\end{lemma}

\begin{proof}---If a size $s$ satisfies $\Delta_A\le \varepsilon_1$ for all $\abs{A}\le s$, then certainly $\Delta_A\le \varepsilon_2$ for all $\abs{A}\le s$ when $\varepsilon_2\ge \varepsilon_1$. Hence the admissible set for $\varepsilon_1$ is contained in that for $\varepsilon_2$, so the maximum cannot decrease. The geometric statement is identical with $A\in\mathfrak{A}(\ell)$.
\end{proof}

Because the physical system contains finitely many regions, $d_\Delta(\varepsilon)$ is integer valued and can change only when $\varepsilon$ crosses one of the finitely many values $\{\Delta_A\}$. It is therefore a nondecreasing step function of the tolerance, not a continuous interpolation. The same statement holds for any finite geometric family.

\begin{lemma}[Trivial bounds and saturation]\label{lem:S3_trivial_bounds}
For any code on $n$ physical sites,
\begin{eqnarray}
0\ \le\ d_\Delta(\varepsilon)\ \le\ n
\quad \text{for all }\varepsilon\ge 0.
\label{eq:S3_trivial_bounds}
\end{eqnarray}
Moreover, $d_\Delta(\varepsilon)=n$ if and only if $\Delta_A\le \varepsilon$ for all regions $A$.
\end{lemma}

\begin{proof}---The bound $0\le d_\Delta(\varepsilon)$ holds because $s=0$ is always admissible. The upper bound is immediate from the definition since no region has size larger than $n$. The final equivalence is a restatement of the defining condition at $s=n$.
\end{proof}

\begin{lemma}[Invariance under logical and physical local basis changes]\label{lem:S3_invariance}
For a fixed region $A$, if
$\hat{V}'=(\hat{U}_A\otimes\hat{U}_{A^c})\hat{V}\hat{U}_L$,
then $\Delta_A(\hat{V}')=\Delta_A(\hat{V})$. If a single physical unitary factorizes onsite,
$\hat{U}_{\rm phys}=\bigotimes_{i=1}^n\hat{U}_i$, and
$\hat{V}'=\hat{U}_{\rm phys}\hat{V}\hat{U}_L$, then this equality holds simultaneously for every region; consequently $d_\Delta(\varepsilon)$ and every profile over a fixed region family are invariant.
\end{lemma}

\begin{proof}---The logical unitary reparametrizes the Haar integral. For the fixed cut $A|A^c$, the physical unitary conjugates $\hat{\rho}_A(\psi_L)$ by $\hat{U}_A$ and therefore preserves its spectrum and purity. An onsite product unitary factorizes across every physical bipartition, so the fixed-region argument applies simultaneously to all $A$ entering the definitions of the soft distance and the profiles.
\end{proof}

\subsection{Exact soft-distance examples}
\label{sec:S3_soft_distance_example}

We give a complete evaluation of Eq.~(\ref{eq:S3_soft_distance_def}) for the rotated $[[9,1,3]]$ surface-code patch. Label its $3\times3$ data qubits by $1,\dots,9$ in row-major order. The independent $X$- and $Z$-type stabilizer generators are
\begin{eqnarray}
\mathcal{S}&=&\bigl\langle
\hat{X}_1\hat{X}_2\hat{X}_4\hat{X}_5,\quad
\hat{X}_5\hat{X}_6\hat{X}_8\hat{X}_9,\quad
\hat{X}_2\hat{X}_3,\quad \hat{X}_7\hat{X}_8,
\nonumber\\
&&\hat{Z}_2\hat{Z}_3\hat{Z}_5\hat{Z}_6,\quad
\hat{Z}_4\hat{Z}_5\hat{Z}_7\hat{Z}_8,\quad
\hat{Z}_1\hat{Z}_4,\quad \hat{Z}_6\hat{Z}_9
\bigr\rangle,
\label{eq:S3_small_patch_stabilizers}
\end{eqnarray}
and logical representatives can be chosen as
\begin{eqnarray}
\hat{\overline{X}}=\hat{X}_1\hat{X}_4\hat{X}_7,
\quad
\hat{\overline{Z}}=\hat{Z}_1\hat{Z}_2\hat{Z}_3.
\label{eq:S3_small_patch_logicals}
\end{eqnarray}
These generators commute and have binary rank $8$, while the logical representatives commute with the stabilizers and anticommute with each other. Thus the patch encodes one qubit. The binary-rank and affine-coset procedure in Eqs.~(\ref{eq:S5_tA_def})--(\ref{eq:S5_affine_system}) evaluates every one of its $2^9=512$ regions without Haar sampling. In this example, we also checked each count directly against all $2^8$ stabilizer elements and their three logical cosets. For each region, the number $q_A$ of supported nonidentity logical cosets is $0$, $1$, or $3$, and Eq.~(\ref{eq:S5_delta_q}) gives its exact ED.

To recover the radius, define the maximum at \emph{exact} size $j$ and then retain the cumulative maximum,
\begin{eqnarray}
a_j:=\max_{\abs{A}=j}\Delta_A,
\quad
\Delta_{\max}(s)=\max_{0\le j\le s}a_j.
\label{eq:S3_small_patch_size_profile}
\end{eqnarray}
Table~\ref{tab:S3_small_patch_soft} lists the complete support counts and both maxima. In particular, the top-row region $A=\{1,2,3\}$ has $t_A=1$ and $q_A=1$, so $\Delta_A=1/(2\sqrt{45})$. No region has a larger ED. The counts sum to $156$, $200$, and $156$ regions in the $q_A=0$, $1$, and $3$ sectors, respectively. The return of $a_j$ to zero for $j\ge7$ reflects the full-access sector and complement symmetry, whereas $\Delta_{\max}(s)$ retains the ED-active size-$3$ obstruction.

\begin{table}[t]
\centering
\setlength{\tabcolsep}{15pt}
\begin{tabular}{c c c c c c c}
\hline\hline
$j$ & $\binom{9}{j}$ & $q_A=0$ & $q_A=1$ & $q_A=3$ & $\sqrt{45}\,a_j$ & $\sqrt{45}\,\Delta_{\max}(j)$\\
\hline
$0$ & $1$   & $1$  & $0$  & $0$  & $0$   & $0$\\
$1$ & $9$   & $9$  & $0$  & $0$  & $0$   & $0$\\
$2$ & $36$  & $36$ & $0$  & $0$  & $0$   & $0$\\
$3$ & $84$  & $68$ & $16$ & $0$  & $1/2$ & $1/2$\\
$4$ & $126$ & $42$ & $84$ & $0$  & $1/2$ & $1/2$\\
$5$ & $126$ & $0$  & $84$ & $42$ & $1/2$ & $1/2$\\
$6$ & $84$  & $0$  & $16$ & $68$ & $1/2$ & $1/2$\\
$7$ & $36$  & $0$  & $0$  & $36$ & $0$   & $1/2$\\
$8$ & $9$   & $0$  & $0$  & $9$  & $0$   & $1/2$\\
$9$ & $1$   & $0$  & $0$  & $1$  & $0$   & $1/2$\\
\hline\hline
\end{tabular}
\caption{Complete $512$-region evaluation for the rotated $[[9,1,3]]$ patch. The three middle columns count regions of size $j$ in each logical-access sector. Here $a_j$ is the exact-size maximum and $\Delta_{\max}(j)$ is the cumulative maximum entering the soft-distance definition. All entries are exact.}
\label{tab:S3_small_patch_soft}
\end{table}

Writing $\varepsilon_c:=1/(2\sqrt{45})\simeq0.0745356$, the global soft distance is therefore
\begin{eqnarray}
d_\Delta(\varepsilon)=
\begin{cases}
2,&0\le\varepsilon<\varepsilon_c,\\
9,&\varepsilon\ge\varepsilon_c.
\end{cases}
\label{eq:S3_small_patch_soft_exact}
\end{eqnarray}
At zero tolerance this equals $d-1$, while increasing the tolerance to or beyond the exact ED maximum produces saturation at $n=9$. Saturation states that every region meets the specified \emph{purity-fluctuation} tolerance; it does not assert that erasing every region is recoverable. For example, the full patch carries the entire logical state and has $\Delta_A=0$.

For a geometric example using the existing $[[121,1,11]]$ data, take the boundary-anchored width-$3$ corridors
\begin{eqnarray}
A_\ell:=\{(x,y):0\le x<\ell,\ 0\le y<3\},
\quad
\mathfrak{A}(\ell):=\{A_r:0\le r\le\ell\},
\quad \ell\in\{0,\dots,11\}.
\label{eq:S3_corridor_family}
\end{eqnarray}
The family is nested and $A_0=\emptyset$. The rectangle profile of Supplementary Note~S5 gives $\Delta_{A_\ell}=0$ for $\ell\le10$, whereas the spanning $A_{11}$ is a full-width strip of depth $3$. Equation~(\ref{eq:S5_strip_delta}) gives $\Delta_{A_{11}}=2^{-5}/\sqrt{45}\simeq0.00465847$, hence
\begin{eqnarray}
\ell_\Delta(\varepsilon)=
\begin{cases}
10,&0\le\varepsilon<2^{-5}/\sqrt{45},\\
11,&\varepsilon\ge2^{-5}/\sqrt{45}.
\end{cases}
\label{eq:S3_corridor_soft_exact}
\end{eqnarray}
This is a length radius for the specified corridor family, not the global site-count radius of the $121$-qubit patch. Supplementary Note~S7 verifies the retained logical representative and proves that the lower-tolerance branch certifies exact correction of the specified $30$-qubit region $A_{10}$. Its absolute tolerance is also tied to the local purity scale: the spanning strip has $\Delta_A/\overline{P}_A=1/\sqrt{20}$ by Eq.~(\ref{eq:S5_strip_relative}). Reporting the region family and raw or normalized tolerance therefore fixes the meaning of any soft-distance comparison.

\subsection{A useful inequality relating purity fluctuations to state distinguishability}

Although our main results only require the direction ``correctable $\Rightarrow \Delta_A=0$,'' it is sometimes helpful to bound entanglement fluctuations in terms of a direct measure of variation of the reduced states.
For any two density operators $\hat{\rho},\hat{\sigma}$ on $\mathcal{H}_A$, the identity
\begin{eqnarray}
\abs{\tr{(\hat{\rho}^2)}-\tr{(\hat{\sigma}^2)}}
 =
\abs{\tr{\bigl((\hat{\rho}-\hat{\sigma})(\hat{\rho}+\hat{\sigma}-\hat{\mathds{1}})\bigr)}}
\ \le\
\norm{\hat{\rho}-\hat{\sigma}}_1\,\norm{\hat{\rho}+\hat{\sigma}-\hat{\mathds{1}}}_\infty
\ \le\
\norm{\hat{\rho}-\hat{\sigma}}_1
\label{eq:S3_purity_trace_bound}
\end{eqnarray}
where the identity may be subtracted because $\tr(\hat{\rho}-\hat{\sigma})=0$, and the final operator norm is at most one because the spectrum of $\hat{\rho}+\hat{\sigma}$ lies in $[0,2]$. Thus
\begin{eqnarray}
\abs{E(\hat{\rho})-E(\hat{\sigma})}
 =
\abs{\tr{(\hat{\sigma}^2)}-\tr{(\hat{\rho}^2)}}
\ \le\ \norm{\hat{\rho}-\hat{\sigma}}_1.
\label{eq:S3_E_Lipschitz}
\end{eqnarray}
Consequently, if the reduced states $\hat{\rho}_A(\psi_L)$ vary little in trace distance over logical inputs, then $E_A(\psi_L)$ has small fluctuations and $\Delta_A$ is small. This provides a bridge between logical-ED and stronger state-based deviation measures (such as mean-square Hilbert--Schmidt deviation of $\hat{\rho}_A$), which may be useful when developing quantitative connections to approximate correctability.


\section{Region families and geometric lemmas for surface and local codes}

\subsection{Geometric codes, metrics, and basic region geometry}

A \emph{geometric code} is specified by a set of physical degrees of freedom arranged on a finite interaction graph (or lattice) $G=(V,E)$ with $\abs{V}=n$ sites, together with an encoding isometry $\hat{V}:\mathcal{H}_L\to\mathcal{H}_{\rm phys}$ whose image is a code subspace $\mathcal{C}\subset \mathcal{H}_{\rm phys}$. In most architectural settings one assumes:
(i) a notion of locality encoded by $G$; and
(ii) check operators (stabilizers, gauge generators, or commuting projectors) supported on bounded-diameter subsets in this metric. The logical-ED framework does not require stabilizer structure; however, many of the geometric lemmas below are most transparent for stabilizer (in particular CSS) codes such as the surface code~\cite{kitaev2003fault,dennis2002topological,fowler2012surface}.

We use the graph distance ${\rm dist}(\cdot,\cdot)$ on $V$ induced by shortest paths in $G$. For a region (subset) $A\subseteq V$ we define:
\begin{eqnarray}
{\rm diam}(A)  := \max_{i,j\in A}{\rm dist}(i,j),
\quad
{\rm dist}(i,A)  := \min_{j\in A}{\rm dist}(i,j).
\label{eq:S4_diam_dist}
\end{eqnarray}
We also define the $r$-neighborhood (thickening) and $r$-boundary of a region:
\begin{eqnarray}
N_r(A)  := \{i\in V:\ {\rm dist}(i,A)\le r\},
\quad
\partial_r A  := N_r(A)\setminus A.
\label{eq:S4_thickening}
\end{eqnarray}
When $r=1$ we write $N(A):=N_1(A)$ and $\partial A:=\partial_1 A$.

These geometric notions allow one to define \emph{region families} parametrized by a length scale (diameter, radius, width), which is the natural language for geometric ED profiles.

\subsection{Canonical region families for ED profiles}\label{sec:S4_canonical}

Given a code on $G=(V,E)$, we will often scan logical-ED over structured families of regions, rather than over all $2^n$ subsets. Below we record several families that are particularly natural for geometric codes. For the geometric soft distance, take $\ell_0$ to be a separate initial scale whose family is $\{\emptyset\}$ and adjoin $\emptyset$ to every later family. If a raw shape parameter has a meaningful zero value, such as a radius-zero ball or an unthickened path, include it at the first positive scale of the cumulative profile. For families with two parameters, fix an ordered scan and accumulate all regions included at earlier scales.

\smallskip\noindent
{\bf (i) Metric balls and connected patches.}
For $x\in V$ and an integer $\ell\ge 0$ define the (closed) ball
\begin{eqnarray}
B_\ell(x) := \{i\in V:\ {\rm dist}(i,x)\le \ell\}.
\label{eq:S4_ball}
\end{eqnarray}
At positive scales the cumulative ball family contains every ball with radius at most $\ell$, namely $\mathfrak{B}(\ell):=\{B_r(x):x\in V,\ 0\le r\le \ell\}$ before adjoining the empty region.
More generally, we define the connected-diameter family
\begin{eqnarray}
\mathfrak{A}_{\rm conn}(\ell)
 :=
\Bigl\{
A\subseteq V:\ A\ {\rm connected\ in}\ G,\ {\rm diam}(A)\le \ell
\Bigr\}.
\label{eq:S4_conn_family}
\end{eqnarray}
These families probe ``bulk'' locality and allow one to define a radius-type geometric soft distance
$\ell_\Delta(\varepsilon)$ as in Eq.~(\ref{eq:S3_geo_soft_distance_def}).

\smallskip\noindent
{\bf (ii) Boundary neighborhoods.}
If the code is defined on a manifold with boundary (e.g., planar surface code), then one has distinguished boundary components (rough/smooth, or more generally different boundary types). Let $\mathcal{B}_\alpha\subseteq V$ denote the set of sites adjacent to boundary component $\alpha$. A natural family is
\begin{eqnarray}
\mathfrak{A}_{\rm bdry}^{(\alpha)}(\ell)
 :=
\Bigl\{
A\subseteq N_\ell(\mathcal{B}_\alpha)
\Bigr\},
\label{eq:S4_boundary_family}
\end{eqnarray}
which probes how logical information approaches and interacts with boundaries.

\smallskip\noindent
{\bf (iii) Strips and boundary-to-boundary corridors.}
For planar codes encoding one logical qubit, the most informative non-local geometries are narrow corridors connecting boundary components. Fix two boundary components $\alpha,\beta$. A \emph{core path} $\gamma$ from $\alpha$ to $\beta$ is a path in $G$ whose endpoints are in $\mathcal{B}_\alpha$ and $\mathcal{B}_\beta$. For an integer width $w\ge 0$, define the $w$-thick strip around $\gamma$ as $N_w(\gamma)$. We define the strip family
\begin{eqnarray}
\mathfrak{A}_{\rm strip}^{(\alpha,\beta)}(w)
 :=
\Bigl\{
N_{w'}(\gamma)\cap V:\ \gamma:\alpha\to\beta,\ 0\leq w'\leq w
\Bigr\}.
\label{eq:S4_strip_family}
\end{eqnarray}
When the code has multiple boundary types, one can refine strip families by requiring $\gamma$ to be a primal or dual path, as discussed in Sec.~\ref{sec:S4_surface_geometry}.

\smallskip\noindent
{\bf (iv) Annuli and hole-neighborhoods (punctures).}
\begingroup
For punctured codes, let $\mathcal{B}_{\rm hole}$ denote a puncture boundary. A family of losses confined to an annular band is
\begin{eqnarray}
\mathfrak{A}_{\rm annulus}(r_{\rm in},r_{\rm out})
&:=&
\Bigl\{A\subseteq N_{r_{\rm out}}(\mathcal{B}_{\rm hole})
\setminus N_{r_{\rm in}}(\mathcal{B}_{\rm hole})\Bigr\}.
\label{eq:S4_annulus_family}
\end{eqnarray}
The full band may also be evaluated as a single region to test for loop-like logical support. For fixed $r_{\rm in}$ these families are nested as $r_{\rm out}$ increases.
\endgroup

\smallskip\noindent
{\bf ED profile.}
Given any family $\mathfrak{A}(\lambda)$ (with parameter $\lambda$ standing for $\ell,w,$ etc.), a concise ``profile'' statistic is the worst-case logical-ED over that family,
\begin{eqnarray}
\Delta_{\max}^{(\mathfrak{A})}(\lambda)
 :=
\max_{A\in\mathfrak{A}(\lambda)}\Delta_A.
\label{eq:S4_profile_def}
\end{eqnarray}
By Theorem~1, correctability of every region in a family implies a vanishing profile. Conversely, a nonzero profile certifies that at least one region in the family is not correctable. A vanishing profile certifies correctability if every scanned region meets the retained logical unitary condition of Theorem~\ref{thm:S7_retained_unitary}. Without such a condition, different reduced states can have identical purity.

\subsection{General lemmas: monotonicity and a stabilizer ``cleanability'' criterion}

We now record two general structural lemmas that clarify how region families interface with erasure correctability and, via Theorem~1 (main text), with logical-ED.

\begin{lemma}[Monotonicity of erasure correctability]\label{lem:S4_mono_correctable}
Let $A,B\subseteq V$.
\begin{itemize}
\item If $A$ is exactly erasure-correctable, then every subset $B\subseteq A$ is exactly erasure-correctable.
\item If $A$ is \emph{not} erasure-correctable, then every superset $B\supseteq A$ is \emph{not} erasure-correctable.
\end{itemize}
\end{lemma}

\begin{proof}---If erasure of $A$ is correctable, there exists a recovery $\mathcal{R}_{A^c}$ acting on $A^c$ such that $\mathcal{R}_{A^c}\circ \tr_A$ is the identity on the code. For $B\subseteq A$, the output after erasing $B$ is $B^c=A^c\cup(A\setminus B)$, which contains at least as much information as $A^c$. A valid recovery is therefore obtained by first discarding the extra subsystem $A\setminus B$ and then applying $\mathcal{R}_{A^c}$; equivalently, $\tr_A=\tr_{A\setminus B}\circ\tr_B$. This proves the first statement.

For the second statement, suppose $A$ is not correctable but some $B\supseteq A$ were correctable. Then by the first part (applied to $A\subseteq B$), $A$ would also be correctable, a contradiction. \end{proof}

\medskip
For stabilizer codes, correctability admits a sharp criterion in terms of the absence/presence of logical operators supported on $A$~\cite{bravyi2009nogo}. We state it in a form convenient for geometric reasoning.

\begin{lemma}[Cleanability criterion for stabilizer codes]\label{lem:S4_cleanability}
Let $\mathcal{C}$ be an exact stabilizer code with stabilizer group $\mathsf{S}\subset\mathsf{P}_n$ and code projector $\hat{P}$.
For a region $A\subseteq V$, the following are equivalent:
\begin{enumerate}
\item[(i)] The erasure of $A$ is exactly correctable on $\mathcal{C}$.
\item[(ii)] Every Pauli operator $\hat{M}$ supported on $A$ that preserves the code space acts trivially on it:
\begin{eqnarray}
\hat{M}\in {\rm Cent}(\mathsf{S}),\ {\rm supp}(\hat{M})\subseteq A
\ \ \Longrightarrow\ \ 
\hat{M}\hat{P}= \pm \hat{P} \ \ \ (\text{equivalently } \hat{M}\in \langle \mathsf{S},\pm\hat{\mathds{1}}\rangle).
\label{eq:S4_cleanability_statement}
\end{eqnarray}
\item[(iii)] There is no nontrivial logical Pauli supported on $A$, i.e., $\bigl({\rm Cent}(\mathsf{S})\cap \mathsf{P}_A\bigr)/\bigl(\mathsf{S}\cap \mathsf{P}_A\bigr)$ is trivial.
\end{enumerate}
\end{lemma}

\begin{proof}---For stabilizer codes it suffices to check the Knill--Laflamme error-detection condition on a Pauli operator basis. Recall that erasure of $A$ is correctable if and only if the code detects all errors supported on $A$, i.e.,
\begin{eqnarray}
\hat{P}\hat{E}\hat{P}  = c_E \hat{P}
\quad
\text{for all Pauli operators }\hat{E}\in \mathsf{P}_A,
\label{eq:S4_detect_all}
\end{eqnarray}
because for an erasure the relevant products $\hat{E}^\dagger \hat{F}$ remain supported on $A$.

Fix $\hat{E}\in\mathsf{P}_A$. If $\hat{E}$ anticommutes with some stabilizer generator, then $\hat{E}$ maps the code space to an orthogonal syndrome sector and hence $\hat{P}\hat{E}\hat{P}=0$, which is of the form $c_E \hat{P}$ with $c_E=0$. If $\hat{E}$ commutes with all stabilizers, then $\hat{E}$ preserves the code space and acts as a logical Pauli on $\mathcal{C}$; in that case $\hat{P}\hat{E}\hat{P}=\hat{P}\hat{E}=\hat{E}\hat{P}$, and Eq.~(\ref{eq:S4_detect_all}) forces $\hat{E}$ to be proportional to $\hat{P}$ on the code. For Pauli operators the only way to act as a scalar on the entire code space is to lie in the stabilizer group up to phase, i.e., $\hat{E}\in\langle \mathsf{S},\pm\hat{\mathds{1}}\rangle$. This proves that Eq.~(\ref{eq:S4_detect_all}) holds for all $\hat{E}\in \mathsf{P}_A$ if and only if there is no nontrivial logical Pauli supported on $A$, which is precisely (ii)--(iii). The equivalence of (i) with Eq.~(\ref{eq:S4_detect_all}) is the KL theorem specialized to erasure (see Supplementary Note~S2).
\end{proof}

\medskip\noindent
{\bf Consequence for logical-ED.}
By Lemma~\ref{lem:S4_cleanability} and Theorem~1 (main text), any region $A$ that is cleanable (i.e., supports no nontrivial logical Pauli) satisfies $\Delta_A=0$. Thus, geometric criteria for (non)existence of logical operators on region families immediately translate into vanishing statements for logical-ED profiles.

\subsection{Surface code: disk, strip, and annulus families}\label{sec:S4_surface_geometry}

We now instantiate the above notions for the planar surface code, emphasizing geometric region families that are particularly informative for logical-ED profiles. We focus on the standard planar surface code encoding one logical qubit; extensions to punctured codes (holes) are outlined at the end.

\smallskip\noindent
{\bf Surface code geometry.}
Consider a planar cellulation with primal graph $G$ and dual graph $G^\ast$; physical qubits are placed on primal edges (equivalently, dual edges). The stabilizer group is generated by:
(i) $X$-type star operators $\hat{A}_v$ (product of $\hat{X}$ on edges incident to a primal vertex $v$);
(ii) $Z$-type plaquette operators $\hat{B}_f$ (product of $\hat{Z}$ on edges along the boundary of a primal face $f$).
Boundary conditions are chosen so that the code encodes one logical qubit, with two \emph{rough} boundary components and two \emph{smooth} boundary components~\cite{kitaev2003fault,dennis2002topological,fowler2012surface}. We denote the two rough boundary components by $\mathcal{B}_{\rm r}^{(1)},\mathcal{B}_{\rm r}^{(2)}$ and the two smooth ones by $\mathcal{B}_{\rm s}^{(1)},\mathcal{B}_{\rm s}^{(2)}$ (each understood as a set of boundary-adjacent qubits or boundary-adjacent vertices/faces; the precise convention does not affect the statements below as long as it matches the standard string-operator picture).

\smallskip\noindent
{\bf Primal and dual strings.}
A \emph{primal path} $\gamma$ is a path in the primal graph $G$, i.e., a sequence of edges (qubits) connecting two boundary components. Its associated $Z$-string operator is
\begin{eqnarray}
\hat{Z}(\gamma) := \prod_{e\in \gamma} \hat{Z}_e.
\label{eq:S4_Z_string}
\end{eqnarray}
Similarly, a \emph{dual path} $\gamma^\ast$ is a path in the dual graph $G^\ast$; it crosses a set of primal edges, and the associated $X$-string operator is
\begin{eqnarray}
\hat{X}(\gamma^\ast) := \prod_{e\in \gamma^\ast} \hat{X}_e,
\label{eq:S4_X_string}
\end{eqnarray}
where the product is over the crossed primal edges (qubits).

It is standard (and follows directly from the commutation structure of stars/plaquettes) that:
(i) $\hat{Z}(\gamma)$ commutes with all plaquettes, and it commutes with all stars except at its endpoints (where it may create $X$-type syndromes unless the endpoint lies on a rough boundary);
(ii) $\hat{X}(\gamma^\ast)$ commutes with all stars, and it commutes with all plaquettes except at its endpoints (unless the endpoint lies on a smooth boundary).
Thus, \emph{boundary-to-boundary} strings of the appropriate type define logical operators: a primal path connecting the two rough boundaries yields a $Z$-type logical operator, while a dual path connecting the two smooth boundaries yields an $X$-type logical operator.

We now encode this in geometric lemmas suitable for our region-family language.

\begin{lemma}[Non-correctability of spanning strips]\label{lem:S4_strip_noncorrectable}
Let $A$ be a region (set of qubits/edges) in the planar surface code.
\begin{itemize}
\item If $A$ contains a primal path $\gamma\subseteq A$ connecting $\mathcal{B}_{\rm r}^{(1)}$ to $\mathcal{B}_{\rm r}^{(2)}$, then $A$ supports a nontrivial $Z$-type logical operator and the erasure of $A$ is not correctable.
\item If $A$ contains a dual path $\gamma^\ast\subseteq A$ connecting $\mathcal{B}_{\rm s}^{(1)}$ to $\mathcal{B}_{\rm s}^{(2)}$, then $A$ supports a nontrivial $X$-type logical operator and the erasure of $A$ is not correctable.
\end{itemize}
\end{lemma}

\begingroup
\begin{proof}
A primal path $\gamma$ connecting the two rough components is a relative $1$-cycle for the pair consisting of the patch and its rough boundary. The connecting map to the reduced zeroth homology of the rough boundary sends its class to the sum of its two endpoints, which belong to distinct components and represent a nonzero element over $\mathbb{F}_2$. Thus $\gamma$ is a nontrivial relative homology class. The corresponding $\hat{Z}(\gamma)$ commutes with the checks and cannot be a product of plaquette checks, which represent relative boundaries. The dual argument exchanges rough and smooth components and gives a nontrivial logical $\hat{X}$ for the dual path. In both cases Lemma~\ref{lem:S4_cleanability} implies that erasure of $A$ is uncorrectable.
\end{proof}

\medskip
The spanning members of the strip families in Sec.~\ref{sec:S4_canonical} therefore fail exact erasure correction. Their reduced states on $A$ depend on the logical input, but those states may still share a common purity. If the complementary region supports a non-scalar logical unitary, Theorem~\ref{thm:S7_retained_unitary} excludes that possibility, and the spanning region has $\Delta_A>0$.
\endgroup

\medskip
Conversely, we can give a clean geometric sufficient condition for correctability, phrased in terms of \emph{non-spanning} regions.

\begin{definition}[Non-spanning regions in the planar surface code]\label{def:S4_nonspanning}
A region $A$ is called \emph{$Z$-nonspanning} if it contains no primal path connecting $\mathcal{B}_{\rm r}^{(1)}$ to $\mathcal{B}_{\rm r}^{(2)}$.
Similarly, $A$ is \emph{$X$-nonspanning} if it contains no dual path connecting $\mathcal{B}_{\rm s}^{(1)}$ to $\mathcal{B}_{\rm s}^{(2)}$.
We say $A$ is \emph{nonspanning} if it is both $Z$-nonspanning and $X$-nonspanning.
\end{definition}

\begin{lemma}[Nonspanning regions are cleanable for the planar code with $k=1$]\label{lem:S4_nonspanning_cleanable}
Consider the planar surface code encoding one logical qubit.
If a region $A$ is nonspanning in the sense of Definition~\ref{def:S4_nonspanning}, then $A$ supports no nontrivial logical Pauli operator. Consequently, the erasure of $A$ is exactly correctable and $\Delta_A=0$.
\end{lemma}

\begingroup
\begin{proof}
The code is CSS, so a Pauli in the centralizer supported on $A$ decomposes into an $X$-type part and a $Z$-type part, each supported within $A$ and commuting with the checks. In particular, the $X$ and $Z$ parts of any logical $\hat{Y}$ representative can be considered separately. A nontrivial $Z$-type logical class is a relative cycle with nonzero boundary-component parity. Decomposing the cycle into paths and loops therefore gives at least one rough-to-rough path in its support. The dual statement gives a smooth-to-smooth path for a nontrivial $X$-type logical class~\cite{kitaev2003fault,dennis2002topological,fowler2012surface,stace2009thresholds}. If $A$ is nonspanning, both parts must be stabilizer-trivial. The claim follows from Lemma~\ref{lem:S4_cleanability} and Theorem~1 of the main text.
\end{proof}
\endgroup

\medskip\noindent
{\bf A geometric sufficient condition (disk family).}
A practically useful sufficient condition for nonspanning is containment in a region that is topologically a disk and does not intersect both members of a boundary pair. One convenient formulation uses thickened boundary neighborhoods.

\begin{corollary}[Disk-like and single-boundary regions are correctable]\label{cor:S4_disk_correctable}
In the planar surface code encoding one logical qubit, suppose $A$ is contained in a simply connected patch $\Omega$ whose boundary intersects at most one rough boundary component and at most one smooth boundary component. Then $A$ is nonspanning, hence erasure-correctable and $\Delta_A=0$.
\end{corollary}

\begin{proof}---If $\Omega$ intersects at most one rough boundary component, then no path entirely contained in $\Omega$ can connect $\mathcal{B}_{\rm r}^{(1)}$ to $\mathcal{B}_{\rm r}^{(2)}$; thus $A\subseteq \Omega$ is $Z$-nonspanning. The same reasoning with smooth boundaries implies $A$ is $X$-nonspanning. Lemma~\ref{lem:S4_nonspanning_cleanable} then applies.
\end{proof}

\subsection{Punctures and annuli: detecting loop-like logical operators}

Many surface-code variants encode additional logical qubits using holes (punctures) or higher-genus manifolds. In these cases, in addition to boundary-to-boundary strings, there exist nontrivial logical operators represented by \emph{loops} encircling a hole (or a handle). The annulus family $\mathfrak{A}_{\rm annulus}(r_{\rm in},r_{\rm out})$ in Eq.~(\ref{eq:S4_annulus_family}) is designed to probe precisely these loop-like supports.

For concreteness, consider a planar surface code with one hole encoding an additional logical qubit. One of the logical operators can be chosen as a $Z$-loop around the hole, supported on a cycle of edges surrounding the puncture. If a region $A$ contains such a cycle (or a thickened neighborhood containing it), then by the same reasoning as Lemma~\ref{lem:S4_strip_noncorrectable} the region supports a nontrivial logical Pauli and is not correctable. Conversely, regions that are disk-like and do not wrap around holes are typically cleanable (and therefore have $\Delta_A=0$), while annular regions that wrap a hole are not. The precise characterization can be formulated using (relative) homology of the primal/dual complexes; we omit these topological details here since our main goal is to provide families that isolate the relevant geometries for ED profiling.

\subsection{General local codes: a geometric ``support-diameter'' bound}

Finally, we record a general geometric lemma that links the \emph{geometric} scale of a region family to the standard weight distance. This statement is not specific to surface codes; it applies to any code defined on a bounded-degree graph.

\begin{lemma}[From diameter to weight on bounded-degree graphs]\label{lem:S4_diam_to_weight}
Let $G=(V,E)$ have maximum degree $\Delta_{\rm max}$. Then any connected region $A\subseteq V$ with $D:={\rm diam}(A)$ satisfies
\begin{eqnarray}
\abs{A}
\ \le\
1+\Delta_{\rm max}\sum_{t=0}^{D-1}(\Delta_{\rm max}-1)^t .
\label{eq:S4_diam_weight_bound}
\end{eqnarray}
For $\Delta_{\rm max}>2$, the right-hand side is $1+\Delta_{\rm max}\bigl((\Delta_{\rm max}-1)^D-1\bigr)/(\Delta_{\rm max}-2)$; for $\Delta_{\rm max}=2$, it is $1+2D$.
\end{lemma}

\begin{proof}---Fix any site $x\in A$. Because $A$ is connected and has diameter $D$, every site in $A$ lies within distance $D$ of $x$. The number of sites within graph distance $t$ of $x$ is at most $\Delta_{\rm max}(\Delta_{\rm max}-1)^{t-1}$ for $t\ge 1$ (a tree-like upper bound: from $x$ there are at most $\Delta_{\rm max}$ choices for the first step and at most $\Delta_{\rm max}-1$ choices thereafter). Summing over shells $t=0,\dots,D$ yields Eq.~(\ref{eq:S4_diam_weight_bound}).
\end{proof}

\medskip
Lemma~\ref{lem:S4_diam_to_weight} yields a conservative implication: if the code distance is $d$ (weight distance), then any connected region with ${\rm diam}(A)$ small enough so that the RHS of Eq.~(\ref{eq:S4_diam_weight_bound}) is $\le d-1$ is necessarily erasure-correctable (by Lemma~\ref{lem:S3_distance_erasure}) and hence has $\Delta_A=0$ (Theorem~1). On a Euclidean lattice of spatial dimension $D_{\rm sp}$, polynomial volume growth gives $\abs{A}\le C(1+\ell)^{D_{\rm sp}}$ for $\operatorname{diam}(A)\le\ell$. For a connected-diameter family, this yields a guaranteed zero-ED range up to scales of order $d^{1/D_{\rm sp}}$, subject to the scales included in the scan. While such bounds are not tight for topological codes (where correctable regions can be much larger than $d$ in weight), they provide a general bridge between weight-based and geometry-based scales that can be useful when comparing different architectures.



\section{Exact large-scale surface-code simulation and numerical details}
\label{sec:S5_surface_sim}

This section gives the derivation, implementation details, and additional numerical results underlying Fig.~2 of the main text. The central computational observation for Fig.~2 is that, for a one-logical-qubit stabilizer code, logical-ED can be obtained exactly from binary ranks and affine row-space tests. The method therefore avoids both a $2^n$-dimensional state vector and enumeration of the $2^{n-1}$ stabilizer elements. We use it for odd-distance rotated planar surface-code patches with parameters $[[d^2,1,d]]$ for $d=5,7,9,11,13,15$, i.e., from $25$ to $225$ data qubits. The stochastic calculations concern static erasure masks and are intentionally separated from decoder-threshold or repeated-syndrome simulations. We also derive the relation between supported logical operators and the ambiguity of ideal syndromes for Pauli errors at unknown locations.

\subsection{Rotated planar patch and binary symplectic representation}

Data qubits occupy the vertices $(x,y)$ of a $d\times d$ square array, with $x,y\in\{0,\ldots,d-1\}$ and odd $d$. Interior weight-$4$ plaquettes are assigned alternating $X$- and $Z$-type checks. Weight-$2$ boundary checks terminate the corresponding checkerboard plaquettes along the top/bottom and left/right edges. This construction has $n=d^2$ data qubits, $m=n-1$ independent stabilizer generators, and one encoded qubit. We choose a vertical data column as $\hat{\overline{X}}$ and a horizontal data row as $\hat{\overline{Z}}$; their single intersection gives the required odd symplectic product.

A Pauli operator is represented, up to phase, by a binary symplectic vector~\cite{hostens2005stabilizer}
\begin{eqnarray}
\bm{v}=(\bm{x}\mid\bm{z})\in\mathbb{F}_2^{2n},
\label{eq:S5_symplectic_vector}
\end{eqnarray}
where $x_j=1$ (respectively, $z_j=1$) denotes an $X$ (respectively, $Z$) component on qubit $j$. Let
\begin{eqnarray}
G\in\mathbb{F}_2^{m\times 2n}
\label{eq:S5_G_def}
\end{eqnarray}
contain the independent stabilizer generators as rows, and let $\bm{\ell}_\alpha$ represent $\hat{\overline{L}}_\alpha$ for $\alpha\in\{x,y,z\}$, with $\hat{\overline{L}}_y=\hat{\overline{L}}_x\hat{\overline{L}}_z$ up to phase. The implementation stores every $2n$-bit row as one arbitrary-precision Python integer. XOR performs row addition over $\mathbb{F}_2$, and the most significant occupied bit serves as the pivot during elimination.

\subsection{Stabilizer-coset formula for the reduced purity}

Let $\mathcal{S}$ be the stabilizer group and let $\bm{r}=(r_x,r_y,r_z)$ be the Bloch vector of the encoded logical qubit. The corresponding physical density operator can be written as
\begin{eqnarray}
\hat{\rho}(\bm{r})
=2^{-n}\sum_{\hat{g}\in\mathcal{S}}\hat{g}\left(\hat{\mathds{1}}+r_x\hat{\overline{L}}_x+r_y\hat{\overline{L}}_y+r_z\hat{\overline{L}}_z\right).
\label{eq:S5_global_density_expansion}
\end{eqnarray}
Only Pauli terms acting trivially on $A^c$ survive $\tr_{A^c}$. Define
\begin{eqnarray}
N_0(A)&:=&\abs{\{\hat{g}\in\mathcal{S}:\operatorname{supp}(\hat{g})\subseteq A\}},
\nonumber\\
N_\alpha(A)&:=&\abs{\{\hat{g}\in\mathcal{S}:\operatorname{supp}(\hat{g}\hat{\overline{L}}_\alpha)\subseteq A\}},\quad \alpha=x,y,z.
\label{eq:S5_coset_counts}
\end{eqnarray}
Pauli orthogonality on $A$ then gives
\begin{eqnarray}
\tr{\bigl(\hat{\rho}_A(\bm{r})^2\bigr)}
=2^{-\abs{A}}\left[N_0(A)+N_x(A)r_x^2+N_y(A)r_y^2+N_z(A)r_z^2\right].
\label{eq:S5_surface_purity_counts}
\end{eqnarray}
For a Haar-random logical qubit,
\begin{eqnarray}
\mathbb{E}(r_\alpha^2)=\frac{1}{3},\quad
\mathbb{E}(r_\alpha^4)=\frac{1}{5},\quad
\mathbb{E}(r_\alpha^2r_\beta^2)=\frac{1}{15}\quad(\alpha\neq\beta).
\label{eq:S5_bloch_moments}
\end{eqnarray}
Since $E_A=1-\tr(\hat{\rho}_A^2)$, Eqs.~(\ref{eq:S5_surface_purity_counts})--(\ref{eq:S5_bloch_moments}) yield
\begin{eqnarray}
\overline{E}_A
&=&1-2^{-\abs{A}}\left[N_0(A)+\frac{N_x(A)+N_y(A)+N_z(A)}{3}\right],
\label{eq:S5_surface_mean_counts}\\
\Delta_A
&=&2^{-\abs{A}}\sqrt{\frac{2}{45}\left[3\sum_{\alpha=x,y,z}N_\alpha(A)^2-
\left(\sum_{\alpha=x,y,z}N_\alpha(A)\right)^2\right]}.
\label{eq:S5_surface_delta_counts}
\end{eqnarray}

The counts in Eq.~(\ref{eq:S5_coset_counts}) have a particularly simple algebraic form. Let $G_{A^c}$ be obtained by retaining the $X$ and $Z$ columns belonging to qubits in $A^c$, and define
\begin{eqnarray}
t_A:=m-\operatorname{rank}_{\mathbb{F}_2}(G_{A^c}).
\label{eq:S5_tA_def}
\end{eqnarray}
The homogeneous system $\bm{u}G_{A^c}=0$ has $2^{t_A}$ solutions, so
\begin{eqnarray}
N_0(A)=2^{t_A}.
\label{eq:S5_N0}
\end{eqnarray}
For a logical coset, the support condition is the affine system
\begin{eqnarray}
\bm{u}G_{A^c}=\bm{\ell}_{\alpha,A^c}.
\label{eq:S5_affine_system}
\end{eqnarray}
It is either inconsistent, giving $N_\alpha(A)=0$, or it is an affine translate of the homogeneous kernel, giving $N_\alpha(A)=2^{t_A}$. Equivalently, consistency is decided by whether $\bm{\ell}_{\alpha,A^c}$ belongs to the row span of $G_{A^c}$.

Let $\mathsf{P}_A$ denote the subgroup of Pauli operators modulo phase whose support is contained in $A$, and define the subgroup of logical Pauli classes supported in $A$ by
\begin{eqnarray}
\mathsf{L}(A)
:=
\frac{{\rm Cent}(\mathcal{S})\cap\mathsf{P}_A}{\mathcal{S}\cap\mathsf{P}_A}.
\label{eq:S5_supported_logical_group}
\end{eqnarray}
For a one-logical-qubit stabilizer code, $\mathsf{L}(A)$ is a binary vector subspace of the two-dimensional logical Pauli space. Let $\kappa_A:=\dim_{\mathbb{F}_2}\mathsf{L}(A)$ and let $q_A$ be the number of its nonidentity elements.

\begin{proposition}[Exact logical-access trichotomy for one-logical-qubit stabilizer codes]\label{prop:S5_trichotomy}
For every region $A$,
\begin{eqnarray}
\kappa_A\in\{0,1,2\},
\quad
q_A=2^{\kappa_A}-1\in\{0,1,3\}.
\label{eq:S5_q_values}
\end{eqnarray}
Moreover:
\begin{enumerate}
\item[(i)] $q_A=0$ if and only if $A$ supports no nontrivial logical Pauli; equivalently, erasure of $A$ is exactly correctable.
\item[(ii)] $q_A=1$ if and only if $A$ supports exactly one logical Pauli axis. The erasure is not correctable and $\Delta_A>0$.
\item[(iii)] $q_A=3$ if and only if $A$ supports representatives of the full logical Pauli algebra. The erasure of $A$ is not correctable, whereas erasure of $A^c$ is correctable, so the logical state is recoverable from $A$; nevertheless, $\Delta_A=0$.
\end{enumerate}
The Haar mean and logical-ED are
\begin{eqnarray}
\overline{E}_A
&=&1-2^{t_A-\abs{A}}\left(1+\frac{q_A}{3}\right),
\label{eq:S5_mean_q}\\
\Delta_A
&=&
\begin{cases}
\displaystyle \frac{2^{t_A-\abs{A}+1}}{\sqrt{45}},&q_A=1,\\[2mm]
0,&q_A=0\ \text{or}\ 3.
\end{cases}
\label{eq:S5_delta_q}
\end{eqnarray}
\end{proposition}

\begin{proof}---Multiplication of two Pauli representatives supported in $A$ produces another representative supported in $A$, so $\mathsf{L}(A)$ is a subgroup, equivalently an $\mathbb{F}_2$-linear subspace, of the logical Pauli space. The latter has dimension two for $k=1$, which proves Eq.~(\ref{eq:S5_q_values}). The case $q_A=0$ is equivalent to the absence of a nontrivial logical Pauli in $A$ and hence, by Lemma~\ref{lem:S4_cleanability}, to exact erasure correctability. The cases $q_A=1$ and $q_A=3$ contain nontrivial logical Paulis and are therefore not correctable. In the $q_A=3$ case, representatives of two anticommuting logical generators are supported in $A$. Any nontrivial logical Pauli supported in the disjoint region $A^c$ would commute with both representatives, but every nonidentity one-qubit Pauli anticommutes with at least one of two independent generators. Hence $A^c$ supports no nontrivial logical Pauli and is erasure-correctable by Lemma~\ref{lem:S4_cleanability}.

For every supported logical coset, the affine solution set in Eq.~(\ref{eq:S5_affine_system}) is a translate of the same $2^{t_A}$-element kernel. Hence exactly $q_A$ of $N_x,N_y,N_z$ equal $2^{t_A}$ and the others vanish. Substitution into Eqs.~(\ref{eq:S5_surface_mean_counts})--(\ref{eq:S5_surface_delta_counts}) gives Eqs.~(\ref{eq:S5_mean_q})--(\ref{eq:S5_delta_q}). One supported axis produces a purity proportional to one component $r_\alpha^2$ and hence a nonzero Haar variance, whereas three supported axes give a term proportional to $r_x^2+r_y^2+r_z^2=1$ and hence zero variance.
\end{proof}

Proposition~\ref{prop:S5_trichotomy} gives the precise operational scope of logical-ED in this setting: it detects the intermediate, one-axis logical-access sector $q_A=1$. It is not a test for the presence of logical information, because it is blind to the full-access sector $q_A=3$. In that sector $A^c$ is correctable, so $\hat{\rho}_{A^c}(\psi_L)$ is fixed; complement symmetry then makes the full nonzero spectrum of $\hat{\rho}_A(\psi_L)$ fixed as well. Thus no reduced-state spectral moment can distinguish logical inputs in the $q_A=3$ sector. Accordingly, $\Pr(\Delta_A>0)=\Pr(q_A=1)$ below is an \emph{ED-active probability}, not an erasure-failure probability $\Pr(q_A>0)$.

\medskip
\noindent\textbf{Algorithm 1: Exact logical-ED of a region in an $[[n,1]]$ stabilizer code.}\label{alg:S5_exact_epd}
\begin{algorithmic}[1]
\Require Stabilizer rows $G$, logical rows $\bm{\ell}_x,\bm{\ell}_y,\bm{\ell}_z$, region mask $A$
\State Restrict every stabilizer row to the $2\abs{A^c}$ symplectic columns of $A^c$
\State Build a GF$(2)$ echelon basis $B$ for the restricted rows
\State $t_A\gets (n-1)-\abs{B}$
\For{$\alpha\in\{x,y,z\}$}
    \State $b_\alpha\gets\bm{\ell}_{\alpha,A^c}$
    \State $s_\alpha\gets 1$ if $b_\alpha\in\operatorname{span}_{\mathbb{F}_2}(B)$, and $0$ otherwise
\EndFor
\State $q_A\gets s_x+s_y+s_z$
\State $\overline{E}_A\gets 1-2^{t_A-\abs{A}}(1+q_A/3)$
\State $\Delta_A\gets 2^{t_A-\abs{A}+1}/\sqrt{45}$ if $q_A=1$, and $0$ otherwise
\State \textbf{return} $t_A$, $(s_x,s_y,s_z)$, $\overline{E}_A$, and $\Delta_A$
\end{algorithmic}
\medskip

\subsection{Deterministic region families and exact scaling}

We first scan structured region families. A compact $3\times3$ block in the bulk is nonspanning and has $q_A=0$. A proper full-width horizontal strip contains a logical $\hat{\overline{Z}}$ representative but no independent $\hat{\overline{X}}$ representative, and therefore has $q_A=1$. A horizontal--vertical cross contains $\hat{\overline{X}}$, $\hat{\overline{Z}}$, and their product, so $q_A=3$ and realizes the blind spot described above.

For a full-width horizontal strip of depth $1\le w<d$ anchored at the top or bottom edge, direct rank reduction gives
\begin{eqnarray}
\abs{A}=dw,\quad t_A=dw-\frac{d+1}{2},\quad q_A=1.
\label{eq:S5_strip_rank}
\end{eqnarray}
Consequently,
\begin{eqnarray}
\Delta_{\rm strip}(d)=\frac{2^{-(d-1)/2}}{\sqrt{45}},
\label{eq:S5_strip_delta}
\end{eqnarray}
which is independent of $w$ within this boundary-anchored family. The stabilizer rank of an interior strip need not be the same, even when it supports the same logical axis. The corresponding Haar-mean purity is $1-\overline{E}_{\rm strip}=4\,2^{-(d+1)/2}/3$, so the relative purity fluctuation remains finite and universal,
\begin{eqnarray}
\frac{\Delta_{\rm strip}}{1-\overline{E}_{\rm strip}}=\frac{1}{\sqrt{20}}.
\label{eq:S5_strip_relative}
\end{eqnarray}
Thus the exponential scale in Eq.~(\ref{eq:S5_strip_delta}) reflects the small absolute purity of a large strip, rather than a loss of relative logical anisotropy. Table~\ref{tab:S5_strip} lists the exact values used in Fig.~2(b) of the main text.

\begin{table}[t]
\centering
\setlength{\tabcolsep}{15pt}
\begin{tabular}{c c c c c}
\hline\hline
$d$ & $n=d^2$ & $\abs{A}$ for $w=1$ & $t_A$ for $w=1$ & $\Delta_{\rm strip}$\\
\hline
$5$  & $25$  & $5$  & $2$ & $3.72678\times10^{-2}$\\
$7$  & $49$  & $7$  & $3$ & $1.86339\times10^{-2}$\\
$9$  & $81$  & $9$  & $4$ & $9.31695\times10^{-3}$\\
$11$ & $121$ & $11$ & $5$ & $4.65847\times10^{-3}$\\
$13$ & $169$ & $13$ & $6$ & $2.32924\times10^{-3}$\\
$15$ & $225$ & $15$ & $7$ & $1.16462\times10^{-3}$\\
\hline\hline
\end{tabular}
\caption{Exact logical-ED of boundary-anchored full-width strips. The value is constant over depths $1\le w<d$ within this family.}
\label{tab:S5_strip}
\end{table}

\begin{figure*}[t]
\centering
\includegraphics[width=0.90\linewidth]{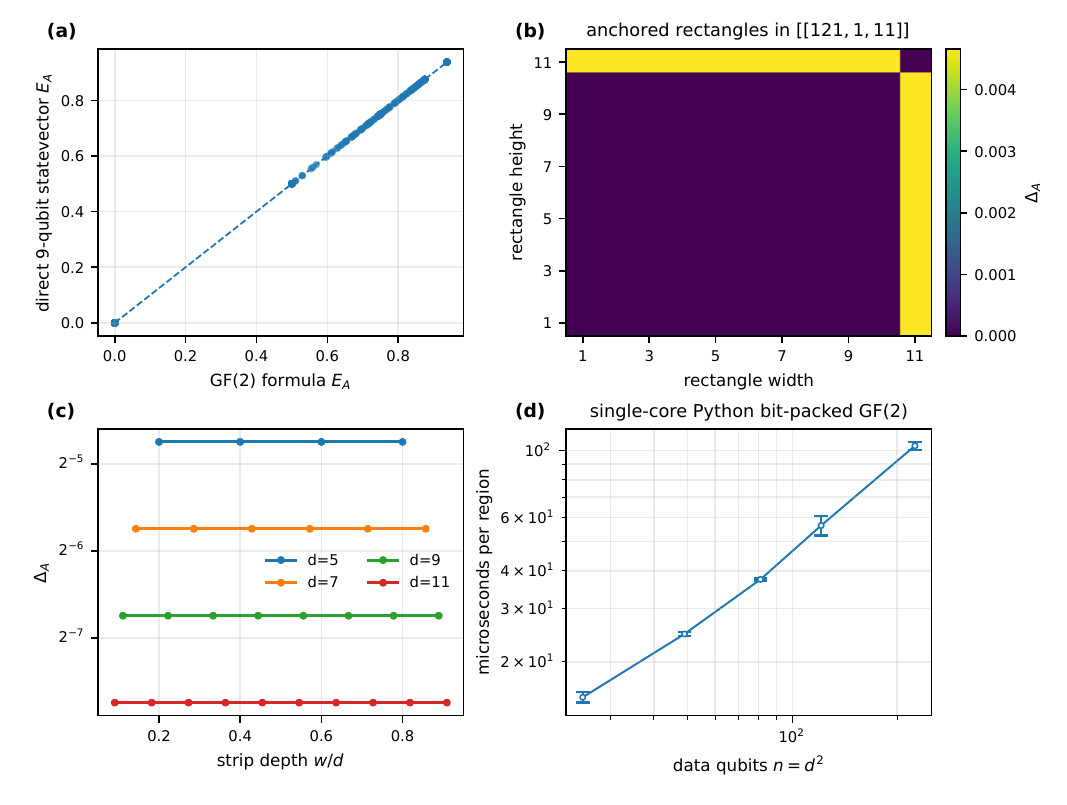}
\caption{\textbf{Deterministic profiles, direct validation, and computational scaling.} \textbf{(a)} Direct $9$-qubit state-vector values of $E_A(\psi_L)$ for individual logical inputs versus the exact GF$(2)$ formula for structured and random regions of the $d=3$ patch. The dashed line is equality. \textbf{(b)} Logical-ED for all lower-left-anchored rectangles in the $[[121,1,11]]$ code. Rectangles with neither dimension spanning the patch are ED-inactive; an ED signal appears exactly when one dimension first spans the patch. The full $11\times11$ region supports all three logical Pauli cosets and therefore returns to $\Delta_A=0$. \textbf{(c)} Boundary-anchored full-width-strip logical-ED versus normalized strip depth for $d=5,7,9,11$. Each profile is constant for $1\le w<d$, in agreement with Eq.~(\ref{eq:S5_strip_delta}). \textbf{(d)} Median single-region runtime of the bit-packed Python implementation. Each point is the median of three runs of $2000$ independently generated masks. Timings are implementation- and hardware-dependent and are reported only to document the numerical workload.}
\label{fig:S5_profiles}
\end{figure*}

A complementary scan uses boundary-anchored rectangles $A_{\ell,h}=\{(x,y):0\le x<\ell,\ 0\le y<h\}$ with $1\le\ell,h\le d$. For $\ell<d$ and $h<d$, the rectangle is nonspanning and $\Delta_A=0$. At $\ell=d$ with $h<d$ it first contains the horizontal logical string and $\Delta_A$ turns on; at $h=d$ with $\ell<d$ the analogous vertical onset occurs. At $(\ell,h)=(d,d)$, the region is the full code block and $q_A=3$, causing $\Delta_A$ to return to zero. This two-dimensional onset map is shown in Fig.~\ref{fig:S5_profiles}(b).

\subsection{Static erasure-mask ensembles}

To model spatially different loss mechanisms at equal single-qubit marginal rate, we use two mask ensembles. In the independent ensemble,
\begin{eqnarray}
\Pr(j\in A)=p\quad\text{independently for every data qubit }j.
\label{eq:S5_iid_model}
\end{eqnarray}
In the line-correlated ensemble, a horizontal or vertical orientation is selected with probability $1/2$, and then each of the $d$ complete lines of that orientation is included independently with probability $p$. Every qubit again obeys $\Pr(j\in A)=p$, but qubits on the same selected control line are perfectly correlated.

For the line model, if $K$ selected lines satisfies $0<K<d$, the mask supports exactly one logical axis and is ED-active. The cases $K=0$ and $K=d$ have $q_A=0$ and $q_A=3$, respectively. Therefore the exact active probability is
\begin{eqnarray}
\Pr_{\rm line}(\Delta_A>0)=1-(1-p)^d-p^d.
\label{eq:S5_line_active_exact}
\end{eqnarray}
For comparison, the probability that the mask is erasure-uncorrectable is
\begin{eqnarray}
\Pr_{\rm line}(q_A>0)=1-(1-p)^d,
\quad
\Pr_{\rm line}(q_A=3)=p^d.
\label{eq:S5_line_uncorrectable_exact}
\end{eqnarray}
Thus Eq.~(\ref{eq:S5_line_active_exact}) omits the full-access event $K=d$: this event is uncorrectable but ED-inactive.
The Monte Carlo points fluctuate around Eq.~(\ref{eq:S5_line_active_exact}); their uncertainty is reported with the two-sided $95\%$ Wilson score interval. For each $d\in\{5,7,9,11\}$, each model, and each $p\in\{0,0.025,\ldots,0.5\}$, we generated $5000$ masks using independent deterministic random-number streams.

\begin{figure*}[t]
\centering
\includegraphics[width=0.90\linewidth]{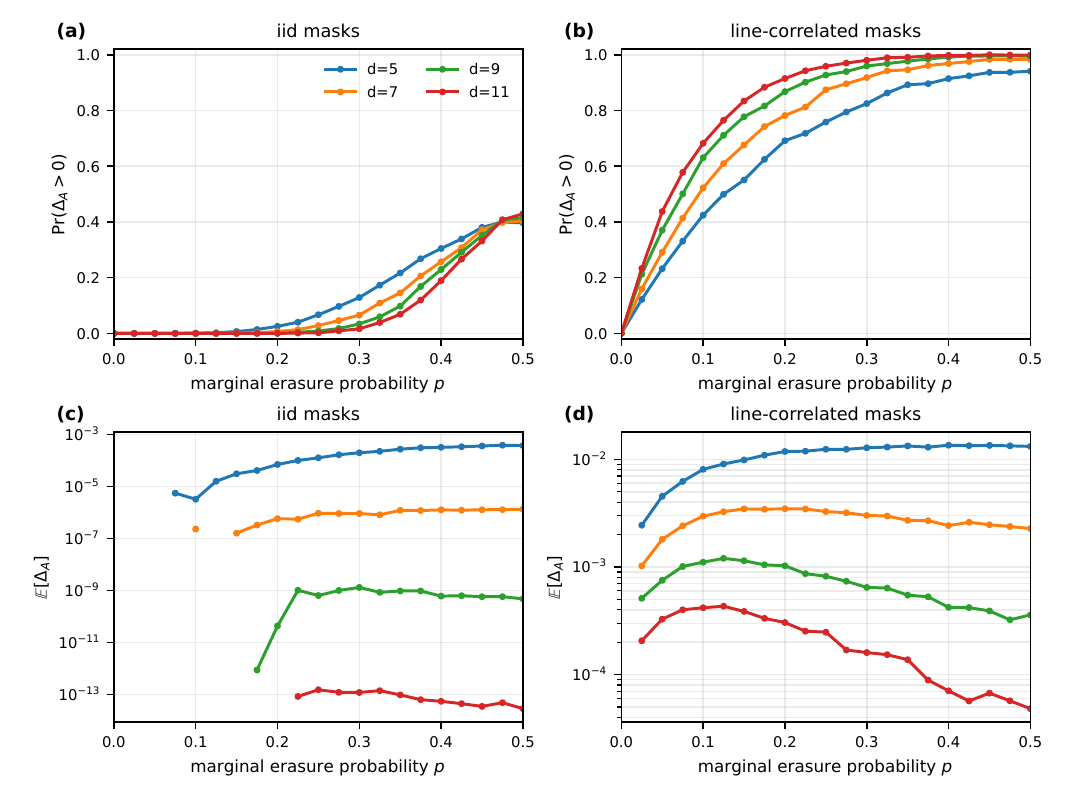}
\caption{\textbf{Distance dependence of the erasure-mask ensembles.} \textbf{(a,b)} ED-active fraction $\Pr(q_A=1)$ for independent and line-correlated masks, respectively, for $d=5,7,9,11$. \textbf{(c,d)} Corresponding ensemble means $\mathbb{E}[\Delta_A]$ on logarithmic axes. Every point uses $5000$ masks. Zero sample means are omitted from logarithmic panels. The same one-qubit marginal $p$ leads to markedly different one-axis logical-access anisotropy because complete line erasures directly create logical corridors.}
\label{fig:S5_erasure}
\end{figure*}

Figure~\ref{fig:S5_erasure} expands the main-text $d=11$ comparison to four code distances. In the independent ensemble, ED-active masks become rarer at fixed $p$ as $d$ increases, because an unstructured mask must accidentally assemble a spanning logical support. In the line ensemble, the active fraction instead increases with $d$ at small and intermediate $p$, as follows directly from Eq.~(\ref{eq:S5_line_active_exact}). The mean ED combines this activation probability with the factor $2^{t_A-\abs{A}+1}$ and therefore need not be monotone in either $d$ or $p$.

Table~\ref{tab:S5_erasure_key} gives representative $d=11$ values. At $p=0.30$, the point-estimate ratio of the line-correlated mean to the independent mean is approximately $1.32\times10^9$; the uncertainty of each mean is reported in the table. This ratio should not be interpreted as a decoder-performance or erasure-failure ratio; it compares logical-ED in static mask ensembles.

\begin{table}[t]
\centering
\setlength{\tabcolsep}{15pt}
\begin{tabular}{c c c c}
\hline\hline
$p$ & model & $\Pr(\Delta_A>0)$ & $\mathbb{E}[\Delta_A]$\\
\hline
$0.10$ & independent & $0.0000\ (0,\,7.68\times10^{-4})$ & $0$\\
$0.10$ & line-correlated & $0.6822\ (0.6692,\,0.6950)$ & $(4.1846\pm0.1758)\times10^{-4}$\\
$0.30$ & independent & $0.0164\ (0.0132,\,0.0203)$ & $(1.2132\pm0.5760)\times10^{-13}$\\
$0.30$ & line-correlated & $0.9804\ (0.9762,\,0.9839)$ & $(1.6047\pm0.1121)\times10^{-4}$\\
\hline\hline
\end{tabular}
\caption{Representative $[[121,1,11]]$ erasure-mask statistics. Parentheses give the $95\%$ Wilson interval for the ED-active fraction $\Pr(q_A=1)$. The uncertainty on the mean is its Monte Carlo standard error.}
\label{tab:S5_erasure_key}
\end{table}

\subsection{\texorpdfstring{Decoder-free Pauli-error distinguishability}{Decoder-free Pauli-error distinguishability}}
\label{sec:S5_decoder_free}

The static erasure-mask calculation above asks whether a \emph{known} region $A$ supports logical operators and how its reduced-state purity varies over logical inputs. Here we ask which Pauli errors at \emph{unknown locations} are unambiguous given an ideal stabilizer syndrome after stabilizer degeneracy is taken into account. We define the exact-layer and cumulative fractions and relate the underlying ambiguity to the supported-logical geometry probed by logical-ED.

Work with Pauli operators modulo the phases $\{\pm1,\pm i\}$. For an $[[n,k,d]]$ stabilizer code, let $\mathcal{S}$ be the stabilizer group, let $\mathcal{N}(\mathcal{S})$ be its Pauli normalizer, and let ${\rm syn}(\hat{E})$ denote the binary stabilizer syndrome of a Pauli $\hat{E}$. Define the exact-weight and accumulated candidate sets by
\begin{eqnarray}
\mathcal{E}_a&:=&\{\hat{E}\in\mathsf{P}_n:\operatorname{wt}(\hat{E})=a\},
\quad
\abs{\mathcal{E}_a}=\binom{n}{a}3^a,
\nonumber\\
\mathcal{E}_{\le w}&:=&\bigcup_{a=0}^{w}\mathcal{E}_a,
\quad
\abs{\mathcal{E}_{\le w}}=\sum_{a=0}^{w}\binom{n}{a}3^a.
\label{eq:S5_error_candidate_sets}
\end{eqnarray}
For $\hat{E}\in\mathcal{E}_{\le w}$, introduce the strict, order-independent indicator
\begin{eqnarray}
u_w(\hat{E}):=
\mathbf{1}\!\left[
\forall \hat{F}\in\mathcal{E}_{\le w},\ 
{\rm syn}(\hat{F})={\rm syn}(\hat{E})
\Rightarrow \hat{E}^\dagger \hat{F}\in\mathcal{S}
\right].
\label{eq:S5_unambiguous_indicator}
\end{eqnarray}
Thus $u_w(\hat{E})=1$ exactly when every enumerated candidate in the same syndrome fiber belongs to the same stabilizer coset as $\hat{E}$. A stabilizer-degenerate pair is jointly correctable and does not spoil uniqueness, whereas the presence of a second logical coset makes the fiber ambiguous. The exact-layer fraction is
\begin{eqnarray}
R_{\rm exact}(w):=
\frac{1}{\abs{\mathcal{E}_w}}
\sum_{\hat{E}\in\mathcal{E}_w}u_w(\hat{E}).
\label{eq:S5_R_exact_def}
\end{eqnarray}
\begingroup
The cumulative fraction uses the same indicator and candidate library, with the average taken over every included layer:
\begin{eqnarray}
R_{\rm cum}(w)
&:=&
\frac{1}{\abs{\mathcal{E}_{\le w}}}
\sum_{\hat{E}\in\mathcal{E}_{\le w}}u_w(\hat{E}).
\label{eq:S5_R_cum_def}
\end{eqnarray}
A larger candidate library can make a lower-weight error ambiguous, so all included errors must be evaluated at the common cutoff $w$. Both fractions quantify uniqueness under ideal syndromes. A decoder success probability additionally depends on which logical coset is chosen and how the errors are weighted.
\endgroup

\begin{proposition}[Syndrome--logical-support bridge]
\label{prop:S5_decoder_bridge}
For any exact stabilizer code, the following statements hold.
\begin{enumerate}
\item[(i)] Two Pauli errors have the same syndrome if and only if their relative product belongs to the Pauli normalizer:
\begin{eqnarray}
{\rm syn}(\hat{E})={\rm syn}(\hat{F})
\quad\Longleftrightarrow\quad
\hat{E}^\dagger \hat{F}\in\mathcal{N}(\mathcal{S}).
\label{eq:S5_same_syndrome_normalizer}
\end{eqnarray}
The pair is stabilizer-degenerate if $\hat{E}^\dagger \hat{F}\in\mathcal{S}$ and is logically ambiguous if $\hat{E}^\dagger \hat{F}\in\mathcal{N}(\mathcal{S})\setminus\mathcal{S}$.

\item[(ii)] Let $\mathsf{P}_A$ be the Pauli subgroup supported in a fixed region $A$. Then
\begin{eqnarray}
&&\left[
{\rm syn}(\hat{E})={\rm syn}(\hat{F})
\Rightarrow \hat{E}^\dagger \hat{F}\in\mathcal{S}
\ \text{for all }\hat{E},\hat{F}\in\mathsf{P}_A
\right]
\nonumber\\
&&\hspace{25mm}\Longleftrightarrow\quad
\mathcal{N}(\mathcal{S})\cap\mathsf{P}_A
=\mathcal{S}\cap\mathsf{P}_A
\nonumber\\
&&\hspace{25mm}\Longleftrightarrow\quad
A\ \text{is erasure-correctable}.
\label{eq:S5_fixed_support_equivalence}
\end{eqnarray}

\item[(iii)] Define the first cumulative weight at which a logically ambiguous pair exists by
\begin{eqnarray}
w_\star:=\min\left\{w:\ \exists \hat{E},\hat{F}\in\mathcal{E}_{\le w},\ 
{\rm syn}(\hat{E})={\rm syn}(\hat{F}),\ \hat{E}^\dagger \hat{F}\notin\mathcal{S}\right\}.
\label{eq:S5_wstar_def}
\end{eqnarray}
Then
\begin{eqnarray}
w_\star=\left\lceil\frac{d}{2}\right\rceil
=\left\lfloor\frac{d-1}{2}\right\rfloor+1.
\label{eq:S5_wstar_distance}
\end{eqnarray}
In particular, $R_{\rm exact}(w)=1$ for every $w\le\lfloor(d-1)/2\rfloor$.
Moreover, under the accumulated-library convention of Eq.~(\ref{eq:S5_unambiguous_indicator}), at least one member of the exact layer $\mathcal{E}_{w_\star}$ belongs to an ambiguous fiber, so $R_{\rm exact}(w_\star)<1$.

\item[(iv)] For a harmful pair let
\begin{eqnarray}
A_{EF}:=\operatorname{supp}(\hat{E}^\dagger \hat{F})
\subseteq\operatorname{supp}(\hat{E})\cup\operatorname{supp}(\hat{F}).
\label{eq:S5_pair_support}
\end{eqnarray}
Then $A_{EF}$ supports a nontrivial logical Pauli, is not erasure-correctable, and obeys $\abs{A_{EF}}\le\operatorname{wt}(\hat{E})+\operatorname{wt}(\hat{F})$. For an $[[n,1]]$ code, $q_{A_{EF}}\in\{1,3\}$: the $q_{A_{EF}}=1$ collisions are logical-ED active, whereas the $q_{A_{EF}}=3$ collisions are logical-ED blind.
\end{enumerate}
\end{proposition}

\begin{proof}---Write the syndrome bit associated with a stabilizer generator $\hat{g}_j$ as the binary symplectic commutation value between $\hat{E}$ and $\hat{g}_j$. The syndrome map is additive under Pauli multiplication, so ${\rm syn}(\hat{E})={\rm syn}(\hat{F})$ if and only if ${\rm syn}(\hat{E}^\dagger \hat{F})=0$. A Pauli has zero syndrome precisely when it commutes with every stabilizer generator, which is membership in $\mathcal{N}(\mathcal{S})$. If the relative product lies in $\mathcal{S}$, it acts trivially on the code space and the two errors are degenerate. If it lies in $\mathcal{N}(\mathcal{S})\setminus\mathcal{S}$, it acts as a nontrivial logical Pauli, so applying a recovery appropriate to one error implements a logical fault on the other. This proves (i).

For (ii), the absence of an element of $\mathcal{N}(\mathcal{S})\setminus\mathcal{S}$ supported in $A$ is exactly the middle equality in Eq.~(\ref{eq:S5_fixed_support_equivalence}). If the equality fails, take such a logical operator $\hat{L}$ together with the identity: they lie in $\mathsf{P}_A$, have the same syndrome, and are not stabilizer-degenerate. Conversely, any harmful pair in $\mathsf{P}_A$ has $\hat{E}^\dagger \hat{F}\in\mathcal{N}(\mathcal{S})\setminus\mathcal{S}$ supported in $A$. The equivalence with erasure correctability is the stabilizer cleanability criterion of Lemma~\ref{lem:S4_cleanability}.

For the lower bound in (iii), if $\hat{E},\hat{F}\in\mathcal{E}_{\le w}$, then $\operatorname{wt}(\hat{E}^\dagger \hat{F})\le2w$. When $2w<d$, the relative product cannot be a nontrivial logical Pauli by the definition of distance, so every same-syndrome pair is stabilizer-degenerate. Hence $w_\star\ge\lceil d/2\rceil$. For the reverse inequality, choose a minimum-weight logical Pauli $\hat{L}$ of weight $d$ and partition its support into disjoint sets $B$ and $B^c\cap\operatorname{supp}(\hat{L})$ of sizes at most $\lceil d/2\rceil$. Let $\hat{E}$ and $\hat{F}$ be the restrictions of $\hat{L}$ to these two sets. Their supports are disjoint, $\hat{E}^\dagger \hat{F}=\hat{L}$ up to phase, and $\hat{L}\in\mathcal{N}(\mathcal{S})\setminus\mathcal{S}$; consequently $\hat{E}$ and $\hat{F}$ have the same syndrome and weights at most $\lceil d/2\rceil$. This proves Eq.~(\ref{eq:S5_wstar_distance}).

Finally, the relative product of a harmful pair is itself a nontrivial logical representative supported in $A_{EF}$, proving (iv) and its size bound. For $k=1$, Proposition~\ref{prop:S5_trichotomy} gives $q_A\in\{0,1,3\}$, with noncorrectability excluding $q_A=0$; Eq.~(\ref{eq:S5_delta_q}) then gives the stated logical-ED classification.
\end{proof}

The one-axis activation statement has a useful limited extension beyond $k=1$, which is relevant because a toric code can encode more than one logical qubit.

\begin{lemma}[One supported logical class activates logical-ED]
\label{lem:S5_one_class_multik}
Let an $[[n,k]]$ stabilizer code have logical dimension $D=2^k$, and let
\begin{eqnarray}
h_A:=\log_2\abs{\mathcal{S}\cap\mathsf{P}_A}.
\label{eq:S5_hA_multik}
\end{eqnarray}
Suppose that, modulo stabilizers, the only logical Pauli classes admitting representatives in $A$ are the identity and one nonidentity class $\hat{\overline{P}}$. Then, for a Haar-random logical pure state,
\begin{eqnarray}
\tr{\bigl(\hat{\rho}_A(\psi_L)^2\bigr)}
&=&2^{h_A-\abs{A}}\left(1+\expt{\hat{\overline{P}}}_{\psi_L}^{2}\right),
\label{eq:S5_oneclass_purity_multik}\\
\Delta_A
&=&2^{h_A-\abs{A}}
\sqrt{\frac{2D}{(D+1)^2(D+3)}}>0.
\label{eq:S5_oneclass_delta_multik}
\end{eqnarray}
The relative purity fluctuation is
\begin{eqnarray}
\frac{\Delta_A}{\overline{P}_A}
=\frac{\sqrt{2D/(D+3)}}{D+2},
\quad
\overline{P}_A:=1-\overline{E}_A.
\label{eq:S5_oneclass_relative_multik}
\end{eqnarray}
For $k=1$, Eq.~(\ref{eq:S5_oneclass_relative_multik}) reduces to $1/\sqrt{20}$.
\end{lemma}

\begin{proof}---Expand the encoded logical state in the logical Pauli basis and retain only physical Pauli terms supported in $A$. By hypothesis there are two supported logical cosets, the stabilizer coset and the coset of $\hat{\overline{P}}$, and each contains $2^{h_A}$ representatives supported in $A$. Pauli orthogonality on $A$ gives Eq.~(\ref{eq:S5_oneclass_purity_multik}). A nonidentity logical Pauli is a traceless Hermitian unitary on the $D$-dimensional logical space. For $x:=\expt{\hat{\overline{P}}}_{\psi_L}$, the Haar moments are
\begin{eqnarray}
\mathbb{E}(x^2)=\frac{1}{D+1},
\quad
\mathbb{E}(x^4)=\frac{3}{(D+1)(D+3)}.
\label{eq:S5_oneclass_Haar_moments}
\end{eqnarray}
Consequently ${\rm Var}(x^2)=2D/[(D+1)^2(D+3)]$, proving Eq.~(\ref{eq:S5_oneclass_delta_multik}). The mean purity is $\overline{P}_A=2^{h_A-\abs{A}}(D+2)/(D+1)$, which gives Eq.~(\ref{eq:S5_oneclass_relative_multik}).
\end{proof}

Proposition~\ref{prop:S5_decoder_bridge} connects the regional and error-space diagnostics through an exact fixed-support criterion. The ED profile scans physical regions and assigns a real-valued purity fluctuation. By contrast, $R_{\rm exact}(w)$ scans Pauli labels over many unknown supports and counts syndrome fibers. Logical-ED detects only the one-axis part of the normalizer obstruction for $k=1$, while $R_{\rm exact}$ treats both one-axis and full-access logical-coset collisions as ambiguous.

At the level of region sizes, define the hard support-conditioned syndrome radius by
\begin{eqnarray}
d_{\rm syn}:=\max\left\{s:\ 
\mathcal{N}(\mathcal{S})\cap\mathsf{P}_A
=\mathcal{S}\cap\mathsf{P}_A
\ \text{for every }\abs{A}\le s\right\}.
\label{eq:S5_syndrome_radius}
\end{eqnarray}
The definition of the code distance and Eq.~(\ref{eq:S5_fixed_support_equivalence}) give the exact identity
\begin{eqnarray}
d_{\rm syn}=d-1\le d_\Delta(0).
\label{eq:S5_syndrome_soft_relation}
\end{eqnarray}
The inequality can be strict because $d_\Delta(0)$ also accepts ED-inactive, noncorrectable regions. If a minimum-weight logical support is ED-active, then the first violation of the zero-tolerance ED condition occurs at size $d$, and Eq.~(\ref{eq:S5_syndrome_soft_relation}) is saturated. This support-radius identity is the sharpest direct relation between exact syndrome distinguishability and the zero-tolerance ED-soft distance.

The relation to the ED-soft distance can be made explicit at arbitrary tolerance. Provided the superlevel set is nonempty, define
\begin{eqnarray}
r_\Delta(\varepsilon)
&:=&\min\{\abs{A}:\Delta_A>\varepsilon\}
=d_\Delta(\varepsilon)+1,
\nonumber\\
w_\Delta(\varepsilon)
&:=&\left\lceil\frac{r_\Delta(\varepsilon)}{2}\right\rceil.
\label{eq:S5_soft_ambiguity_scale}
\end{eqnarray}
Because $\Delta_A>\varepsilon\ge0$ implies that $A$ is not erasure-correctable, a stabilizer code has a nontrivial logical Pauli supported in that region. Splitting one such representative as in the proof above produces a harmful same-syndrome pair of weights at most $w_\Delta(\varepsilon)$. Therefore
\begin{eqnarray}
w_\star\le w_\Delta(\varepsilon)
=\left\lceil\frac{d_\Delta(\varepsilon)+1}{2}\right\rceil.
\label{eq:S5_soft_collision_bound}
\end{eqnarray}
At zero tolerance, if at least one minimum-weight logical support is ED-active, then $r_\Delta(0)=d$, $d_\Delta(0)=d-1$, and Eq.~(\ref{eq:S5_soft_collision_bound}) is saturated:
\begin{eqnarray}
w_\star
=\left\lceil\frac{d}{2}\right\rceil
=\left\lceil\frac{d_\Delta(0)+1}{2}\right\rceil.
\label{eq:S5_zero_soft_collision_equality}
\end{eqnarray}
This is the relevant situation for a minimum one-axis logical string or loop. If every minimum logical support visible at the first obstruction were ED-inactive (for example, a full-access support), only the one-sided bound survives; the decoder-free collision still exists, but logical-ED does not locate it.

For future joint analyses it is useful to isolate the ambiguity that is explicitly witnessed by logical-ED. Define
\begin{eqnarray}
R_{\rm ED\text{-}amb}(w)
:=\frac{1}{\abs{\mathcal{E}_w}}
\sum_{\hat{E}\in\mathcal{E}_w}
\mathbf{1}\!\left[
\begin{array}{l}
\exists \hat{F}\in\mathcal{E}_{\le w}:\ {\rm syn}(\hat{F})={\rm syn}(\hat{E}),\ 
\hat{E}^\dagger \hat{F}\notin\mathcal{S},\\[-1mm]
\hfill \Delta_{A_{EF}}>0
\end{array}
\right].
\label{eq:S5_R_ED_amb_def}
\end{eqnarray}
Every event counted here is ambiguous under Eq.~(\ref{eq:S5_R_exact_def}), and hence
\begin{eqnarray}
0\le R_{\rm ED\text{-}amb}(w)\le1-R_{\rm exact}(w).
\label{eq:S5_R_ED_amb_bound}
\end{eqnarray}
For $k=1$, Eq.~(\ref{eq:S5_R_ED_amb_def}) selects collisions admitting a $q_{A_{EF}}=1$ witness. The possible gap in Eq.~(\ref{eq:S5_R_ED_amb_bound}) contains logical-coset ambiguity not witnessed by regional purity, including the $q_A=3$ blind sector. The inequality separates the part of syndrome ambiguity witnessed by logical-ED from the full count.

\begingroup
The quantities in Eqs.~(\ref{eq:S5_R_exact_def}) and (\ref{eq:S5_R_cum_def}) can be evaluated by counting the errors of each weight in each syndrome and stabilizer-coset pair. At cutoff $w$ a syndrome fiber is ambiguous if it contains at least two occupied cosets in $\mathcal{E}_{\le w}$. The numerator of $1-R_{\rm cum}(w)$ includes \emph{all} errors in every such fiber, including the identity if it belongs to one. The numerator of $1-R_{\rm exact}(w)$ includes only its weight-$w$ errors. Proposition~\ref{prop:S5_decoder_bridge} fixes the first ambiguous cutoff but does not quantify the fraction beyond that cutoff.

For example, an unrotated planar surface-code patch uses $n=d^2+(d-1)^2$ data qubits, giving $85$ at $d=7$ and $313$ at $d=13$. The distance guarantee for arbitrary errors on six unknown sites requires $d\ge13$. The fractions above quantify combinatorial distinguishability under ideal syndrome extraction and depend on the code and candidate library. Logical failure probabilities and fault-tolerance thresholds require a specified noise model and recovery procedure.
\endgroup

\subsection{Direct validation, runtime, and reproducibility}

We validated the coset formula against a direct state-vector construction of the $[[9,1,3]]$ patch. A random vector was successively projected into the $+1$ eigenspaces of all stabilizers and logical $\hat{\overline{Z}}$ to obtain $\ket{0_L}$; applying $\hat{\overline{X}}$ produced $\ket{1_L}$. For each of $12$ structured or random regions and $32$ independently sampled logical qubit states, we formed $\alpha\ket{0_L}+\beta\ket{1_L}$, explicitly traced out $A^c$, and compared the individual linear entropy $E_A(\psi_L)=1-\tr(\hat{\rho}_A^2)$ with one minus the purity from Eq.~(\ref{eq:S5_surface_purity_counts}). The largest absolute discrepancy over all $384$ comparisons was
\begin{eqnarray}
8.88\times10^{-16},
\label{eq:S5_validation_error}
\end{eqnarray}
consistent with double-precision roundoff. The Haar-mean and deviation formulas follow separately from the analytic moments in Eq.~(\ref{eq:S5_bloch_moments}).

The exact region calculation is sufficiently light to scan thousands of masks on codes far beyond state-vector reach. Table~\ref{tab:S5_runtime} reports the median time per region for the single-core, bit-packed Python implementation used for the surface-code calculations. The benchmark used three repeats of $2000$ random masks at each distance; it is not intended as a hardware-independent complexity claim. The dominant operation is elimination of $m=n-1$ integers whose occupied bit length is at most $2n$.

\begin{table}[t]
\centering
\setlength{\tabcolsep}{15pt}
\begin{tabular}{c c c}
\hline\hline
$d$ & $n=d^2$ & median time per region ($\mu$s)\\
\hline
$5$ & $25$ & $15.0$\\
$7$ & $49$ & $24.6$\\
$9$ & $81$ & $37.3$\\
$11$ & $121$ & $57.3$\\
$15$ & $225$ & $103.0$\\
\hline\hline
\end{tabular}
\caption{Median runtime per region in the bit-packed implementation. The values are the medians of three $2000$-mask runs in the benchmark environment.}
\label{tab:S5_runtime}
\end{table}

The input check conventions, rank tests and sampling parameters are specified in this note. See the Data Availability and Code Availability statements in the main text for access to the source code and underlying data.


\section{Generalizations and further comments}

\subsection{Qudit and non-uniform local dimensions}

The definition of logical-ED and the exact-correctability theorem of Supplementary Note~S2 extend directly from qubits to \emph{qudits} and non-uniform local dimensions. Stabilizer- and surface-code-specific geometric statements require the corresponding generalized-Pauli conventions~\cite{hostens2005stabilizer} and are not asserted here. Concretely, let the physical Hilbert space be
\begin{eqnarray}
\mathcal{H}_{\rm phys}  = \bigotimes_{i=1}^{n}\mathbb{C}^{q_i},
\label{eq:S6_phys_qudit}
\end{eqnarray}
allowing even site-dependent local dimension $q_i$. For a region $A\subseteq\{1,\dots,n\}$ define
\begin{eqnarray}
\mathcal{H}_A := \bigotimes_{i\in A}\mathbb{C}^{q_i},
\quad
d_A := \dim(\mathcal{H}_A)=\prod_{i\in A}q_i.
\label{eq:S6_dimA}
\end{eqnarray}
Let $\hat{V}:\mathcal{H}_L\to\mathcal{H}_{\rm phys}$ be an encoding isometry and $\hat{\rho}_A(\psi_L)$ be the reduced state on $A$ for the encoded pure state $\hat{V}\ket{\psi_L}$. We define
\begin{eqnarray}
E_A(\psi_L) := 1-\tr{\bigl(\hat{\rho}_A(\psi_L)^2\bigr)},
\quad
\Delta_A := \sqrt{\int d\psi_L\,E_A(\psi_L)^2-\left(\int d\psi_L\,E_A(\psi_L)\right)^2},
\label{eq:S6_EPD_qudit_def}
\end{eqnarray}
with Haar measure $d\psi_L$ on $\mathcal{H}_L$.
The bounds in Eq.~(\ref{eq:S1_linear_entropy_bounds}) generalize immediately:
\begin{eqnarray}
0\ \le\ E_A(\psi_L)\ \le\ 1-\frac{1}{d_A}.
\label{eq:S6_E_bounds_qudit}
\end{eqnarray}
The proof of Theorem~1 (main text) in Supplementary Note~S2 uses only the Knill--Laflamme condition and the Kraus representation of $\tr_A$ in terms of matrix units on $\mathcal{H}_A$, both of which are dimension-agnostic. Hence the conclusion
\begin{eqnarray}
A \ \text{erasure-correctable}
\quad\Longrightarrow\quad
\hat{\rho}_A(\psi_L)\ \text{independent of }\psi_L
\quad\Longrightarrow\quad
\Delta_A=0
\label{eq:S6_qudit_thm1}
\end{eqnarray}
holds for arbitrary qudit codes and arbitrary region Hilbert space dimensions.

\subsection{Subsystem codes and gauge degrees of freedom}

Many fault-tolerant architectures employ \emph{subsystem} codes (including gauge codes) in which only a \emph{logical subsystem} must be protected, while an additional \emph{gauge subsystem} may fluctuate~\cite{bacon2006operator,poulin2005stabilizer,kribs2005operator}. In this setting, the natural generalization of Theorem~1 (main text) is that a correctable region cannot carry \emph{logical} information, although it may carry gauge information. This section makes that statement precise and explains how logical-ED can be defined in a gauge-consistent way.

\smallskip\noindent
{\bf Subsystem encoding.}
Let the logical and gauge Hilbert spaces be $\mathcal{H}_L$ and $\mathcal{H}_G$. An encoding isometry is now
\begin{eqnarray}
\hat{V}:\ \mathcal{H}_L\otimes \mathcal{H}_G\ \longrightarrow\ \mathcal{H}_{\rm phys},
\quad
\hat{V}^\dagger \hat{V}=\hat{\mathds{1}}_{LG}.
\label{eq:S6_subsystem_isometry}
\end{eqnarray}
The image $\mathcal{C}:={\rm Im}(\hat{V})$ is isomorphic to $\mathcal{H}_L\otimes\mathcal{H}_G$. We denote the projector onto $\mathcal{C}$ by
\begin{eqnarray}
\hat{P} := \hat{V}\hat{V}^\dagger.
\label{eq:S6_subsystem_P}
\end{eqnarray}
For product logical--gauge inputs, the encoded state is $\hat{V}(\hat{\sigma}_L\otimes\hat{\tau}_G)\hat{V}^\dagger$.

\smallskip\noindent
{\bf Exact correctability for the logical subsystem.}
For an erasure of region $A$, we say that $A$ is \emph{exactly correctable for the logical subsystem} if there exists a recovery channel $\mathcal{R}$ acting on $A^c$ and a CPTP map $\Lambda$ on $\mathcal{H}_G$ such that for all $\hat{\sigma}_L,\hat{\tau}_G$,
\begin{eqnarray}
(\mathcal{R}\circ \tr_A)\Bigl(\hat{V}(\hat{\sigma}_L\otimes \hat{\tau}_G)\hat{V}^\dagger\Bigr)
 =
\hat{V}\Bigl(\hat{\sigma}_L\otimes \Lambda(\hat{\tau}_G)\Bigr)\hat{V}^\dagger .
\label{eq:S6_subsystem_correctability}
\end{eqnarray}
This expresses that $\mathcal{R}$ restores the logical state perfectly, while the gauge subsystem may undergo a known transformation $\Lambda$.

\smallskip\noindent
{\bf Generalized Knill--Laflamme condition.}
A standard characterization of subsystem QEC states that Eq.~(\ref{eq:S6_subsystem_correctability}) is equivalent to a generalized Knill--Laflamme condition in which error overlaps act trivially on $\mathcal{H}_L$ but may act nontrivially on $\mathcal{H}_G$~\cite{kribs2005operator,bacon2006operator}. Specializing to the erasure channel, one obtains the following (we include a direct proof sketch tailored to our setting).

\begin{lemma}[Generalized KL condition for erasure on a subsystem code]\label{lem:S6_subsystem_KL}
Assume $A$ is exactly correctable for the logical subsystem in the sense of Eq.~(\ref{eq:S6_subsystem_correctability}). Let $\{\ket{a}\}_{a=1}^{d_A}$ be an orthonormal basis of $\mathcal{H}_A$ and define the Kraus operators of $\tr_A$ by $\hat{K}_a=\bra{a}_A\otimes \hat{\mathds{1}}_{A^c}$ as in Eq.~(\ref{eq:S2_Kraus_trace}). Then there exist operators $\hat{\alpha}_{ab}$ on $\mathcal{H}_G$ such that
\begin{eqnarray}
\hat{V}^\dagger\Bigl(\hat{K}_a^\dagger \hat{K}_b\Bigr)\hat{V}
 =
\hat{\mathds{1}}_L\otimes \hat{\alpha}_{ab}
\quad
\text{for all }a,b.
\label{eq:S6_subsystem_KL_form}
\end{eqnarray}
Equivalently, using $\hat{K}_a^\dagger \hat{K}_b=\ket{a}\bra{b}_A\otimes \hat{\mathds{1}}_{A^c}$,
\begin{eqnarray}
\hat{V}^\dagger\Bigl(\ket{a}\bra{b}_A\otimes \hat{\mathds{1}}_{A^c}\Bigr)\hat{V}
 =
\hat{\mathds{1}}_L\otimes \hat{\alpha}_{ab}.
\label{eq:S6_subsystem_matrix_units}
\end{eqnarray}
\end{lemma}

\begin{proof}---We give a detailed proof specialized to the erasure channel and the tensor-factor decomposition $\mathcal{H}_L\otimes \mathcal{H}_G$.

\smallskip\noindent
{\bf Step 1: A fixed gauge input implies no logical dependence on $A$.}
Fix an arbitrary gauge state $\hat{\tau}_G$.
Define the induced encoding channel on logical inputs by
\begin{eqnarray}
\mathcal{V}_{\hat{\tau}_G}(\hat{\sigma}_L)
 :=
\hat{V}(\hat{\sigma}_L\otimes \hat{\tau}_G)\hat{V}^\dagger,
\label{eq:S6_V_tau_def}
\end{eqnarray}
and consider the induced channel to the unerased subsystem,
$\mathcal{N}_{A^c}^{(\hat{\tau}_G)}:=\tr_A\circ \mathcal{V}_{\hat{\tau}_G}$.
By assumption~(\ref{eq:S6_subsystem_correctability}), there exists a recovery $\mathcal{R}$ such that for all $\hat{\sigma}_L$
\begin{eqnarray}
(\mathcal{R}\circ \mathcal{N}_{A^c}^{(\hat{\tau}_G)})(\hat{\sigma}_L)
 =
\hat{V}\bigl(\hat{\sigma}_L\otimes \Lambda(\hat{\tau}_G)\bigr)\hat{V}^\dagger.
\label{eq:S6_recover_for_fixed_tau}
\end{eqnarray}
From the recovered encoded state one can decode the logical subsystem by applying the (extended) inverse isometry $\hat{V}^\dagger$ and tracing out the gauge:
define a CPTP map $\mathcal{D}$ on physical states by
\begin{eqnarray}
\mathcal{D}(\hat{\varrho})
 :=
\tr_G\!\Bigl(\hat{V}^\dagger \hat{P}\hat{\varrho}\hat{P} \hat{V}\Bigr)
\ +\
\tr{\bigl((\hat{\mathds{1}}-\hat{P})\hat{\varrho}\bigr)}\,\ketbra{0}{0},
\label{eq:S6_decode_map}
\end{eqnarray}
where $\hat{P}=\hat{V}\hat{V}^\dagger$ is the code projector and $\ket{0}$ is a fixed reference state on $\mathcal{H}_L$.
On states supported on the code, $\mathcal{D}$ reduces to $\tr_G(\hat{V}^\dagger(\cdot)\hat{V})$.
Combining Eqs.~(\ref{eq:S6_recover_for_fixed_tau}) and (\ref{eq:S6_decode_map}) gives
\begin{eqnarray}
(\mathcal{D}\circ \mathcal{R}\circ \mathcal{N}_{A^c}^{(\hat{\tau}_G)})(\hat{\sigma}_L)
 =
\hat{\sigma}_L
\quad \text{for all }\hat{\sigma}_L.
\label{eq:S6_left_inverse_channel}
\end{eqnarray}
Thus, for each fixed $\hat{\tau}_G$, $\mathcal{N}_{A^c}^{(\hat{\tau}_G)}$ is reversible on the logical input.

Now view $\mathcal{N}_{A^c}^{(\hat{\tau}_G)}$ as a channel with input system $\mathcal{H}_L$ and output system $\mathcal{H}_{A^c}$. Choose a purification $\ket{\tau}_{GR}$ of $\hat{\tau}_G$. A Stinespring isometry for $\mathcal{N}_{A^c}^{(\hat{\tau}_G)}$ is obtained from $(\hat{V}\otimes\hat{\mathds{1}}_R)(\ket{\psi_L}\otimes\ket{\tau}_{GR})$ by taking $A^c$ as the channel output. Its complementary output is $A\otimes R$. Exact reversibility of the channel to $A^c$ forces this full complementary output to be independent of the logical input. Tracing out $R$ then shows that the physical marginal on $A$ is logical-state independent: there exists a state $\hat{\omega}_A(\hat{\tau}_G)$ such that
\begin{eqnarray}
\tr_{A^c}\!\Bigl(\hat{V}(\ketbra{\psi_L}{\psi_L}\otimes \hat{\tau}_G)\hat{V}^\dagger\Bigr)
 =
\hat{\omega}_A(\hat{\tau}_G)
\quad \text{for all }\ket{\psi_L}.
\label{eq:S6_rhoA_indep_step1}
\end{eqnarray}

\smallskip\noindent
{\bf Step 2: Linear dependence on $\hat{\tau}_G$ and Riesz representation.}
For fixed $a,b$, define the scalar functional
\begin{eqnarray}
f_{ab}(\hat{\tau}_G)
 :=
\bra{b}\hat{\omega}_A(\hat{\tau}_G)\ket{a}.
\label{eq:S6_fab_def}
\end{eqnarray}
Because the map $\hat{\tau}_G\mapsto \hat{\omega}_A(\hat{\tau}_G)$ arises from composing a linear CPTP map with a partial trace, it is linear in $\hat{\tau}_G$; hence $f_{ab}$ is a linear functional on operators on $\mathcal{H}_G$.
By the finite-dimensional Riesz representation theorem, there exists an operator $\hat{\alpha}_{ab}$ on $\mathcal{H}_G$ such that
\begin{eqnarray}
f_{ab}(\hat{\tau}_G)
 =
\tr{\bigl(\hat{\alpha}_{ab}\hat{\tau}_G\bigr)}
\quad \text{for all }\hat{\tau}_G.
\label{eq:S6_Riesz}
\end{eqnarray}

\smallskip\noindent
{\bf Step 3: Upgrading scalar identities to an operator identity on $\mathcal{H}_L\otimes\mathcal{H}_G$.}
Define the operator on $\mathcal{H}_L\otimes\mathcal{H}_G$
\begin{eqnarray}
\hat{X}_{ab}
 :=
\hat{V}^\dagger\Bigl(\ket{a}\bra{b}_A\otimes \hat{\mathds{1}}_{A^c}\Bigr)\hat{V}.
\label{eq:S6_Xab_def}
\end{eqnarray}
For any pure $\ket{\psi_L}$ and any $\hat{\tau}_G$, we have
\begin{eqnarray}
\tr{\Bigl((\ketbra{\psi_L}{\psi_L}\otimes \hat{\tau}_G)\,\hat{X}_{ab}\Bigr)}
&=&
\tr{\Bigl(\bigl(\ket{a}\bra{b}_A\otimes \hat{\mathds{1}}_{A^c}\bigr)\,\hat{V}(\ketbra{\psi_L}{\psi_L}\otimes \hat{\tau}_G)\hat{V}^\dagger\Bigr)}
\nonumber\\
&=&
\bra{b}\,\tr_{A^c}\!\Bigl(\hat{V}(\ketbra{\psi_L}{\psi_L}\otimes \hat{\tau}_G)\hat{V}^\dagger\Bigr)\,\ket{a}
 =
f_{ab}(\hat{\tau}_G),
\label{eq:S6_expect_Xab}
\end{eqnarray}
which is independent of $\ket{\psi_L}$ by Eq.~(\ref{eq:S6_rhoA_indep_step1}).
Write $\mathcal{H}_L$ in an orthonormal basis $\{\ket{i}\}_{i=1}^{d_L}$ and expand
\begin{eqnarray}
\hat{X}_{ab}
 =
\sum_{i,j=1}^{d_L} \ketbra{i}{j}\otimes \hat{X}_{ab}^{(ij)},
\quad
\hat{X}_{ab}^{(ij)}\in \mathcal{B}(\mathcal{H}_G).
\label{eq:S6_Xab_block}
\end{eqnarray}
For fixed $\hat{\tau}_G$, Eq.~(\ref{eq:S6_expect_Xab}) becomes
\begin{eqnarray}
\sum_{i,j}\braket{\psi_L}{i}\braket{j}{\psi_L}\ \tr{\bigl(\hat{X}_{ab}^{(ij)}\hat{\tau}_G\bigr)}
 =
\tr{\bigl(\hat{\alpha}_{ab}\hat{\tau}_G\bigr)}
\quad \text{for all }\ket{\psi_L}.
\label{eq:S6_quadratic_form}
\end{eqnarray}
The left-hand side is the quadratic form $\bra{\psi_L} \hat{M}(\hat{\tau}_G)\ket{\psi_L}$ of the matrix $\hat{M}(\hat{\tau}_G)$ with entries $M_{ij}(\hat{\tau}_G):=\tr{\bigl(\hat{X}_{ab}^{(ij)}\hat{\tau}_G\bigr)}$.
Since Eq.~(\ref{eq:S6_quadratic_form}) holds for all unit vectors $\ket{\psi_L}$, it follows that $\hat{M}(\hat{\tau}_G)$ is proportional to the identity:
\begin{eqnarray}
M_{ij}(\hat{\tau}_G)=0\ (i\neq j),
\quad
M_{ii}(\hat{\tau}_G)=\tr{\bigl(\hat{\alpha}_{ab}\hat{\tau}_G\bigr)}\ \text{for all }i.
\label{eq:S6_M_identity}
\end{eqnarray}
Because Eq.~(\ref{eq:S6_M_identity}) holds for all density operators $\hat{\tau}_G$, we conclude
\begin{eqnarray}
\hat{X}_{ab}^{(ij)}=0\ (i\neq j),
\quad
\hat{X}_{ab}^{(ii)}=\hat{\alpha}_{ab}\ \text{for all }i.
\label{eq:S6_blocks_identified}
\end{eqnarray}
Substituting Eq.~(\ref{eq:S6_blocks_identified}) into Eq.~(\ref{eq:S6_Xab_block}) gives
\begin{eqnarray}
\hat{X}_{ab}
 =
\hat{\mathds{1}}_L\otimes \hat{\alpha}_{ab},
\label{eq:S6_Xab_tensor}
\end{eqnarray}
which is precisely Eq.~(\ref{eq:S6_subsystem_KL_form}) (and its equivalent form Eq.~(\ref{eq:S6_subsystem_matrix_units})).
\end{proof}

\smallskip\noindent
{\bf Logical indistinguishability on $A$ (subsystem form).}
Lemma~\ref{lem:S6_subsystem_KL} implies that operators supported on $A$ cannot distinguish different \emph{logical} states, although they may distinguish different gauge states. This is the direct subsystem analogue of Eq.~(\ref{eq:S2_scalar_matrix_units}). We now prove the corresponding reduced-state independence.

\begin{theorem}[Subsystem analogue of Theorem~1]\label{thm:S6_subsystem_thm}
Assume $A$ is exactly correctable for the logical subsystem in the sense of Eq.~(\ref{eq:S6_subsystem_correctability}). Then for every gauge state $\hat{\tau}_G$ there exists a density operator $\hat{\omega}_A(\hat{\tau}_G)$ on $\mathcal{H}_A$ such that for all logical pure states $\ket{\psi_L}$,
\begin{eqnarray}
\hat{\rho}_A(\psi_L;\hat{\tau}_G)
 :=
\tr_{A^c}\!\Bigl(\hat{V}(\ketbra{\psi_L}{\psi_L}\otimes \hat{\tau}_G)\hat{V}^\dagger\Bigr)
 =
\hat{\omega}_A(\hat{\tau}_G),
\label{eq:S6_subsystem_rhoA_indep}
\end{eqnarray}
i.e., the reduced state on $A$ is independent of the logical input.
\end{theorem}

\begin{proof}---Fix a basis $\{\ket{a}\}$ of $\mathcal{H}_A$ and consider the matrix elements
$\bra{b}\hat{\rho}_A(\psi_L;\hat{\tau}_G)\ket{a}$.
As in Eq.~(\ref{eq:S2_rhoA_matrix_2}),
\begin{eqnarray}
\bra{b}\hat{\rho}_A(\psi_L;\hat{\tau}_G)\ket{a}
&=&
\tr{\Bigl(\bigl(\ket{a}\bra{b}_A\otimes \hat{\mathds{1}}_{A^c}\bigr)\,
\hat{V}(\ketbra{\psi_L}{\psi_L}\otimes \hat{\tau}_G)\hat{V}^\dagger\Bigr)}
\nonumber\\
&=&
\tr{\Bigl(
\hat{V}^\dagger(\ket{a}\bra{b}_A\otimes \hat{\mathds{1}}_{A^c})\hat{V}\,
(\ketbra{\psi_L}{\psi_L}\otimes \hat{\tau}_G)
\Bigr)}.
\label{eq:S6_subsystem_matrixel_1}
\end{eqnarray}
Now apply Eq.~(\ref{eq:S6_subsystem_matrix_units}):
\begin{eqnarray}
\bra{b}\hat{\rho}_A(\psi_L;\hat{\tau}_G)\ket{a}
&=&
\tr{\Bigl(
(\hat{\mathds{1}}_L\otimes \hat{\alpha}_{ab})\,
(\ketbra{\psi_L}{\psi_L}\otimes \hat{\tau}_G)
\Bigr)}
 =
\tr{\bigl(\ketbra{\psi_L}{\psi_L}\bigr)}\ \tr{\bigl(\hat{\alpha}_{ab}\hat{\tau}_G\bigr)}
\nonumber\\
&=&
\tr{\bigl(\hat{\alpha}_{ab}\hat{\tau}_G\bigr)}.
\label{eq:S6_subsystem_matrixel_2}
\end{eqnarray}
The right-hand side is independent of $\psi_L$, hence $\hat{\rho}_A(\psi_L;\hat{\tau}_G)$ is independent of the logical input. Define $\hat{\omega}_A(\hat{\tau}_G)$ by its matrix elements $\bra{b}\hat{\omega}_A(\hat{\tau}_G)\ket{a}:=\tr{(\hat{\alpha}_{ab}\hat{\tau}_G)}$. Positivity and unit trace follow because $\hat{\omega}_A(\hat{\tau}_G)$ is a reduced state of a physical density operator.
\end{proof}

\smallskip\noindent
{\bf Consequence for logical-ED in subsystem codes.}
There are several natural ways to define logical-ED in the presence of gauge degrees of freedom. Two convenient choices are:

\smallskip\noindent
(i) \emph{Gauge-fixed logical-ED.} Fix a reference gauge state $\ket{g}\in\mathcal{H}_G$ and define the induced isometry
\begin{eqnarray}
\hat{V}_g:\ \mathcal{H}_L\to\mathcal{H}_{\rm phys},
\quad
\hat{V}_g\ket{\psi_L}:=\hat{V}\bigl(\ket{\psi_L}\otimes\ket{g}\bigr).
\label{eq:S6_Vg_def}
\end{eqnarray}
Then define $\Delta_A^{(g)}$ from $\hat{V}_g$ exactly as in the main text. Theorem~\ref{thm:S6_subsystem_thm} implies that if $A$ is logically correctable then $\hat{\rho}_A^{(g)}(\psi_L)$ is independent of $\psi_L$, hence $\Delta_A^{(g)}=0$.

\smallskip\noindent
(ii) \emph{Fixed mixed-gauge purity fluctuation.} Choose a fixed gauge state $\hat{\tau}_G$ (e.g., maximally mixed $\hat{\tau}_G=\hat{\mathds{1}}_G/d_G$) and define
$E_A(\psi_L;\hat{\tau}_G)=1-\tr{(\hat{\rho}_A(\psi_L;\hat{\tau}_G)^2)}$ and its deviation $\Delta_A(\hat{\tau}_G)$ over Haar-random $\psi_L$. For mixed $\hat{\tau}_G$ this is a fluctuation of regional purity, since the physical encoded state can be mixed. Again, Theorem~\ref{thm:S6_subsystem_thm} yields $\Delta_A(\hat{\tau}_G)=0$ for all $\hat{\tau}_G$ when $A$ is logically correctable.

These extensions show that the fundamental implication of our framework,
\begin{eqnarray}
\text{(logical) correctability of }A\quad\Longrightarrow\quad \text{vanishing ED on }A,
\label{eq:S6_subsystem_implication}
\end{eqnarray}
persists for subsystem codes, with the natural modification that reduced states on $A$ may depend on the gauge sector but not on the logical sector.

\subsection{Complementary channel viewpoint and general noise channels}

Theorem~1 (main text) can be understood as a special case of a general ``decoupling to the environment'' principle: if a noise channel is exactly correctable on a code, then its complementary channel carries no information about the logical input~\cite{kretschmann2008information}.

\smallskip\noindent
{\bf Stinespring dilation.}
Let $\mathcal{N}$ be a channel acting on the physical system with a Stinespring isometry
\begin{eqnarray}
\hat{U}_{\mathcal{N}}:\ \mathcal{H}_{\rm phys}\ \to\ \mathcal{H}_{\rm out}\otimes \mathcal{H}_{\rm env},
\quad
\mathcal{N}(\hat{\rho})=\tr_{\rm env}\!\left(\hat{U}_{\mathcal{N}}\hat{\rho} \hat{U}_{\mathcal{N}}^\dagger\right).
\label{eq:S6_Stinespring}
\end{eqnarray}
The \emph{complementary channel} is
\begin{eqnarray}
\mathcal{N}^c(\hat{\rho})=\tr_{\rm out}\!\left(\hat{U}_{\mathcal{N}}\hat{\rho} \hat{U}_{\mathcal{N}}^\dagger\right).
\label{eq:S6_complementary}
\end{eqnarray}

\begin{lemma}[Exact correction implies constant complementary output on the code]\label{lem:S6_complementary_constant}
Let $\hat{V}:\mathcal{H}_L\to\mathcal{H}_{\rm phys}$ be an encoding isometry. If there exists a recovery channel $\mathcal{R}$ such that $(\mathcal{R}\circ\mathcal{N})(\hat{V}\hat{\sigma}_L\hat{V}^\dagger)=\hat{V}\hat{\sigma}_L\hat{V}^\dagger$ for all $\hat{\sigma}_L$, then there exists a fixed environment state $\hat{\omega}_{\rm env}$ such that
\begin{eqnarray}
\mathcal{N}^c\bigl(\hat{V}\hat{\sigma}_L\hat{V}^\dagger\bigr)=\hat{\omega}_{\rm env}
\quad \text{for all }\hat{\sigma}_L.
\label{eq:S6_complementary_constant_eq}
\end{eqnarray}
\end{lemma}

\begin{proof}---Fix Kraus operators $\{\hat{E}_\mu\}$ of $\mathcal{N}$ and let $\hat{P}$ be the code projector. By Lemma~\ref{lem:S2_KL}, exact correctability of $\mathcal{N}$ on the code implies the Knill--Laflamme relations
\begin{eqnarray}
\hat{P}\hat{E}_\mu^\dagger \hat{E}_\nu \hat{P}  = \alpha_{\mu\nu}\,\hat{P}
\quad \text{for all }\mu,\nu
\label{eq:S6_KL_recall}
\end{eqnarray}
for some matrix $\alpha=(\alpha_{\mu\nu})$.

Choose the standard Stinespring dilation associated with this Kraus family:
let $\mathcal{H}_{\rm env}$ have orthonormal basis $\{\ket{\mu}\}$ and define the isometry
\begin{eqnarray}
\hat{U}_{\mathcal{N}}
 :=
\sum_\mu \hat{E}_\mu\otimes \ket{\mu}.
\label{eq:S6_Stinespring_isom}
\end{eqnarray}
Then the complementary channel is
\begin{eqnarray}
\mathcal{N}^c(\hat{\rho})
 =
\tr_{\rm out}\!\Bigl(\hat{U}_{\mathcal{N}}\,\hat{\rho}\,\hat{U}_{\mathcal{N}}^\dagger\Bigr)
 =
\sum_{\mu,\nu}\tr{\bigl(\hat{\rho}\,\hat{E}_\nu^\dagger \hat{E}_\mu\bigr)}\,\ketbra{\mu}{\nu}.
\label{eq:S6_comp_explicit}
\end{eqnarray}
If $\hat{\rho}$ is supported on the code ($\hat{\rho}=\hat{P}\hat{\rho}\hat{P}$), then Eq.~(\ref{eq:S6_KL_recall}) implies
\begin{eqnarray}
\tr{\bigl(\hat{\rho}\,\hat{E}_\nu^\dagger \hat{E}_\mu\bigr)}
=
\alpha_{\nu\mu}\tr{(\hat{\rho})}
=
\alpha_{\nu\mu}.
\end{eqnarray}
Substituting into Eq.~(\ref{eq:S6_comp_explicit}) yields
\begin{eqnarray}
\mathcal{N}^c(\hat{\rho})
 =
\sum_{\mu,\nu}\alpha_{\nu\mu}\,\ketbra{\mu}{\nu}
 =: \hat{\omega}_{\rm env},
\label{eq:S6_comp_constant}
\end{eqnarray}
which is independent of the code input state $\hat{\rho}$.
The operator $\hat{\omega}_{\rm env}$ is a valid density operator because it equals $\mathcal{N}^c(\hat{\rho}_0)$ for any fixed code state $\hat{\rho}_0$.
\end{proof}

\smallskip\noindent
{\bf Theorem~1 as an instance.}
For the erasure channel $\mathcal{N}=\tr_A$ (output on $A^c$), one can choose a Stinespring isometry that coherently stores the discarded subsystem $A$ in the environment. Then the complementary channel is precisely $\mathcal{N}^c(\hat{\rho})=\tr_{A^c}(\hat{\rho})$, i.e., it outputs the reduced state on $A$. Lemma~\ref{lem:S6_complementary_constant} therefore reproduces Theorem~1:
exact correctability of $\tr_A$ implies $\hat{\rho}_A(\psi_L)$ is constant over logical inputs.

\smallskip\noindent
{\bf Comment: channel-ED.}
Lemma~\ref{lem:S6_complementary_constant} suggests a broader extension beyond erasures. Given a noise channel $\mathcal{N}$ and code isometry $\hat{V}$, one may apply the deviation construction to the complementary output $\hat{\rho}_{\rm env}(\psi_L):=\mathcal{N}^c(\hat{V}\ketbra{\psi_L}{\psi_L}\hat{V}^\dagger)$. The resulting environment ED detects logical-input dependence of the environment-output purity. Nonzero environment ED witnesses a nonconstant complementary output and therefore rules out exact correctability. As for erasures, zero environment ED is not sufficient for decoupling because isospectral environment outputs can remain logically distinguishable.

\subsection{Alternative entanglement functionals and design order}

The choice of linear entropy in Eq.~(\ref{eq:S6_EPD_qudit_def}) is guided by two considerations emphasized in Supplementary Note~S1: (i) analytic tractability via low-order Haar moments; and (ii) a finite-copy swap-operator representation. Estimation cost still depends on the subsystem and measurement protocol.

Nevertheless, the logical-ED philosophy applies to many other functionals.
For example, for an integer $m\ge 2$ one may consider
\begin{eqnarray}
E^{(m)}_A(\psi_L) := 1-\tr{\bigl(\hat{\rho}_A(\psi_L)^m\bigr)}.
\label{eq:S6_m_purity}
\end{eqnarray}
This probes higher Rényi moments of the reduced state.
The mean of $E^{(m)}_A$ over Haar-random $\psi_L$ requires the $m$-th Haar moment, while the second moment (hence the deviation) requires the $2m$-th Haar moment. Thus, if one replaces Haar sampling by unitary designs on $\mathcal{H}_L$, then a $(2m)$-design suffices to evaluate the deviation exactly.
In contrast, the von Neumann entropy is not a polynomial functional and typically does not admit closed expressions from finite moments, making it less convenient for fluctuation-based analytics.

Higher spectral moments can resolve cases in which purity is constant but the reduced-state spectrum changes. They do not, however, remove the general blind spot. If $\hat{\rho}_A(\psi_L)=\hat{U}_A(\psi_L)\hat{\omega}_A\hat{U}_A(\psi_L)^\dagger$, then every $\tr{(\hat{\rho}_A(\psi_L)^m)}$ is constant. Equation~(\ref{eq:S1_identity_encoding_counterexample}) is the extreme example: all moments equal one although $A$ contains the complete logical state. A complete diagnostic must therefore include state-sensitive distinguishability, not spectral moments alone.

\subsection{A one-way bound from approximate decoupling to small logical-ED}

Our main theorem is exact. For approximate QEC, the operational object is the distance of the complementary erasure channel from a constant channel, or equivalently up to standard information--disturbance bounds, an optimal recovery error~\cite{beny2010general}. We first record an elementary state-level bound and then state the channel-level implication.

\begin{lemma}[Trace-distance variation bounds logical-ED]\label{lem:S6_trace_to_EPD}
Let $\hat{\rho}_A(\psi_L)$ be the reduced state on $A$ for encoded pure inputs $\ket{\psi_L}$. Suppose there exists $\delta\ge 0$ such that
\begin{eqnarray}
\sup_{\psi_L,\phi_L}\ \norm{\hat{\rho}_A(\psi_L)-\hat{\rho}_A(\phi_L)}_1 \ \le\ \delta.
\label{eq:S6_trace_variation_assumption}
\end{eqnarray}
Then the linear-entropy functional satisfies
\begin{eqnarray}
\sup_{\psi_L,\phi_L}\ \abs{E_A(\psi_L)-E_A(\phi_L)}\ \le\ \delta,
\label{eq:S6_range_bound}
\end{eqnarray}
and consequently the logical-ED obeys
\begin{eqnarray}
\Delta_A\ \le\ \frac{\delta}{2}.
\label{eq:S6_Delta_bound}
\end{eqnarray}
\end{lemma}

\begin{proof}---By Eq.~(\ref{eq:S3_E_Lipschitz}) in Supplementary Note~S3 (a Lipschitz bound for linear entropy),
\begin{eqnarray}
\abs{E(\hat{\rho})-E(\hat{\sigma})}\ \le\ \norm{\hat{\rho}-\hat{\sigma}}_1.
\label{eq:S6_Lipschitz_recall}
\end{eqnarray}
Applying this with $\hat{\rho}=\hat{\rho}_A(\psi_L)$ and $\hat{\sigma}=\hat{\rho}_A(\phi_L)$ and using Eq.~(\ref{eq:S6_trace_variation_assumption}) yields Eq.~(\ref{eq:S6_range_bound}). For any real-valued random variable $X$ supported on an interval of width $w$, the variance satisfies ${\rm Var}(X)\le w^2/4$ (Popoviciu's inequality). Here $X=E_A(\psi_L)$ has width $w\le \delta$, hence ${\rm Var}(E_A)\le \delta^2/4$ and $\Delta_A\le \delta/2$.
\end{proof}

To state the operational content precisely, define the complementary erasure channel from the logical input to the discarded region,
\begin{eqnarray}
\mathcal{N}_A(\hat{\sigma}_L)
:=
\tr_{A^c}\!\left(\hat{V}\hat{\sigma}_L \hat{V}^\dagger\right),
\label{eq:S6_NA_channel}
\end{eqnarray}
and, for a density operator $\hat{\omega}_A$, the constant channel
\begin{eqnarray}
\mathcal{C}_{\hat{\omega}_A}(\hat{\sigma}_L)
:=
\tr{(\hat{\sigma}_L)}\hat{\omega}_A.
\label{eq:S6_constant_channel}
\end{eqnarray}
We use the decoupling error
\begin{eqnarray}
\eta_A
:=
\inf_{\hat{\omega}_A}
\norm{\mathcal{N}_A-\mathcal{C}_{\hat{\omega}_A}}_\diamond.
\label{eq:S6_eta_def}
\end{eqnarray}
Unlike logical-ED, $\eta_A$ controls distinguishability even for logical inputs entangled with a reference system and is the standard type of complementary-channel quantity related to optimal recovery by information--disturbance theorems.

\begin{proposition}[Operational approximate decoupling implies small logical-ED]\label{prop:S6_eta_to_ED}
For every pure-state encoding and every region $A$,
\begin{eqnarray}
\Delta_A\leq \eta_A.
\label{eq:S6_eta_ED_bound}
\end{eqnarray}
No converse bound forcing $\eta_A\to0$ as $\Delta_A\to0$ exists for general encodings.
\end{proposition}

\begin{proof}---Let $\hat{\omega}_A$ attain the infimum in Eq.~(\ref{eq:S6_eta_def}), which exists in finite dimension. For every logical pure state,
\begin{eqnarray}
\norm{\hat{\rho}_A(\psi_L)-\hat{\omega}_A}_1
\leq \eta_A.
\label{eq:S6_pointwise_eta}
\end{eqnarray}
By Eq.~(\ref{eq:S3_E_Lipschitz}), with $e_0:=1-\tr{(\hat{\omega}_A^2)}$,
\begin{eqnarray}
\abs{E_A(\psi_L)-e_0}\leq \eta_A.
\label{eq:S6_E_eta}
\end{eqnarray}
The variance about the mean is no larger than the mean-square deviation about any fixed constant; hence
\begin{eqnarray}
\Delta_A^2
=\min_c\int d\psi_L\,[E_A(\psi_L)-c]^2
\leq\int d\psi_L\,[E_A(\psi_L)-e_0]^2
\leq\eta_A^2,
\end{eqnarray}
which proves Eq.~(\ref{eq:S6_eta_ED_bound}).

For the nonconverse, use Eq.~(\ref{eq:S1_identity_encoding_counterexample}). There $\Delta_A=0$, whereas $\mathcal{N}_A$ is the identity channel up to an isometric embedding. For two orthogonal logical states, the triangle inequality implies $\eta_A\geq1$ in the present, unhalved diamond-norm convention. Thus vanishing logical-ED is compatible with order-one decoupling error and complete loss of recoverability after erasing $A$.
\end{proof}

Standard information--disturbance continuity results relate small $\eta_A$ to small optimal recovery error for erasure of $A$. Proposition~\ref{prop:S6_eta_to_ED} therefore establishes the valid direction
\begin{eqnarray}
\text{approximate correctability}
\ \Longrightarrow\ 
\text{approximate decoupling}
\ \Longrightarrow\ 
\text{small logical-ED},
\label{eq:S6_one_way_chain}
\end{eqnarray}
but logical-ED alone cannot be used in the reverse direction to certify approximate correctability. Consequently, the unrestricted $d_\Delta(\varepsilon)$ is a tolerance-dependent ED-uniformity scale, not an operationally certified recovery radius. The retained logical unitary certificate in Supplementary Note~S7 gives a converse at exactly zero deviation under an exact implementation condition. A quantitative estimate for imperfect implementations would require additional bounds.

\subsection{Multi-logical systems, restricted ensembles, and symmetry sectors}

For $k>1$ logical qudits, our definitions apply with $\mathcal{H}_L\cong(\mathbb{C}^q)^{\otimes k}$ and Haar measure on a $q^k$-dimensional space. Logical-ED then probes uniformity of region entanglement over the full logical state manifold. In some applications it may be natural to restrict the logical ensemble:
for example, to a symmetry sector, to stabilizer (Clifford) states, or to states preparable within a given circuit depth.
Any such restriction defines a \emph{conditional} logical-ED that quantifies uniformity only over the chosen family. While such conditional diagnostics are not invariants of the full code, they may be the appropriate objects in hardware- or protocol-specific analyses.

\section{\texorpdfstring{Retained logical unitaries and operational erasure certificates}{Retained logical unitaries and operational erasure certificates}}
\label{sec:S7_protected_axis}

This note first establishes an exact erasure certificate for arbitrary finite-dimensional subspace encodings. We then use the one-logical-qubit stabilizer structure to obtain a finite tolerance gap, explicit recovery fidelities, and surface-code examples. Throughout, $B=A^c$ is the surviving subsystem for a known erasure of $A$, and logical-ED is defined over Haar-random pure logical states.

\subsection{\texorpdfstring{Exact certificate for subspace encodings}{Exact certificate for subspace encodings}}
\begingroup
\begin{theorem}[Retained logical unitary]
\label{thm:S7_retained_unitary}
Let $\hat V:\mathcal H_L\longrightarrow\mathcal H_A\otimes\mathcal H_B$ be a finite-dimensional encoding isometry. Suppose there exist a unitary $\hat U_B$ on $B$ and a nonscalar unitary $\hat U_L$ on the logical space such that
\begin{eqnarray}
(\hat{\mathds{1}}_A\otimes \hat U_B)\hat V=\hat V\hat U_L.
\label{eq:S7_retained_unitary}
\end{eqnarray}
Then $\Delta_A=0$ if and only if erasure of $A$ is exactly correctable.
\end{theorem}

\begin{proof}
Define the channel to the erased subsystem by
$\mathcal N_{A\leftarrow L}(\hat X)=\tr_B(\hat V\hat X\hat V^\dagger)$.
Invariance of the partial trace under $\hat U_B$ gives
\begin{eqnarray}
\mathcal N_{A\leftarrow L}(\hat U_L\hat X\hat U_L^\dagger)
=\mathcal N_{A\leftarrow L}(\hat X)
\qquad\text{for every operator }\hat X.
\label{eq:S7_erased_channel_invariance}
\end{eqnarray}
Let $\hat U_L=\sum_j e^{i\phi_j}\hat Q_j$ be its spectral resolution, with distinct eigenvalues and nonzero projectors $\hat Q_j$. There are at least two projectors because $\hat U_L$ is nonscalar. If $j\ne k$, the input $\hat Q_j\hat X\hat Q_k$ is multiplied by $e^{i(\phi_j-\phi_k)}\ne 1$ under conjugation by $\hat U_L$. Thus
$\mathcal N_{A\leftarrow L}(\hat Q_j\hat X\hat Q_k)=0$ for $j\ne k$.

Suppose $\Delta_A=0$. The output purity $P_A(\psi)=\tr(\hat\rho_A(\psi)^2)$ is continuous on the pure logical states. Haar measure has full support, so a zero variance makes $P_A(\psi)$ constant on that set. For unit vectors $\ket u,\ket v$ in distinct eigenspaces, put $\ket{+_{uv}}=(\ket u+\ket v)/\sqrt 2$. Since cross terms vanish under $\mathcal N_{A\leftarrow L}$, its output on this superposition is $(\hat\rho_A(u)+\hat\rho_A(v))/2$. Expanding the square gives
\begin{eqnarray}
\frac{P_A(u)+P_A(v)}{2}-P_A(+_{uv})
=\frac14\tr\bigl[(\hat\rho_A(u)-\hat\rho_A(v))^2\bigr].
\label{eq:S7_three_purity_identity}
\end{eqnarray}
The left side vanishes by constant purity, and the right side is one quarter of the squared Hilbert--Schmidt norm. Hence $\hat\rho_A(u)=\hat\rho_A(v)$. Comparing any two pure states in the same eigenspace with a fixed state in another eigenspace shows that every unit vector in every eigenspace has the same output $\hat\omega_A$. Any pure logical state can be written as a weighted superposition of unit vectors from these eigenspaces. Its output is the weighted mixture of their identical outputs, since the cross terms vanish. By linearity, every logical density operator therefore has output $\hat\omega_A$.

For any operator $\hat O_A$, constancy of this channel gives
$\hat V^\dagger(\hat O_A\otimes\hat{\mathds{1}}_B)\hat V
=\tr(\hat\omega_A\hat O_A)\hat{\mathds{1}}_L$.
Taking $\hat O_A$ to be each matrix unit on $A$ gives the Knill--Laflamme conditions for erasure of $A$~\cite{knill1997theory,nielsen_chuang}. Conversely, exact correctability gives a constant reduced state on $A$ by Theorem~\ref{thm:S2_main}, so $\Delta_A=0$.
\end{proof}

For a loss family, the condition is pointwise. Both unitaries may vary with $A$. A single physical representative with support disjoint from every candidate region is one way to meet it. In particular, a physical Pauli supported on $B$ that preserves the code and induces a nonidentity logical Pauli satisfies Eq.~(\ref{eq:S7_retained_unitary}). The exact spectral argument uses the code isometry and exact logical implementation. A small observed deviation alone does not furnish a finite-shot or approximate-recovery guarantee.

\smallskip\noindent
\textbf{Three-purity test for a logical qubit.}
If $\dim\mathcal H_L=2$, the two eigenspaces of $\hat U_L$ are one-dimensional. Choose their unit eigenvectors $\ket{u}$ and $\ket{v}$. The single gap
$G_A=[P_A(u)+P_A(v)]/2-P_A(+_{uv})$
in Eq.~(\ref{eq:S7_three_purity_identity}) is nonnegative. It vanishes precisely when the two eigenstate outputs coincide, which by Eq.~(\ref{eq:S7_erased_channel_invariance}) is equivalent to a constant erased-subsystem channel. Thus three exact purity values suffice to test exact erasure correctability under the retained-unitary condition.

\smallskip\noindent
\textbf{Proof of the three-purity decoupling bound.}
Write $\hat D_A=\hat\rho_A(u)-\hat\rho_A(v)$ and let $\hat\omega_A=[\hat\rho_A(u)+\hat\rho_A(v)]/2$. For any constant output $\hat\tau_A$, testing the channel difference on the two eigenstates and applying the triangle inequality gives
\begin{eqnarray}
\norm{\mathcal N_A-\mathcal C_{\hat\tau_A}}_\diamond
\ge\max\{\norm{\hat\rho_A(u)-\hat\tau_A}_1,\norm{\hat\rho_A(v)-\hat\tau_A}_1\}
\ge\tfrac12\norm{\hat D_A}_1 .
\label{eq:S7_three_purity_lower}
\end{eqnarray}
For the converse bound, take any operator $\hat X$ on $L\otimes R$, with $R$ an arbitrary reference, and write its diagonal blocks on $R$ in the $(u,v)$ basis as $\hat X_{uu}$ and $\hat X_{vv}$. Equation~(\ref{eq:S7_erased_channel_invariance}) makes the erased-subsystem channel insensitive to the off-diagonal blocks. Consequently
\begin{eqnarray}
\bigl[(\mathcal N_A-\mathcal C_{\hat\omega_A})\otimes{\rm id}_R\bigr](\hat X)
=\tfrac12\hat D_A\otimes(\hat X_{uu}-\hat X_{vv}).
\label{eq:S7_three_purity_extension}
\end{eqnarray}
The trace norm of the block-diagonal part of $\hat X$ is $\norm{\hat X_{uu}}_1+\norm{\hat X_{vv}}_1\le\norm{\hat X}_1$. This part is an average of $\hat X$ and its conjugate by the unitary that changes the sign of the $v$ component on $L$. The triangle inequality and multiplicativity of the trace norm on tensor products bound the diamond norm in Eq.~(\ref{eq:S7_three_purity_extension}) by $\norm{\hat D_A}_1/2$. The midpoint therefore attains the infimum in Eq.~(\ref{eq:S6_eta_def}). Cauchy--Schwarz on the nonzero singular values of $\hat D_A$, together with Eq.~(\ref{eq:S7_three_purity_identity}), gives
\begin{eqnarray}
\eta_A=\tfrac12\norm{\hat D_A}_1\le\sqrt{\operatorname{rank}(\hat D_A)\,G_A}.
\label{eq:S7_three_purity_bound}
\end{eqnarray}
This bound assumes exact implementation of the retained logical unitary. For finite-shot purity estimates it applies using an upper confidence bound on $G_A$ and an independently justified upper bound on $\operatorname{rank}(\hat D_A)$.

\smallskip\noindent
\textbf{Analytic example beyond stabilizer codes.}
Consider two physical qubits and the encoding
\begin{eqnarray}
\hat V_\theta\ket{0_L}
&=&\ket 0_A\ket 0_B,\nonumber\\
\hat V_\theta\ket{1_L}
&=&(\cos\theta\ket 0_A+\sin\theta\ket 1_A)\ket 1_B,
\qquad 0\le\theta\le\frac{\pi}{2}.
\label{eq:S7_nonstabilizer_encoding}
\end{eqnarray}
A physical $\hat Z_B$ implements logical $\hat Z_L$ for every $\theta$. For $0<\theta<\pi/2$ the two $A$-side factors are neither collinear nor orthogonal. A Pauli stabilizing both product codewords would require them to be eigenvectors of its $A$ factor, forcing that factor to be the identity. The same eigenvalue must then hold for both orthogonal $B$-side factors, forcing the $B$ factor to be the identity as well. Thus the code is nonstabilizer in this interval.

The outputs on $A$ of the logical eigenstates are pure, and their squared overlap is $\cos^2\theta$. Equation~(\ref{eq:S7_three_purity_identity}) gives $G_A=\sin^2\theta/2$. For a general pure logical input with probability $p$ on $\ket{0_L}$, the linear entropy is $E_A=2p(1-p)\sin^2\theta$. Under the logical-qubit Haar measure, $z=2p-1$ is uniform on $[-1,1]$, with $\mathrm{Var}(z^2)=4/45$. This yields
\begin{eqnarray}
\Delta_A=\frac{\sin^2\theta}{\sqrt{45}}.
\label{eq:S7_nonstabilizer_delta}
\end{eqnarray}
The case $\theta=0$ is exactly correctable because both eigenstates have the same erased-subsystem output. All $\theta>0$ give a strictly positive gap and are uncorrectable, despite the surviving logical $\hat Z$ operation.
\endgroup

\subsection{Complementary logical access and the retained-axis condition}

For the stabilizer results below, specialize to an exact $[[n,1]]$ subspace stabilizer code. The logical-access calculation distinguishes the two ED-inactive sectors and quantifies the one-axis sector.

Work modulo Pauli phases. Let $\mathsf{E}=\mathbb{F}_2^{2n}$ be the physical symplectic space, $\mathsf{S}$ its isotropic stabilizer subspace of dimension $m=n-1$, and $\mathsf{L}=\mathsf{S}^{\perp}/\mathsf{S}$ the two-dimensional logical symplectic space. Write $\mathsf{E}_A$ for vectors supported in $A$, $t_A=\dim(\mathsf{S}\cap\mathsf{E}_A)$, and $\mathsf{L}(A)$ for the image of $\mathsf{S}^{\perp}\cap\mathsf{E}_A$ in $\mathsf{L}$. Thus $\kappa_A=\dim\mathsf{L}(A)$ and $q_A=2^{\kappa_A}-1$.

\begin{lemma}[Complementary logical-access identity]
\label{lem:S7_complementary_access}
For every bipartition $A|B$,
\begin{eqnarray}
\kappa_A+\kappa_B=2,
\quad
\mathsf{L}(B)=\mathsf{L}(A)^{\perp}.
\label{eq:S7_complementary_access}
\end{eqnarray}
In particular, if $\kappa_A=1$, both regions support the same logical Pauli axis.
\end{lemma}

\begin{proof}---Let $\pi_A$ be coordinate restriction to $A$. Its kernel on $\mathsf{S}$ is $\mathsf{S}\cap\mathsf{E}_B$, so $\dim\pi_A\mathsf{S}=m-t_B$. Commutation with a vector supported in $A$ depends only on $\pi_A\mathsf{S}$. Nondegeneracy of the symplectic form on $\mathsf{E}_A$ therefore gives
\begin{eqnarray}
\kappa_A
&=&\dim(\mathsf{S}^{\perp}\cap\mathsf{E}_A)-t_A
=2\abs{A}-m+t_B-t_A.
\label{eq:S7_kappa_rank}
\end{eqnarray}
Exchanging $A,B$ and adding proves the first identity. Disjointly supported representatives commute, giving $\mathsf{L}(B)\subseteq\mathsf{L}(A)^{\perp}$; their equal dimensions make this inclusion an equality. A line in a two-dimensional symplectic space equals its own orthogonal complement.
\end{proof}

The dimension identity follows from the supported-subgroup framework used in Note~S5. Its role here is to distinguish the two ED-inactive sectors, rather than introduce a new complement duality.

\begin{theorem}[Exact certificate with a retained logical axis]
\label{thm:S7_protected_axis}
Let $\mathfrak{F}$ be a family of physical regions. Suppose a fixed physical representative $\hat{\overline{P}}$ of a nontrivial logical Pauli satisfies
\begin{eqnarray}
\operatorname{supp}(\hat{\overline{P}})\subseteq A^c
\quad\text{for every }A\in\mathfrak{F}.
\label{eq:S7_fixed_representative}
\end{eqnarray}
Then $\mathsf{L}(A)\subseteq\operatorname{span}([\hat{\overline{P}}])$, so $q_A\in\{0,1\}$, and
\begin{eqnarray}
\Delta_A=0
\quad\Longleftrightarrow\quad
A\text{ is exactly erasure-correctable}
\quad(A\in\mathfrak{F}).
\label{eq:S7_ED_correctability}
\end{eqnarray}
The same equivalence holds under the weaker pointwise condition $\kappa_{A^c}>0$ for every region in the family; the retained class and its physical representative may then vary with $A$.
\end{theorem}

\begin{proof}---Every logical representative supported in $A$ commutes with $\hat{\overline{P}}$ by disjoint support. The centralizer of a nonzero logical class in $\mathsf{L}\cong\mathbb{F}_2^2$ is its one-dimensional span. Thus the full-access sector $q_A=3$ is excluded. The trichotomy of Note~S5 then gives $\Delta_A=0$ precisely in the correctable sector $q_A=0$. More generally, Lemma~\ref{lem:S7_complementary_access} shows that $\kappa_{A^c}>0$ is equivalent to $\kappa_A<2$, which excludes exactly the same blind sector.
\end{proof}

The weaker condition is necessary and sufficient to exclude $q_A=3$ within this setting. A common physical representative is a convenient sufficient certificate that can be checked from geometry before evaluating ED. It does not assume that $A$ is correctable: a region with $q_A=1$ still satisfies the condition. A common logical class with region-dependent representatives is also sufficient. When ED is computed by the exact coset algorithm, $q_A$ is already available; the theorem supplies an interpretation, rather than a computational advantage over that classification.

\subsection{Exact family radii and finite tolerance gaps}

Let $\mathfrak{A}(\ell)$ be nested cumulative families on the finite ordered scale set $\mathcal{L}$, with $\mathfrak{A}(\ell_0)=\{\emptyset\}$. Assume the condition of Theorem~\ref{thm:S7_retained_unitary} holds for every $A\in\mathfrak{F}=\bigcup_{\ell\in\mathcal{L}}\mathfrak{A}(\ell)$, and define
\begin{eqnarray}
\ell_{\rm er}
&:=&\max\{\ell\in\mathcal{L}:\text{ every }A\in\mathfrak{A}(\ell)
\text{ is erasure-correctable}\}.
\label{eq:S7_erasure_radius}
\end{eqnarray}
\begin{corollary}[Operational meaning of the zero-tolerance family radius]
\label{cor:S7_protected_radius}
Under these assumptions, $\ell_\Delta(0)=\ell_{\rm er}$.
\end{corollary}
\begin{proof}---Theorem~\ref{thm:S7_retained_unitary} makes the feasible scales in the two maximizations identical.
\end{proof}

For a finite family meeting the exact implementation condition, the minimum of the strictly positive deviations is a gap when at least one region has positive deviation. In the one-logical-qubit stabilizer case its value takes the explicit form
\begin{eqnarray}
\gamma_{\mathfrak{F}}
&:=&\min_{A\in\mathfrak{F}:q_A=1}
\frac{2^{t_A-\abs{A}+1}}{\sqrt{45}},
\label{eq:S7_raw_gap}
\end{eqnarray}
with $\gamma_{\mathfrak{F}}=+\infty$ if no active region exists. Then $\ell_\Delta(\varepsilon)=\ell_{\rm er}$ for $0\le\varepsilon<\gamma_{\mathfrak{F}}$. The inequality must be strict because the ED-radius definition accepts equality. If $\abs{A}\le s$ throughout the family, $t_A\ge0$ gives the sufficient bound $\gamma_{\mathfrak{F}}\ge2^{1-s}/\sqrt{45}$.

A scale-independent separation follows by normalizing with mean purity:
\begin{eqnarray}
\delta_A&:=&\frac{\Delta_A}{\overline{P}_A},
\quad \overline{P}_A:=1-\overline{E}_A>0,\\
\delta_A&=&\begin{cases}
1/\sqrt{20},&q_A=1,\\
0,&q_A=0\text{ or }3.
\end{cases}
\label{eq:S7_normalized_gap}
\end{eqnarray}
Indeed, $\overline{P}_A=2^{t_A-\abs{A}}(1+q_A/3)$ by Note~S5. Within an eligible family, any normalized tolerance below $1/\sqrt{20}$ yields the same exact classification and radius. These are discrete-gap statements for exact stabilizer encodings; they do not certify approximate correction for perturbed encoders from small measured ED alone. Nor do they identify the unrestricted size radius $d_\Delta(0)$ with $d-1$ without an ED-active minimum logical support.

\subsection{Optimal recovery fidelity and the one-axis channel}

Define $\mathcal{N}_B(\hat{\rho})=\tr_A(\hat{V}\hat{\rho}\hat{V}^\dagger)$ and let recovery have logical output, $\mathcal{R}:B\to L$. We use the normalized Bell state $\ket{\Phi}_{RL}=(\ket{00}+\ket{11})/\sqrt{2}$ and the conventional entanglement-fidelity overlap~\cite{schumacher1996sending},
\begin{eqnarray}
F_e^\star(A):=\max_{\mathcal{R}\ \mathrm{CPTP}}
\bra{\Phi}\bigl(\operatorname{id}_R\otimes(\mathcal{R}\circ\mathcal{N}_B)\bigr)
(\ketbra{\Phi}{\Phi})\ket{\Phi}.
\label{eq:S7_optimal_entanglement_fidelity}
\end{eqnarray}
Thus the input density operator is $\hat{\mathds{1}}_L/2$ and no square root is taken.

\begin{proposition}[Exact recovery-fidelity trichotomy]
\label{prop:S7_optimal_fidelity}
For known erasure of $A$,
\begin{eqnarray}
F_e^\star(A)=2^{-\kappa_A}
=\begin{cases}
1,&q_A=0,\\
1/2,&q_A=1,\\
1/4,&q_A=3.
\end{cases}
\label{eq:S7_fidelity_trichotomy}
\end{eqnarray}
\end{proposition}

\begin{proof}---For $q_A=0$, exact erasure recovery gives fidelity one. For $q_A=3$, Lemma~\ref{lem:S7_complementary_access} gives $q_B=0$, hence $\mathcal{N}_B$ is constant. After any recovery the reference-output state is $\hat{\mathds{1}}_R/2\otimes\hat{\tau}_L$, whose Bell overlap is $1/4$.

For $q_A=1$, both sides support the same class. Choose a Hermitian representative $\hat{P}_B$ in $B$, let $\hat{Q}=\hat{V}^\dagger(\hat{\mathds{1}}_A\otimes\hat{P}_B)\hat{V}$, and let $\mathcal{S}_B$ be the physical stabilizer subgroup supported entirely in $B$. Put
\begin{eqnarray}
\hat{\Pi}_B:=2^{-t_B}\sum_{\hat{g}\in\mathcal{S}_B}\hat{g}_B,
\quad D_B:=\tr{\hat{\Pi}_B}=2^{\abs{B}-t_B}.
\label{eq:S7_survivor_projector}
\end{eqnarray}
Partial tracing the stabilizer expansion gives
\begin{eqnarray}
\mathcal{N}_B(\hat{\rho})
&=&D_B^{-1}\hat{\Pi}_B
\left[\hat{\mathds{1}}_B+\tr{(\hat{\rho}\hat{Q})}\hat{P}_B\right]\\
&=&\sum_{\lambda=\pm1}\tr{(\hat{\rho}\hat{\Pi}^L_\lambda)}\hat{\sigma}_\lambda,
\label{eq:S7_one_axis_channel}\\
\hat{\Pi}^L_\lambda&:=&(\hat{\mathds{1}}_L+\lambda\hat{Q})/2,
\quad
\hat{\sigma}_\lambda:=\hat{\Pi}_B(\hat{\mathds{1}}_B+\lambda\hat{P}_B)/D_B.
\nonumber
\end{eqnarray}
Here $\tr{(\hat{\Pi}_B\hat{P}_B)}=0$ because $\hat{P}_B$ is nontrivial modulo $\mathcal{S}_B$. Consequently $\hat{\sigma}_\pm$ are normalized states on orthogonal eigenspaces. The channel measures the logical axis and prepares perfectly distinguishable physical states. Its reference-output state is separable, as is the state after any recovery; its Bell overlap is at most $1/2$. Measuring these orthogonal supports and preparing the corresponding logical eigenstates attains complete $\hat{Q}$-dephasing and fidelity $1/2$.
\end{proof}

The channel statement is basis independent. For arbitrary $\hat{V}=\hat{V}_0\hat{U}_L$, physical logical Paulis induce $\hat{U}_L^\dagger\hat{\sigma}_\alpha\hat{U}_L$, which need not be coordinate Paulis when $\hat{U}_L$ is non-Clifford. Bloch components in the coset formulas refer to this induced frame; the one-axis channel dephases in $\hat{Q}$'s eigenbasis. Haar ED and Eq.~(\ref{eq:S7_optimal_entanglement_fidelity}) are unchanged by this reparametrization. The fidelities in Eq.~(\ref{eq:S7_fidelity_trichotomy}) are not average pure-input fidelities, whose corresponding values are $1$, $2/3$, and $1/2$.

Combining the two trichotomies gives the particularly simple relation
\begin{eqnarray}
F_e^\star(A)=1-\sqrt{5}\,\delta_A
\quad\text{if }\kappa_{A^c}>0.
\label{eq:S7_normalized_operational_relation}
\end{eqnarray}
Without eligibility, the $q_A=3$ sector invalidates this relation. Raw ED can become exponentially small while a one-axis erasure retains $F_e^\star=1/2$; its magnitude alone therefore does not quantify recovery quality.

\subsection{Exact surface-code patterns and flagged erasure ensembles}

For the $[[121,1,11]]$ rotated patch, let $A_{\ell,h}=\{(x,y):0\le x<\ell,\ 0\le y<h\}$. The four exact rank calculations in Table~\ref{tab:S7_patterns} illustrate both the certificate and its geometry dependence. Here $G_B$ is the stabilizer matrix restricted to $B$, and $t_A=120-\operatorname{rank}_{\mathbb{F}_2}G_B$.

\begin{table}[t]
\centering
\setlength{\tabcolsep}{15pt}
\begin{tabular}{c c c c c c c c c}
\hline\hline
$(\ell,h)$ & $\abs{A}$ & $\operatorname{rank}G_B$ & $t_A$ & $t_B$ & $(\kappa_A,\kappa_B)$ & $(q_A,q_B)$ & $\Delta_A$ & $F_e^\star$\\
\hline
$(10,3)$  & $30$  & $97$  & $23$ & $83$  & $(0,2)$ & $(0,3)$ & $0$ & $1$\\
$(11,3)$  & $33$  & $93$  & $27$ & $82$  & $(1,1)$ & $(1,1)$ & $1/(32\sqrt{45})$ & $1/2$\\
$(10,10)$ & $100$ & $30$  & $90$ & $10$  & $(0,2)$ & $(0,3)$ & $0$ & $1$\\
$(11,1)$  & $11$  & $115$ & $5$  & $104$ & $(1,1)$ & $(1,1)$ & $1/(32\sqrt{45})$ & $1/2$\\
\hline\hline
\end{tabular}
\caption{Exact logical-access and recovery data for four rectangles in the $[[121,1,11]]$ patch.}
\label{tab:S7_patterns}
\end{table}

The horizontal logical $\hat{\overline{Z}}$ row at $y=10$ lies outside all four regions. For $A_{10,10}$, the vertical $\hat{\overline{X}}$ column at $x=10$ also survives, explaining exact recovery despite erasure of $100$ sites. For the nested width-$3$ corridor family of Note~S3, the same fixed $\hat{\overline{Z}}$ row survives every region. The length radius is therefore operational: $\ell_{\rm er}=\ell_\Delta(0)=10$, with fidelity one through length $10$ and $1/2$ at length $11$. Its raw active gap is $1/(32\sqrt{45})$.

For a classical flag identifying each erased mask, recovery may depend on $A$. Write $p_j=\Pr(q_A=j)$. Equation~(\ref{eq:S7_fidelity_trichotomy}) gives
\begin{eqnarray}
\mathbb{E}[F_e^\star(A)]=1-\frac{p_1}{2}-\frac{3p_3}{4}.
\label{eq:S7_flagged_mean_fidelity}
\end{eqnarray}
In the common-line model, $K\sim\operatorname{Bin}(d,p)$ erased lines give $q_A=0,1,3$ for $K=0$, $0<K<d$, and $K=d$, respectively. Exhaustive checks of all $2048$ line subsets in each orientation confirm this classification for $d=11$. Thus
\begin{eqnarray}
\mathbb{E}_{\rm line}[F_e^\star]
=\frac12+\frac12(1-p)^d-\frac14p^d.
\label{eq:S7_line_mean_fidelity}
\end{eqnarray}
The full-erasure event $p^d$ remains ED-inactive and must be included separately. At $d=11,p=0.01$, Eq.~(\ref{eq:S7_line_mean_fidelity}) is approximately $0.947669127129$. For independent erasures, all masks of size at most $10$ are correctable and every other mask has $F_e^\star\ge1/4$. Consequently
\begin{eqnarray}
\mathbb{E}_{\rm ind}[F_e^\star]
&\ge&1-\frac34\Pr[\operatorname{Bin}(121,0.01)\ge11]
\ge0.999999965097,
\label{eq:S7_independent_fidelity_bound}
\end{eqnarray}
using the binomial tail $4.653729906866\times10^{-8}$. This is a rigorous conservative lower bound, not an estimated independent-erasure fidelity.

These quantities describe optimal recovery for ideal static erasures with known masks. They neither specify a practical decoder nor include faulty encoding, syndrome-extraction circuits, repeated rounds, measurement faults, or temporal correlations. The retained-unitary condition identifies fault families for which the zero-ED profile certifies exact protection. In the one-axis sector of the stabilizer model, the surviving subsystem preserves a classical logical observable and the optimal entanglement fidelity is $1/2$. Extensions to mixed encodings, subsystem gauge ensembles, and noisy circuits require separate operational analysis.

\bibliographystyle{naturemag}

\begin{thebibliography}{}
\expandafter\ifx\csname url\endcsname\relax
  \def\url#1{\texttt{#1}}\fi
\expandafter\ifx\csname urlprefix\endcsname\relax\def\urlprefix{URL }\fi
\providecommand{\bibinfo}[2]{#2}
\providecommand{\eprint}[2][]{\url{#2}}

\end{thebibliography}


\begin{thebibliography}{10}
\expandafter\ifx\csname url\endcsname\relax
  \def\url#1{\texttt{#1}}\fi
\expandafter\ifx\csname urlprefix\endcsname\relax\def\urlprefix{URL }\fi
\providecommand{\bibinfo}[2]{#2}
\providecommand{\eprint}[2][]{\url{#2}}

\bibitem{shor1995scheme}
\bibinfo{author}{Shor, P.~W.}
\newblock \bibinfo{title}{Scheme for reducing decoherence in quantum computer
  memory}.
\newblock \emph{\bibinfo{journal}{Physical Review A}}
  \textbf{\bibinfo{volume}{52}}, \bibinfo{pages}{R2493--R2496}
  (\bibinfo{year}{1995}).

\bibitem{steane1996error}
\bibinfo{author}{Steane, A.~M.}
\newblock \bibinfo{title}{Error correcting codes in quantum theory}.
\newblock \emph{\bibinfo{journal}{Physical Review Letters}}
  \textbf{\bibinfo{volume}{77}}, \bibinfo{pages}{793--797}
  (\bibinfo{year}{1996}).

\bibitem{knill1997theory}
\bibinfo{author}{Knill, E.} \& \bibinfo{author}{Laflamme, R.}
\newblock \bibinfo{title}{Theory of quantum error-correcting codes}.
\newblock \emph{\bibinfo{journal}{Physical Review A}}
  \textbf{\bibinfo{volume}{55}}, \bibinfo{pages}{900--911}
  (\bibinfo{year}{1997}).

\bibitem{kitaev2003fault}
\bibinfo{author}{Kitaev, A.~Y.}
\newblock \bibinfo{title}{Fault-tolerant quantum computation by anyons}.
\newblock \emph{\bibinfo{journal}{Annals of Physics}}
  \textbf{\bibinfo{volume}{303}}, \bibinfo{pages}{2--30}
  (\bibinfo{year}{2003}).

\bibitem{dennis2002topological}
\bibinfo{author}{Dennis, E.}, \bibinfo{author}{Kitaev, A.},
  \bibinfo{author}{Landahl, A.} \& \bibinfo{author}{Preskill, J.}
\newblock \bibinfo{title}{Topological quantum memory}.
\newblock \emph{\bibinfo{journal}{Journal of Mathematical Physics}}
  \textbf{\bibinfo{volume}{43}}, \bibinfo{pages}{4452--4505}
  (\bibinfo{year}{2002}).

\bibitem{fowler2012surface}
\bibinfo{author}{Fowler, A.~G.}, \bibinfo{author}{Mariantoni, M.},
  \bibinfo{author}{Martinis, J.~M.} \& \bibinfo{author}{Cleland, A.~N.}
\newblock \bibinfo{title}{Surface codes: Towards practical large-scale quantum
  computation}.
\newblock \emph{\bibinfo{journal}{Physical Review A}}
  \textbf{\bibinfo{volume}{86}}, \bibinfo{pages}{032324}
  (\bibinfo{year}{2012}).

\bibitem{terhal2015quantum}
\bibinfo{author}{Terhal, B.~M.}
\newblock \bibinfo{title}{Quantum error correction for quantum memories}.
\newblock \emph{\bibinfo{journal}{Reviews of Modern Physics}}
  \textbf{\bibinfo{volume}{87}}, \bibinfo{pages}{307--346}
  (\bibinfo{year}{2015}).

\bibitem{google2025below}
\bibinfo{author}{{Google Quantum AI and Collaborators}}.
\newblock \bibinfo{title}{Quantum error correction below the surface code
  threshold}.
\newblock \emph{\bibinfo{journal}{Nature}} \textbf{\bibinfo{volume}{638}},
  \bibinfo{pages}{920--926} (\bibinfo{year}{2025}).

\bibitem{harrington2025synchronous}
\bibinfo{author}{Harrington, P.~M.} \emph{et~al.}
\newblock \bibinfo{title}{Synchronous detection of cosmic rays and correlated
  errors in superconducting qubit arrays}.
\newblock \emph{\bibinfo{journal}{Nature Communications}}
  \textbf{\bibinfo{volume}{16}}, \bibinfo{pages}{6428} (\bibinfo{year}{2025}).

\bibitem{cleve1999share}
\bibinfo{author}{Cleve, R.}, \bibinfo{author}{Gottesman, D.} \&
  \bibinfo{author}{Lo, H.-K.}
\newblock \bibinfo{title}{How to share a quantum secret}.
\newblock \emph{\bibinfo{journal}{Physical Review Letters}}
  \textbf{\bibinfo{volume}{83}}, \bibinfo{pages}{648--651}
  (\bibinfo{year}{1999}).

\bibitem{ekert2002direct}
\bibinfo{author}{Ekert, A.~K.} \emph{et~al.}
\newblock \bibinfo{title}{Direct estimations of linear and nonlinear
  functionals of a quantum state}.
\newblock \emph{\bibinfo{journal}{Physical Review Letters}}
  \textbf{\bibinfo{volume}{88}}, \bibinfo{pages}{217901}
  (\bibinfo{year}{2002}).

\bibitem{elben2023toolbox}
\bibinfo{author}{Elben, A.} \emph{et~al.}
\newblock \bibinfo{title}{The randomized measurement toolbox}.
\newblock \emph{\bibinfo{journal}{Nature Reviews Physics}}
  \textbf{\bibinfo{volume}{5}}, \bibinfo{pages}{9--24} (\bibinfo{year}{2023}).

\bibitem{kretschmann2008information}
\bibinfo{author}{Kretschmann, D.}, \bibinfo{author}{Schlingemann, D.} \&
  \bibinfo{author}{Werner, R.~F.}
\newblock \bibinfo{title}{The information-disturbance tradeoff and the
  continuity of {Stinespring}'s representation}.
\newblock \emph{\bibinfo{journal}{IEEE Transactions on Information Theory}}
  \textbf{\bibinfo{volume}{54}}, \bibinfo{pages}{1708--1717}
  (\bibinfo{year}{2008}).

\bibitem{beny2010general}
\bibinfo{author}{B{\'e}ny, C.} \& \bibinfo{author}{Oreshkov, O.}
\newblock \bibinfo{title}{General conditions for approximate quantum error
  correction and near-optimal recovery channels}.
\newblock \emph{\bibinfo{journal}{Physical Review Letters}}
  \textbf{\bibinfo{volume}{104}}, \bibinfo{pages}{120501}
  (\bibinfo{year}{2010}).

\bibitem{gottesman1997stabilizer}
\bibinfo{author}{Gottesman, D.}
\newblock \emph{\bibinfo{title}{Stabilizer Codes and Quantum Error
  Correction}}.
\newblock Ph.D. thesis, \bibinfo{school}{California Institute of Technology}
  (\bibinfo{year}{1997}).

\bibitem{fattal2004entanglement}
\bibinfo{author}{Fattal, D.}, \bibinfo{author}{Cubitt, T.~S.},
  \bibinfo{author}{Yamamoto, Y.}, \bibinfo{author}{Bravyi, S.} \&
  \bibinfo{author}{Chuang, I.~L.}
\newblock \bibinfo{title}{Entanglement in the stabilizer formalism}
  (\bibinfo{year}{2004}).
\newblock \eprint{quant-ph/0406168}.

\bibitem{zhao2026graph}
\bibinfo{author}{Zhao, W.}, \bibinfo{author}{Fang, M.} \& \bibinfo{author}{Su,
  D.}
\newblock \bibinfo{title}{Graph-based approach to entanglement entropy of
  quantum error-correcting codes}.
\newblock \emph{\bibinfo{journal}{Physical Review A}}
  \textbf{\bibinfo{volume}{113}}, \bibinfo{pages}{052412}
  (\bibinfo{year}{2026}).

\bibitem{bravyi2009nogo}
\bibinfo{author}{Bravyi, S.} \& \bibinfo{author}{Terhal, B.~M.}
\newblock \bibinfo{title}{A no-go theorem for a two-dimensional self-correcting
  quantum memory based on stabilizer codes}.
\newblock \emph{\bibinfo{journal}{New Journal of Physics}}
  \textbf{\bibinfo{volume}{11}}, \bibinfo{pages}{043029}
  (\bibinfo{year}{2009}).

\bibitem{zhang2026logical}
\bibinfo{author}{Zhang, B.} \emph{et~al.}
\newblock \bibinfo{title}{Logical qubits with erasure conversion using
  metastable neutral atoms}.
\newblock \emph{\bibinfo{journal}{Nature Physics}}
  \textbf{\bibinfo{volume}{22}}, \bibinfo{pages}{910--916}
  (\bibinfo{year}{2026}).

\bibitem{bonillaataides2021xzzx}
\bibinfo{author}{Bonilla~Ataides, J.~P.}, \bibinfo{author}{Tuckett, D.~K.},
  \bibinfo{author}{Bartlett, S.~D.}, \bibinfo{author}{Flammia, S.~T.} \&
  \bibinfo{author}{Brown, B.~J.}
\newblock \bibinfo{title}{The {XZZX} surface code}.
\newblock \emph{\bibinfo{journal}{Nature Communications}}
  \textbf{\bibinfo{volume}{12}}, \bibinfo{pages}{2172} (\bibinfo{year}{2021}).

\bibitem{bombin2006topological}
\bibinfo{author}{Bombin, H.} \& \bibinfo{author}{Martin-Delgado, M.~A.}
\newblock \bibinfo{title}{Topological quantum distillation}.
\newblock \emph{\bibinfo{journal}{Physical Review Letters}}
  \textbf{\bibinfo{volume}{97}}, \bibinfo{pages}{180501}
  (\bibinfo{year}{2006}).

\bibitem{mele2024haar}
\bibinfo{author}{Mele, A.~A.}
\newblock \bibinfo{title}{Introduction to {Haar} measure tools in quantum
  information: A beginner's tutorial}.
\newblock \emph{\bibinfo{journal}{Quantum}} \textbf{\bibinfo{volume}{8}},
  \bibinfo{pages}{1340} (\bibinfo{year}{2024}).

\bibitem{cho2025EPD}
\bibinfo{author}{Cho, K.} \& \bibinfo{author}{Bang, J.}
\newblock \bibinfo{title}{Entangling power and its deviation: A quantitative
  analysis on input-state dependence and variability in entanglement
  generation}.
\newblock \emph{\bibinfo{journal}{Physical Review A}}
  \textbf{\bibinfo{volume}{113}}, \bibinfo{pages}{012442}
  (\bibinfo{year}{2026}).

\bibitem{hu2025shallow}
\bibinfo{author}{Hu, H.-Y.} \emph{et~al.}
\newblock \bibinfo{title}{Demonstration of robust and efficient quantum
  property learning with shallow shadows}.
\newblock \emph{\bibinfo{journal}{Nature Communications}}
  \textbf{\bibinfo{volume}{16}}, \bibinfo{pages}{2943} (\bibinfo{year}{2025}).

\bibitem{brandao2016local}
\bibinfo{author}{Brand{\~a}o, F. G. S.~L.}, \bibinfo{author}{Harrow, A.~W.} \&
  \bibinfo{author}{Horodecki, M.}
\newblock \bibinfo{title}{Local random quantum circuits are approximate
  polynomial-designs}.
\newblock \emph{\bibinfo{journal}{Communications in Mathematical Physics}}
  \textbf{\bibinfo{volume}{346}}, \bibinfo{pages}{397--434}
  (\bibinfo{year}{2016}).

\end{thebibliography}

\begin{thebibliography}{10}
\expandafter\ifx\csname url\endcsname\relax
  \def\url#1{\texttt{#1}}\fi
\expandafter\ifx\csname urlprefix\endcsname\relax\def\urlprefix{URL }\fi
\providecommand{\bibinfo}[2]{#2}
\providecommand{\eprint}[2][]{\url{#2}}

\bibitem{ekert2002direct}
\bibinfo{author}{Ekert, A.~K.} \emph{et~al.}
\newblock \bibinfo{title}{Direct estimations of linear and nonlinear
  functionals of a quantum state}.
\newblock \emph{\bibinfo{journal}{Physical Review Letters}}
  \textbf{\bibinfo{volume}{88}}, \bibinfo{pages}{217901}
  (\bibinfo{year}{2002}).

\bibitem{elben2019statistical}
\bibinfo{author}{Elben, A.}, \bibinfo{author}{Vermersch, B.},
  \bibinfo{author}{Roos, C.~F.} \& \bibinfo{author}{Zoller, P.}
\newblock \bibinfo{title}{Statistical correlations between locally randomized
  measurements: A toolbox for probing entanglement in many-body quantum
  states}.
\newblock \emph{\bibinfo{journal}{Physical Review A}}
  \textbf{\bibinfo{volume}{99}}, \bibinfo{pages}{052323}
  (\bibinfo{year}{2019}).

\bibitem{cho2025EPD}
\bibinfo{author}{Cho, K.} \& \bibinfo{author}{Bang, J.}
\newblock \bibinfo{title}{Entangling power and its deviation: A quantitative
  analysis on input-state dependence and variability in entanglement
  generation}.
\newblock \emph{\bibinfo{journal}{Physical Review A}}
  \textbf{\bibinfo{volume}{113}}, \bibinfo{pages}{012442}
  (\bibinfo{year}{2026}).

\bibitem{zanardi2000entangling}
\bibinfo{author}{Zanardi, P.}, \bibinfo{author}{Zalka, C.} \&
  \bibinfo{author}{Faoro, L.}
\newblock \bibinfo{title}{Entangling power of quantum evolutions}.
\newblock \emph{\bibinfo{journal}{Physical Review A}}
  \textbf{\bibinfo{volume}{62}}, \bibinfo{pages}{030301}
  (\bibinfo{year}{2000}).

\bibitem{mele2024haar}
\bibinfo{author}{Mele, A.~A.}
\newblock \bibinfo{title}{Introduction to {Haar} measure tools in quantum
  information: A beginner's tutorial}.
\newblock \emph{\bibinfo{journal}{Quantum}} \textbf{\bibinfo{volume}{8}},
  \bibinfo{pages}{1340} (\bibinfo{year}{2024}).

\bibitem{cleve1999share}
\bibinfo{author}{Cleve, R.}, \bibinfo{author}{Gottesman, D.} \&
  \bibinfo{author}{Lo, H.-K.}
\newblock \bibinfo{title}{How to share a quantum secret}.
\newblock \emph{\bibinfo{journal}{Physical Review Letters}}
  \textbf{\bibinfo{volume}{83}}, \bibinfo{pages}{648--651}
  (\bibinfo{year}{1999}).

\bibitem{knill1997theory}
\bibinfo{author}{Knill, E.} \& \bibinfo{author}{Laflamme, R.}
\newblock \bibinfo{title}{Theory of quantum error-correcting codes}.
\newblock \emph{\bibinfo{journal}{Physical Review A}}
  \textbf{\bibinfo{volume}{55}}, \bibinfo{pages}{900--911}
  (\bibinfo{year}{1997}).

\bibitem{nielsen_chuang}
\bibinfo{author}{Nielsen, M.~A.} \& \bibinfo{author}{Chuang, I.~L.}
\newblock \emph{\bibinfo{title}{Quantum Computation and Quantum Information}}
  (\bibinfo{publisher}{Cambridge University Press},
  \bibinfo{address}{Cambridge}, \bibinfo{year}{2010}), \bibinfo{edition}{10th
  anniversary} edn.

\bibitem{kitaev2003fault}
\bibinfo{author}{Kitaev, A.~Y.}
\newblock \bibinfo{title}{Fault-tolerant quantum computation by anyons}.
\newblock \emph{\bibinfo{journal}{Annals of Physics}}
  \textbf{\bibinfo{volume}{303}}, \bibinfo{pages}{2--30}
  (\bibinfo{year}{2003}).

\bibitem{dennis2002topological}
\bibinfo{author}{Dennis, E.}, \bibinfo{author}{Kitaev, A.},
  \bibinfo{author}{Landahl, A.} \& \bibinfo{author}{Preskill, J.}
\newblock \bibinfo{title}{Topological quantum memory}.
\newblock \emph{\bibinfo{journal}{Journal of Mathematical Physics}}
  \textbf{\bibinfo{volume}{43}}, \bibinfo{pages}{4452--4505}
  (\bibinfo{year}{2002}).

\bibitem{fowler2012surface}
\bibinfo{author}{Fowler, A.~G.}, \bibinfo{author}{Mariantoni, M.},
  \bibinfo{author}{Martinis, J.~M.} \& \bibinfo{author}{Cleland, A.~N.}
\newblock \bibinfo{title}{Surface codes: Towards practical large-scale quantum
  computation}.
\newblock \emph{\bibinfo{journal}{Physical Review A}}
  \textbf{\bibinfo{volume}{86}}, \bibinfo{pages}{032324}
  (\bibinfo{year}{2012}).

\bibitem{bravyi2009nogo}
\bibinfo{author}{Bravyi, S.} \& \bibinfo{author}{Terhal, B.~M.}
\newblock \bibinfo{title}{A no-go theorem for a two-dimensional self-correcting
  quantum memory based on stabilizer codes}.
\newblock \emph{\bibinfo{journal}{New Journal of Physics}}
  \textbf{\bibinfo{volume}{11}}, \bibinfo{pages}{043029}
  (\bibinfo{year}{2009}).

\bibitem{stace2009thresholds}
\bibinfo{author}{Stace, T.~M.}, \bibinfo{author}{Barrett, S.~D.} \&
  \bibinfo{author}{Doherty, A.~C.}
\newblock \bibinfo{title}{Thresholds for topological codes in the presence of
  loss}.
\newblock \emph{\bibinfo{journal}{Physical Review Letters}}
  \textbf{\bibinfo{volume}{102}}, \bibinfo{pages}{200501}
  (\bibinfo{year}{2009}).

\bibitem{hostens2005stabilizer}
\bibinfo{author}{Hostens, E.}, \bibinfo{author}{Dehaene, J.} \&
  \bibinfo{author}{De~Moor, B.}
\newblock \bibinfo{title}{Stabilizer states and {Clifford} operations for
  systems of arbitrary dimensions and modular arithmetic}.
\newblock \emph{\bibinfo{journal}{Physical Review A}}
  \textbf{\bibinfo{volume}{71}}, \bibinfo{pages}{042315}
  (\bibinfo{year}{2005}).

\bibitem{bacon2006operator}
\bibinfo{author}{Bacon, D.}
\newblock \bibinfo{title}{Operator quantum error-correcting subsystems for
  self-correcting quantum memories}.
\newblock \emph{\bibinfo{journal}{Physical Review A}}
  \textbf{\bibinfo{volume}{73}}, \bibinfo{pages}{012340}
  (\bibinfo{year}{2006}).

\bibitem{poulin2005stabilizer}
\bibinfo{author}{Poulin, D.}
\newblock \bibinfo{title}{Stabilizer formalism for operator quantum error
  correction}.
\newblock \emph{\bibinfo{journal}{Physical Review Letters}}
  \textbf{\bibinfo{volume}{95}}, \bibinfo{pages}{230504}
  (\bibinfo{year}{2005}).

\bibitem{kribs2005operator}
\bibinfo{author}{Kribs, D.~W.}, \bibinfo{author}{Laflamme, R.},
  \bibinfo{author}{Poulin, D.} \& \bibinfo{author}{Lesosky, M.}
\newblock \bibinfo{title}{Operator quantum error correction}.
\newblock \emph{\bibinfo{journal}{Quantum Information and Computation}}
  \textbf{\bibinfo{volume}{6}}, \bibinfo{pages}{382--399}
  (\bibinfo{year}{2006}).

\bibitem{kretschmann2008information}
\bibinfo{author}{Kretschmann, D.}, \bibinfo{author}{Schlingemann, D.} \&
  \bibinfo{author}{Werner, R.~F.}
\newblock \bibinfo{title}{The information-disturbance tradeoff and the
  continuity of {Stinespring}'s representation}.
\newblock \emph{\bibinfo{journal}{IEEE Transactions on Information Theory}}
  \textbf{\bibinfo{volume}{54}}, \bibinfo{pages}{1708--1717}
  (\bibinfo{year}{2008}).

\bibitem{beny2010general}
\bibinfo{author}{B{\'e}ny, C.} \& \bibinfo{author}{Oreshkov, O.}
\newblock \bibinfo{title}{General conditions for approximate quantum error
  correction and near-optimal recovery channels}.
\newblock \emph{\bibinfo{journal}{Physical Review Letters}}
  \textbf{\bibinfo{volume}{104}}, \bibinfo{pages}{120501}
  (\bibinfo{year}{2010}).

\bibitem{schumacher1996sending}
\bibinfo{author}{Schumacher, B.}
\newblock \bibinfo{title}{Sending entanglement through noisy quantum channels}.
\newblock \emph{\bibinfo{journal}{Physical Review A}}
  \textbf{\bibinfo{volume}{54}}, \bibinfo{pages}{2614--2628}
  (\bibinfo{year}{1996}).

\end{thebibliography}

\end{document}